\documentclass[twocolumn,trackchanges]{aastex701}
\usepackage{float}
\usepackage{amsmath}
\usepackage{hyperref}

\usepackage[caption=false]{subfig}
\usepackage{hyperref} 
\usepackage{makecell}
\usepackage{array}
\usepackage{dsfont}
\usepackage{gensymb}
\usepackage{longtable}
\hypersetup{
    colorlinks=true,
    linkcolor=blue,
    filecolor=magenta,      
    urlcolor=cyan,
    pdftitle={Overleaf Example},
    pdfpagemode=FullScreen,
    }

\shorttitle{Multi-Modal Stellar Astronomy}
\shortauthors{Kamai, Bronstein, Perets}

\begin{document}

\title{Machine-learning inference of stellar properties using integrated photometric and spectroscopic data}

\author[0009-0008-5080-496X]{Ilay Kamai}
\affiliation{Physics Department, Technion -- Israel Institute of Technology,
Haifa 32000,
Israel}
\email[show]{ilay.kamai@campus.technion.ac.il}

\author[0000-0001-9699-8730]{Alex M. Bronstein}
\affiliation{Computer Science Department, Technion -- Israel Institute of Technology,
Haifa 32000,
Israel}
\affiliation{
Institute of Science and Technology Austria, Klosterneuburg 3400, Austria
}
\email{bron@cs.technion.ac.il}

\author[0000-0002-5004-199X]{Hagai B. Perets}
\affiliation{Physics Department, Technion -- Israel Institute of Technology,
Haifa 32000,
Israel}
\affiliation{ACRO, Open University of Israel, R'anana, 
Israel}
\email{hperets@technion.ac.il}

\begin{abstract}
Stellar astrophysics relies on diverse observational modalities-primarily photometric light curves and spectroscopic data from which fundamental stellar properties are inferred. While machine learning (ML) has advanced analysis within individual modalities, the complementary information encoded across modalities remains largely underexploited. We present DESA (Dual Embedding model for Stellar Astrophysics), a novel multi-modal foundation model that integrates light curves and spectra to learn a unified, physically meaningful latent space for stars. DESA first trains separate modality-specific encoders using a hybrid supervised/self-supervised scheme, and then aligns them through DualFormer, a Transformer-based cross-modal integration module tailored for astrophysical data. DualFormer combines cross- and self-attention, a novel dual-projection alignment loss, and a projection-space eigendecomposition that yields physically structured embeddings.

We demonstrate that DESA significantly outperforms leading unimodal and self-supervised baselines across a range of tasks. In zero- and few-shot settings, DESA's learned representations recover stellar color–magnitude and Hertzsprung–Russell diagrams with high fidelity ($R^2 = 0.92$ for photometric regressions). In full fine-tuning, DESA achieves state-of-the-art accuracy for binary star detection (AUC = $0.99$, AP = $1.00$) and stellar age prediction (RMSE = $0.94$ Gyr). As a compelling case, DESA naturally separates synchronized binaries from young stars, two populations with nearly identical light curves—purely from their embedded positions in UMAP space, without requiring external kinematic or luminosity information. DESA thus offers a powerful new framework for multimodal, data-driven stellar population analysis, enabling both accurate prediction and novel discovery.
\end{abstract}


\section{Introduction}
Understanding the fundamental properties of stars is central to astrophysics, enabling insights into stellar evolution, galactic structure, and the conditions for planet formation. Traditionally, this effort was done using an analysis of different stellar measurements, such as stellar light curves (photometry) and stellar spectra (spectroscopy). While classical spectra analysis is usually used to predict stellar parameters related to absorption and emission lines such as $T_\mathrm{eff}, \log g, v \sin i$ and metallicity $[\mathrm{Fe} / \mathrm{H}]$ \citep{Garcia2016, Wu2014}, classical light curves analysis usually uses spots modulation to find periodicity signatures \citep{Reinhold2013, McQuillan2014, Santos2019, Lu_2020, Santos2021, Reinhold2023, Hattori2025} and magnetic activity  \citep{Mathur2014, Santos2024}. The revolution of deep learning models also affected stellar astrophysics, with various works using data-driven models that learn to predict or classify stellar parameters from observations and simulations. For example, \cite{Blancato2020}, \cite{Claytor2024},\cite{Kamai2025}, and \cite{Claytor2025} used deep learning models to predict rotational period from light curves. \cite{Pan2024} and \cite{Zuo2025} utilized Transformer-based models to predict $\log g$. The use of machine learning in spectral analysis dates back to the pre-deep learning era, with the work of \cite{Bailer-Jones2000}. Since then, many works have been done combining deep learning and stellar spectra, with \cite{Leung2019}, \cite{Bai_2020}, \cite{Olney_2020}, \cite{Bovy2023}, \cite{Li2023}, and \cite{Koblischke2024} as some examples. For a review on machine learning in astronomy, please refer to \cite{Li2025}. 

Although the richness of works at the intersection between machine learning and stellar astrophysics, most of them use a single modality, and we refer to them as unimodal models.
Multi-modality models, on the contrary, try to combine the information from different modalities of the same object. This approach showed great success in NLP and vision with works like CLIP \citep{Radford2021} and its variants. Multimodality is of great importance in astrophysics -- for example, photometry and spectroscopy reveal partial and very different stellar information; thus, accurate estimation of some stellar properties requires information from both modalities. Despite its importance, there are few works that utilize multi-modal models in astrophysics. AstroCLIP \citep{Parker2024} is one of such few attempts. The authors used galaxy images and spectra to train a CLIP-like model that aligns the representations of pre-trained individual encoders. Maven \citep{Zhang_2024} is another example, in which simulated and real light curves and spectra of supernovae were used to create a multimodal model for supernova classification and redshift estimation. In addition, AstroM3 \citep{Rizhko2025} is a model that uses light curves, spectra, and meta-information for stellar variability classification.

Despite these early efforts, no existing model has yet demonstrated the full potential of multimodal integration for stellar astrophysics. In this work, we present DESA -- a large-scale, multi-modal model that unifies spectroscopic and photometric data to produce robust, physically informative stellar embeddings. DESA introduces a novel alignment mechanism, DualFormer, which combines cross-attention, spectral duality constraints, and an eigenspace projection to produce a shared latent representation that preserves complementary information from both modalities. Importantly, we show that DESA’s embeddings are not just useful for parameter inference but also encode rich astrophysical structure that enables meaningful clustering, similarity search, and physical interpretation.

We validate DESA on a wide range of astrophysical tasks, demonstrating both predictive performance and novel discovery capabilities. In zero- and few-shot experiments, DESA recovers color–magnitude and Hertzsprung–Russell diagrams with high accuracy, outperforming state-of-the-art unimodal and self-supervised baselines. Fine-tuning on challenging problems like binary detection and stellar age estimation yields substantial improvements over existing methods. Moreover, DESA’s latent space naturally separates stellar populations such as synchronized binaries and young stars—two classes with nearly identical observables—without requiring external labels or kinematic measurements. These results illustrate that DESA is not only a high-performing model but also a general-purpose tool for discovering and interpreting stellar populations from large, heterogeneous surveys.
A schematic description of DESA is depicted in the upper panel of Figure \ref{fig:architecture}. In the sequel, we describe the DESA model, compare it with previous studies, and show its many advantages and successes.  

The paper is organized as follows: In Section \ref{sec:related_work}, we give general background on multi modal self-supervised learning; in Section \ref{sec:model} we present the details of the DESA model; in Section \ref{sec:data} we discuss the dataset and pre-processing steps; in Section \ref{sec:implementation} we discuss implementation details; in Section \ref{sec:results} we shows the results of our model on various tasks and compare them to various baselines. Finally, Section \ref{sec:conclusions} summarizes the findings and discusses future directions.

\section{Multi-Modal Self-Supervised Learning}
\label{sec:related_work}
Multi-modal self-supervised learning is a subfield of self-supervised learning that focuses on leveraging multiple observations (modalities) of the same object, without relying on explicit labels. For example, this could involve aligning an image with its corresponding description, where the two modalities—vision and language—represent the same object. In the context of astrophysical measurements, multimodality typically refers to different surveys observing the same objects, such as a light curve and a spectrum (as discussed in this work) or a spectrum and image, as in \cite{Parker2024}. The key advantage of multimodality over unimodality is the ability to uncover inter-modality relationships that may not be evident within each modality individually. These relationships can enhance performance on complex tasks that require information from both modalities. Given the surge in data volume and diversity across many scientific fields, coupled with a lack of corresponding labels, it is unsurprising that multimodal self-supervised learning has gained significant popularity in various domains \citep{Brown2025_alpha_earth, cui_2025}.
The primary challenge in multimodal learning lies in aligning the different modalities. There are numerous approaches to address this challenge—some specifically designed for multimodality, while others are more general self-supervised techniques. In the following, we discuss two of the most widely used self-supervised methods that can also be applied to alignment in multimodal scenarios.

\subsection{Contrastive alignment}
Contrastive methods are a subset of self-supervised learning (SSL) techniques that utilize the idea of `positive' and `negative' pairs, aiming to create an embedding space where the positive pairs are close to each other, while the negative pairs are pushed apart. This is performed by using special architectures and contrastive loss functions. Contrastive methods demonstrated great success in the computer vision domain with particularly influential works like SimCLR \citep{Chen_simclr2020} and MoCo \citep{He2019}.
While very popular, classical contrastive methods have drawbacks: they require a large batch size to sufficiently represent negative samples at each iteration. Also, the distinction between `positive' and `negative' pairs is a simplified assumption over the data. This is especially important in domains like astrophysics, where the transition between positive and negative pairs is often continuous. In addition, there might be more than one pair of the same object. In particular, in astrophysical data, having more than one spectrum for the same object is very common. Another common issue is the collapse phenomenon that occurs when the model ignores the inputs and creates identical and constant, trivially similar output vectors. Many works have tried to mitigate those drawbacks in different ways. \cite{Chen2020} suggested a model called SimSiam with a modification to the classical infoNCE loss used in SimCLR. In SimSiam, the propagation of gradients is performed only through one side of the network. The authors showed experimentally that using this `stop gradient' mechanism allowed the use of the positive pairs only while avoiding the collapse. MoCo uses a dynamical queue and a momentum-based encoder to reduce the need for a large batch size. Similarly, BYOL \citep{Grill2020} applies a slowly-moving average update to one of the encoders. In the context of multimodality, CLIP \citep{Radford2021} is a pioneering work that uses a contrastive approach to align text and images. As of today, contrastive methods are still among the most popular SSL frameworks for both unimodal and multimodal learning. As an example, both AstroCLIP \citep{Parker2024}, Maven \citep{Zhang_2024}, and AstroM3 \citep{Rizhko2025} were trained using CLIP-like contrastive methods.

\subsection{Regularized alignment}
A different line of work focuses on feature-level discrimination rather than instance-level discrimination, as in contrastive methods. This idea is motivated by canonical correlation analysis and was suggested as a self-supervised method by \cite{zhang2021} and \cite{Zbontar2021}. The latter was the motivation of the Variance-Invariance-Covariance Regularization (VicReg) architecture \citep{Bardes2021a}. VicReg applies three different losses to prevent collapse and maintain alignment, ensuring the variance of the embeddings is sufficiently large (variance term), pushing the covariance between features to be the identity matrix (covariance term), and minimizing the $L_2$ distance between pairs of embeddings (invariance term). One advantage of this method is that it does not use negative pairs or asymmetric architectures. In their paper, the authors show that this model outperforms contrastive approaches on both unimodal and multimodal settings. 

\section{Multi-Modal Neural Network for Stellar Astrophysics} \label{sec:model}
We propose a new framework for multimodal learning of stellar astronomy. Our model comprises two parts: individual encoders and an alignment module. In what follows, we present both components. 

\subsection{Hybrid training of individual modalities} \label{subsec:individual_encoders}
Similar to \cite{Parker2024} and \cite{Zhang_2024}, we start by training individual modalities separately. However, instead of only self-supervised training, we use a hybrid framework. The hybrid framework adds a supervised head to the self-supervised framework and trains the model with the following loss function - 
\begin{align}
    \mathcal{L}_\mathrm{hybrid} = (1-\lambda)\mathcal{L}_\mathrm{ssl} + \lambda \mathcal{L}_\mathrm{sup}
    ,
\end{align}
where $\lambda$ is a hyperparameter that balances supervision and self-supervision. The idea of using hybrid training comes from the fact that there are some stellar parameters which are known with good accuracy, and we can use those parameters to `guide' our model into more physical representations. This idea is not new, and was already used in \cite{Walmsley2022} for a unimodal galaxy model. In our case, the main role of supervision is to create good features at the end of each encoder, rather than the accuracy of the labels themselves. Nevertheless, we show that using this approach leads to state-of-the-art performance in nearly all tasks.

\begin{figure}
    \centering
    \begin{minipage}[b]{\columnwidth}
        \centering
        \includegraphics[width=\textwidth]{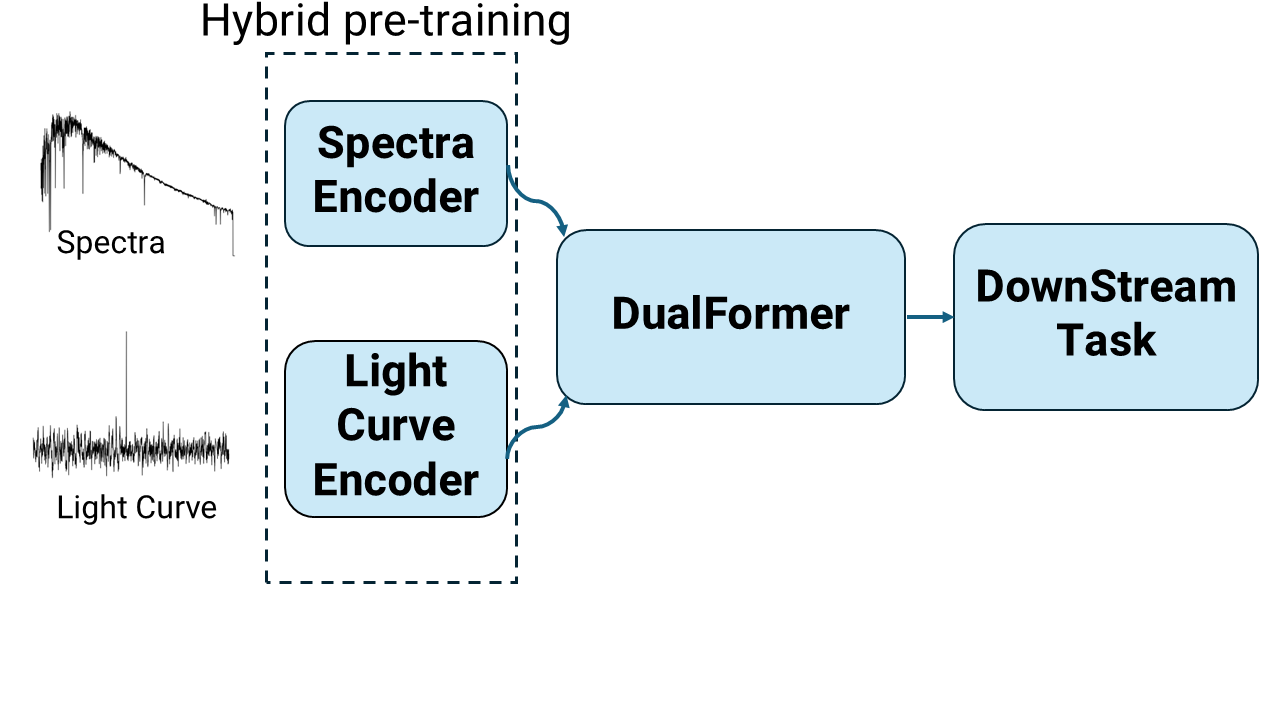}
    \end{minipage}
    \hfill
    \begin{minipage}[b]{\columnwidth}
        \centering
        \includegraphics[width=\textwidth]{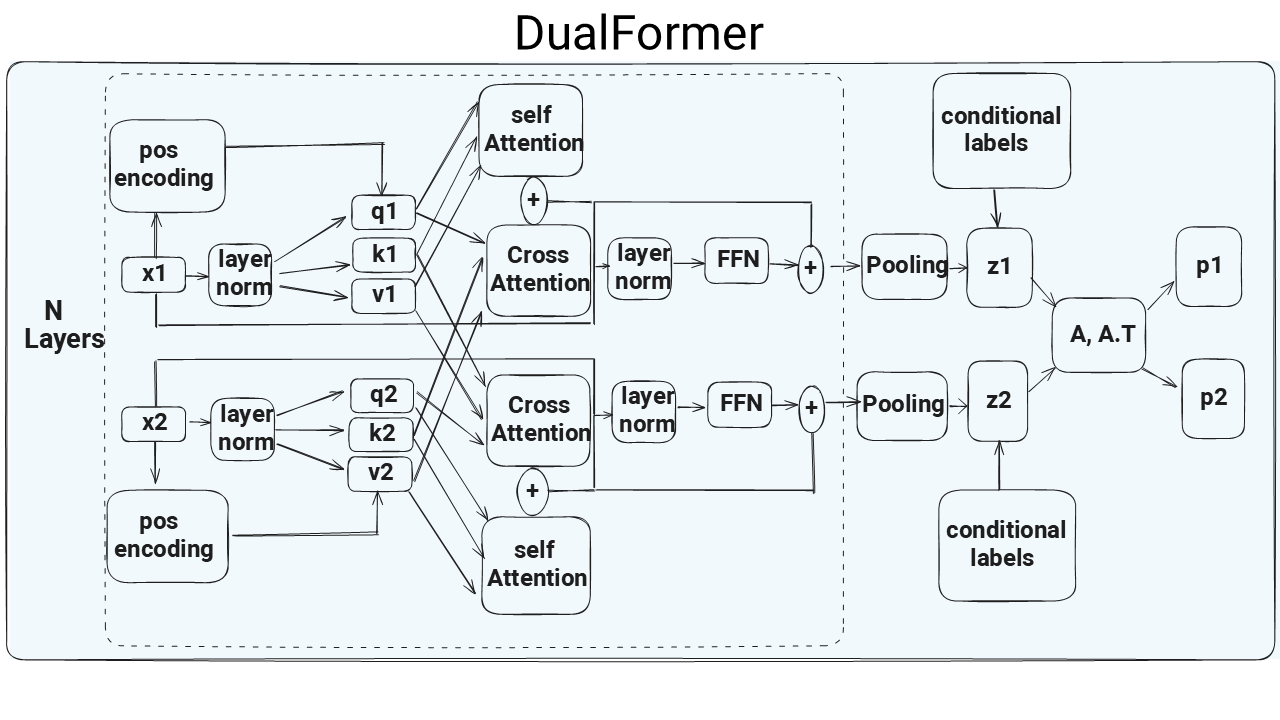}
    \end{minipage}
    \begin{minipage}[b]{\columnwidth}
        \centering
        \includegraphics[width=\textwidth]{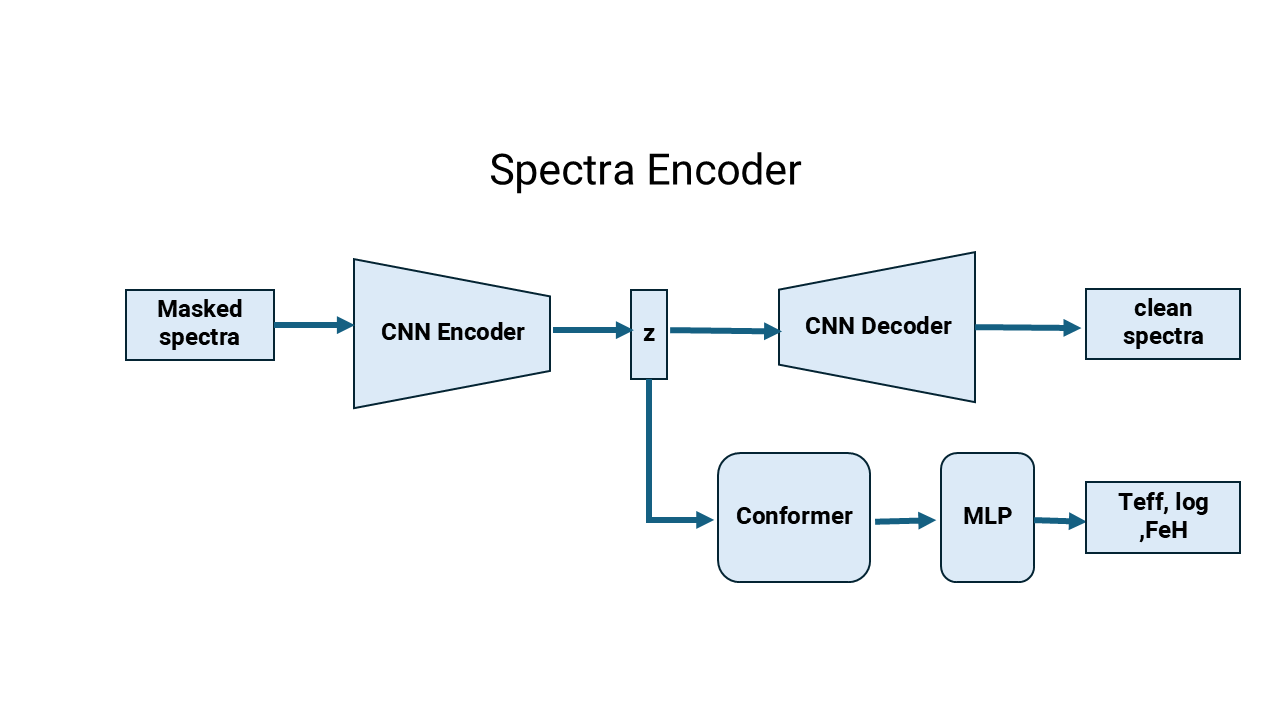}
    \end{minipage}
    \begin{minipage}[b]{\columnwidth}
        \centering
        \includegraphics[width=\textwidth]{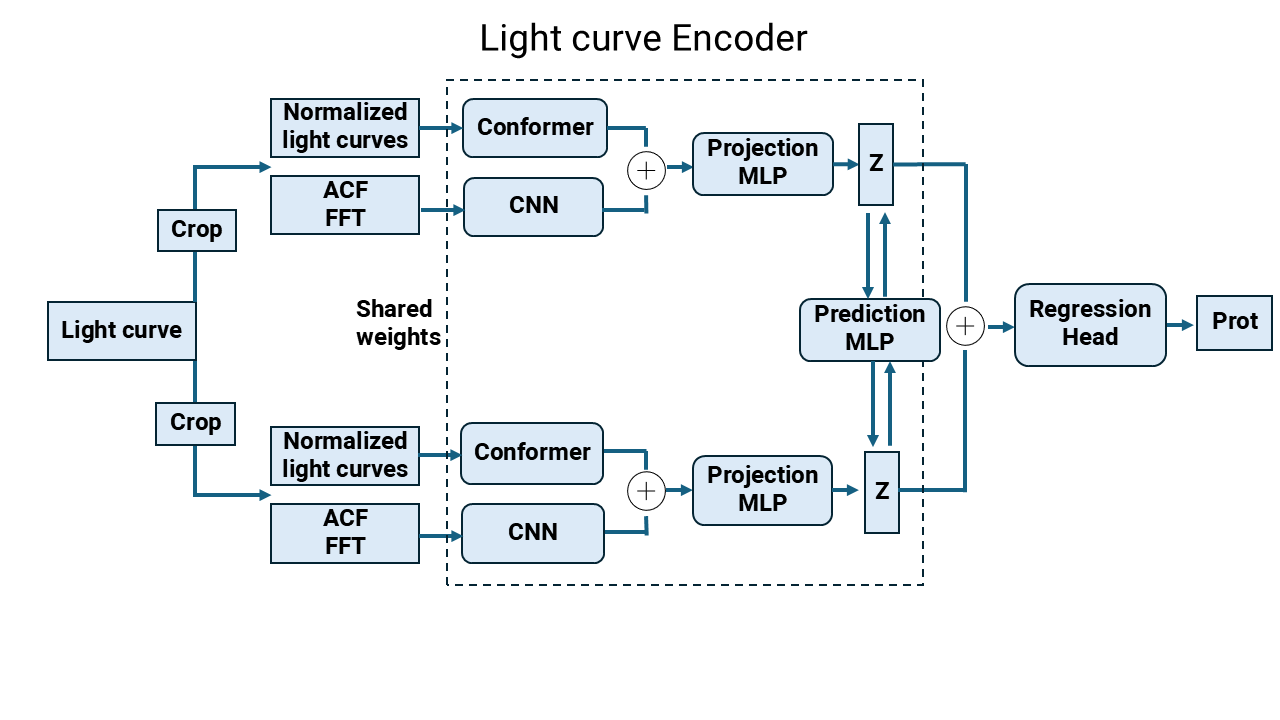}
    \end{minipage}
    \caption{Upper panel - High-level diagram of the entire model. Lower panels - Detailed diagrams of the DualFormer module, the spectra encoder, and the light curve encoder.} 
    \label{fig:architecture}
\end{figure}

\begin{figure}
    \centering
    \begin{minipage}[b]{\columnwidth}
        \centering
        \includegraphics[width=\textwidth]{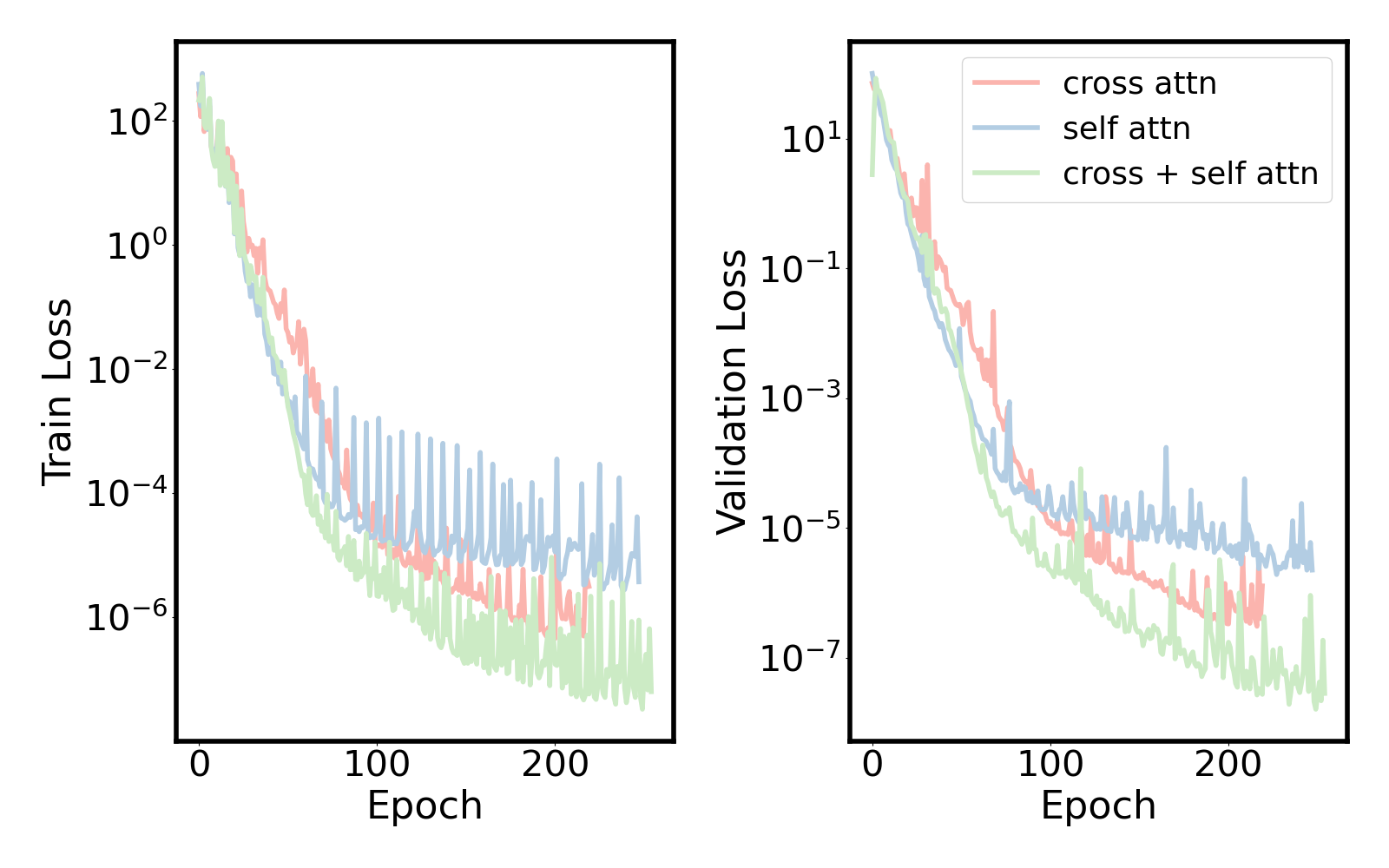}
    \end{minipage}
    \hfill
    \begin{minipage}[b]{\columnwidth}
        \centering
        \includegraphics[width=\textwidth]{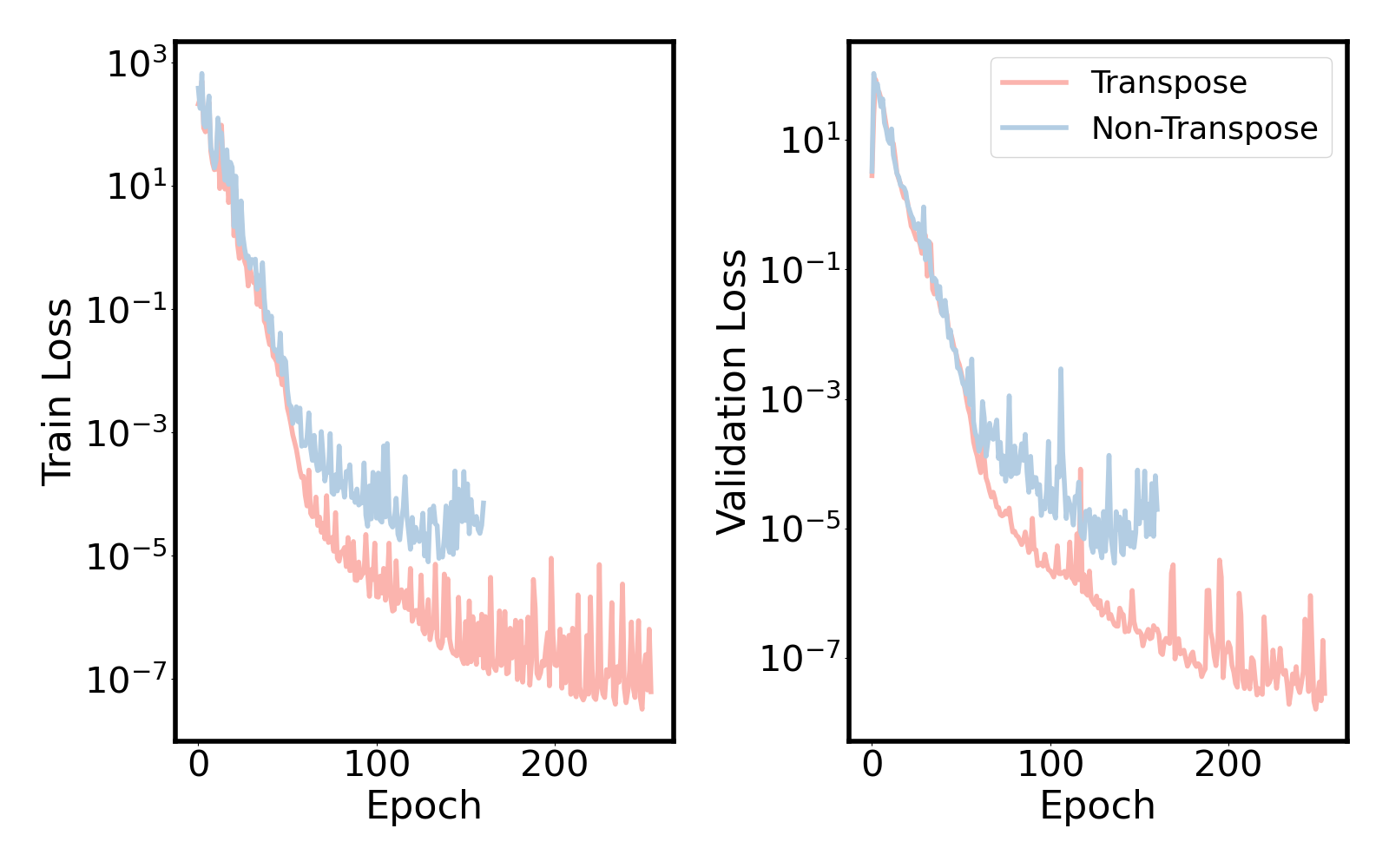}
    \end{minipage}
    \caption{Ablation study results. Upper panel:  different attention mechanisms. Lower panel: different uses of $A$ -- with and without $A^T$ for one of the projections (refer to Section~\ref{subsec:dualformer} for details).}
    \label{fig:ablation}
\end{figure}

\subsubsection{Spectra encoder}\label{subsec:spectra_encoder}
To train the spectra encoder, we start with a CNN encoder followed by two different channels, one for the self-supervised task and one for the supervised task. The self-supervised approach for spectra was chosen to be masked filling. As such, the input is a masked spectrum where $15\%$ of the points are masked and replaced with either a pre-defined value ($80\%$ of the points) or a random value ($20\%$ of the points). The masked spectra are processed through the CNN encoder and then split into two branches: a CNN decoder that produces a filled spectrum, and a branch that consists of a Conformer \citep{Gulati2020} followed by a 2-layer multi-layer perceptron (MLP) which predicts $T_\mathrm{eff}$, $\log g$ and $[\mathrm{Fe} / \mathrm{H}]$. Conformer is a Transformer-based architecture with an added convolution between the multi-head self-attention layers. It was shown to be effective in capturing both global and local information, and was used by \cite{Pan2024} and \cite{Kamai2025} in their models. As suggested in \cite{Pan2024}, we modified the standard Conformer by using rotary position embedding (RoPE) \citep{Su2021} instead of sinusoidal positional encoding. We chose $\mathcal{L}_\mathrm{ssl}$ to be the mean squared error (MSE) between the masked and filled spectra. $\mathcal{L}_\mathrm{sup}$ is a conformalized quantile regression (CQR) loss \citep{Romano2019}. CQR is a form of quantile regression loss that mitigates one of the biggest problems of standard quantile regression -- wrong prediction intervals. In CQR, quantile regression is combined with conformal prediction to create statistically calibrated confidence intervals. This is done by first calculating a `conformity score', which quantifies the error at each interval. This error is then used to calibrate the intervals on the test set. 

One important aspect of stellar measurements is their signal-to-noise ratio (SNR): low-SNR samples are much harder to analyze. Ideally, we would like our model to be aware of those samples and attribute to them lower importance. To incorporate this information, we used the signal-to-noise ratio (SNR) of the spectra as a weight of the final loss during training. This ensures that low-SNR samples have less importance compared to samples with high SNR during training.

\subsubsection{Light curve encoder}\label{subsec:lightcurve_encoder}
First, we preprocessed the light curves. Preprocessing is crucial because the raw Kepler measurements are noisy and uncalibrated. In addition, light curves themselves may not always be the best input for some tasks. For example, the period of a strictly periodic signal would be much more easily detected in the frequency domain than in the time domain (see for example discussion in \cite{Claytor2022}). One complication with light curve pre-processing is that different methods highlight specific information but usually hinder other types of information. For example, standard z-score normalization enhances periodicity but hinders activity-related patterns. Since we aim to create a general encoder, we decided to stack different normalizations and transformations into multiple input channels. We specify the details of light curve preprocessing in section \ref{subsec:lightcurve_preprocessing}. \\
Since we trained the light curve encoder in a contrastive-hybrid method, each light curve was augmented into two different views by cropping it at different times. For each view, we calculate the Autocorrelation Function (ACF) and the Fast Fourier Transform (FFT), which potentially encode important information (e.g., identification of periodic and quasi-periodic behaviors arising from stellar rotation of, e.g., eclipsing binaries), and add them as additional channels. Each view is then processed by a backbone that consists of two separate channels: a CNN encoder and a Conformer module. The ACF and FFT channels are sent to the CNN encoder, and the normalized light curve is sent to the Conformer encoder. The intuition behind this choice is that, in the frequency domain (ACF and FFT channels), periodicity is detected through peak detection. These tasks are 'short-range' and are typical for CNNs. Extracting meaningful information from the light curve requires some level of denoising and is more 'long-range'. Therefore, it is better suited for Transformers \citep{Morvan2022}. This architectural choice is similar to the one presented in \cite{Kamai2025}. They used the same Conformer encoder and an LSTM instead of CNN, and showed that using both the ACF and the normalized light curve together is better than using each one alone for period detection (see Table 2 in their paper). The embeddings are then combined and processed using a SimSiam method. The SimSiam framework can be described as follows: given the two outputs $e_1$ and $e_2$ of the backbone network, we apply an MLP network, $f$, such that $z_1= f(e_1)$, $z_2 = f(e_2)$. We then project $z_1$ onto $z_2$ using another MLP $g$. The loss function is the cosine similarity between the projected and the unprojected features:
\begin{align}
    \mathcal{D}(z_1,z_2) = \frac{\langle g(z_1), z \rangle }{\lVert g(z_1)\rVert\, \lVert z_2\rVert},  
    \label{eq:asym_loss}
\end{align}
where $\langle \cdot , \cdot \rangle$ and $\lVert \cdot \rVert$ denote, respectively, the standard Euclidean inner product and the $\ell_2$ norm. One important aspect of this loss is that the gradients flow only through $z_1$ and not through $z_2$. In the SimSiam paper, the author refer to this as a `stop-gradient' mechanism. The final self-supervised loss is a symmetrized version of (\ref{eq:asym_loss})
\begin{align}
    \mathcal{L}_\mathrm{ssl} = \frac{1}{2}\mathcal{D}(z_1,z_2) + \frac{1}{2}\mathcal{D}(z_2,z_1).
\end{align}
We add a two-layer MLP network on top of the final features to predict the rotational period, $P_\mathrm{rot}$, when available, with $\mathcal{L}_\mathrm{sup}$ as CQR. 

\subsection{DualFormer}\label{subsec:dualformer}
The next step is to combine the embeddings from the pre-trained individual encoders. Here we are using a novel approach specifically tailored for multi-modality in stellar astrophysics. This is motivated by the observation that the information relationships between light curves and spectra of stars are very different from those found in language and vision modalities. While text describes its corresponding image and vice versa, neither the light curve nor the spectra describes the other. They both partially describe the star, each from a very different perspective. Intuitively, light curves and spectra can be seen as orthogonal views of the star, as the former uses the time domain while the latter uses the frequency domain. Of course, the measurements are not mathematically orthogonal, since the measurements come from different surveys, with different bands and sensitivities, and can be taken at very different times, which complicates the relationship. Nevertheless, this intuition does not exist in other sets of modalities, such as image-text or image-spectra. Moreover, in both text and images (as well as spectra and images such as those in AstroCLIP), the dynamics of the system is not manifested in the data. In contrast, light curves measure time-dependent phenomena by design. This creates a time-dependent information relationship in the case of light curve and spectra alignment. 

Another uniqueness of astronomical data, not related to specific modalities, is the importance of prior knowledge. In astrophysics, one usually has some extra information about the observed objects. This can be, for example, stellar parameters that are known with good accuracy (inferred using classical analysis methods or previous unimodal ML models). As mentioned in Section \ref{subsec:individual_encoders}, this information can be used to train individual encoders, but it can also be crucial during the alignment process, since this information is modality-invariant. \\
These differences suggest that standard multi-modal approaches might not be sufficient in our scenario, and that a specific model is needed.

The lower panel of Figure \ref{fig:architecture} presents a diagram of the alignment module which we call DualFormer. The inputs are the final features from the light curve and spectra encoders. They are first processed in a Transformer-like module with a modified multi-head self-attention, but instead of self-attention, we use both self-attention and cross-attention, where the former focuses on intra-modality relationships, while the latter focuses on cross-modality relationships. This results in two feature branches with mixed information.

Next, we aggregate the information using average pooling, add conditional prior information, and project both features through the same linear layer denoted by the matrix $A$. This layer is the effective bottleneck of the network and should store the important shared information. Specifically, we use $A$ to project the features, with one branch projected using $A$ and the other branch using $A^T$. If the features were truly orthogonal, such a transformation would collineate them. 
To align the features while preventing collapse, we follow \cite{zhang2021} and \cite{Bardes2021a} with some important modifications. We adopt the same covariance loss that decorrelates the features,
\begin{align}
    \mathcal{L}_\mathrm{cov}(x, x') = \frac{1}{d} \sum_{k \neq l} [\mathrm{Cov}(x,x')]^2_{kl},
\end{align}
where $d$ denotes the vector dimension.
However, unlike the original authors, we use the loss inside each branch and between branches; specifically, we decorrelate the features after projection, $p_1=Az_1$ and $p_2=A^Tz_2$ in Figure \ref{fig:architecture}. The full covariance loss assumes the form
\begin{align}
    \mathcal{L}_\mathrm{cov} = 
    \mathcal{L}_\mathrm{cov}(p_1, p_1) + \mathcal{L}_\mathrm{cov}(p_2, p_2) + 
    \mathcal{L}_\mathrm{cov}(p_1, p_2).
\end{align}
In addition, instead of a point-wise MSE loss between features, we use the following loss term: 
\begin{align}
    \mathcal{L}_\mathrm{dual} = \| \langle z_1, p_1 \rangle - \langle z_2, p_2 \rangle \|^2.
\end{align}
$\mathcal{L}_\mathrm{dual}$ can be seen as a less constrained version of the invariance term from \cite{Bardes2021a}: note that the loss is minimized when the quadratic forms $z_1^TAz_1$ and $z_2^TA^Tz_2$ are equal.
We see that while the standard invariance term requires $z_1$ and $z_2$ to be identical vectors, $\mathcal{L}_\mathrm{dual}$ does not even require them to lie on the same hyper-surface (since in general $A \neq A^T$). This gives much more freedom for $z_1$ and $z_2$ to be different, but constrains the projections, namely $A$, and $A^T$. This is also why we chose to decorrelate the features after projection in $\mathcal{L}_\mathrm{cov}$. Since $A$ is not necessarily Hermitian, we expect the meaningful information to be stored in a shared vector space of $A$ and $A^T$. Ideally, this would be the eigenspace of $A$. In Section \ref{sec:results}, we show that this is indeed the case and that the eigenspace of $A$ is an effective latent space that stores all the relevant information, as anticipated. Lastly, we do not use the variance loss term, as in \cite{Bardes2021a}, since $\mathcal{L}_\mathrm{dual}$ requires fewer constraints on features, making the high-variance requirement redundant. The full training loss of the DualFormer is therefore 
\begin{align}
    \mathcal{L} = (1-\lambda)\mathcal{L}_\mathrm{dual} + \lambda \mathcal{L}_\mathrm{cov},
\end{align}
where we set $\lambda = 0.5$.

Although this motivation may sound reasonable, it is not guaranteed to perform well in practice. We therefore test the architectural choices of DualFormer using an ablation study. Figure \ref{fig:ablation} shows two such studies. The upper panel shows a comparison of different attention mechanisms: self-attention (blue), cross-attention (red), and a combination of both (green). It can be seen that using both self- and cross-attention results in better performance in training and validation losses, as we might expect. The lower panel compares two cases: the first where, as described before, $A$ and $A^T$ are used to project the two features (red), and the second where $A$ is used to project both features (blue). Again, we see that the chosen architecture (using $A$ and $A^T$) outperforms the alternative. \\
A general concern in multi-modal models is that one modality will obscure the information from another modality. In our case, this might happen because the number of spectra samples used in the pretraining phase is $\sim30$ times larger than the number of light curves. As a consequence, the spectra encoder is larger than the lightcurve encoder by roughly the same factor (see section \ref{sec:implementation} for details), and we might have a situation where all the information in the final features comes effectively from the spectra features. First, we need to remember that for basic stellar properties, a spectrum is more informative than a light curve. Indeed, most basic stellar properties (e.g., Teff, FeH, logg, spectral type) are usually detected using spectral lines rather than light curve features. Therefore, it makes sense that there would be more 'spectra-related' information in the final features. However, we want to make sure that the light curve information is not completely overwhelmed by the spectra information and that using the light curve improves the final performance, especially in tasks that require a sophisticated combination of different stellar parameters rather than basic stellar parameters. In sections \ref{subsec:binary_res} and \ref{subsec:age prediction}, we test such tasks and see that using the combined model significantly improved the performance compared to spectra-only or lightcurve-only encoders. Another possible test is a sensitivity analysis, not related to a specific task, that identifies which features are most influenced by each modality.  For that, we performed the following experiment: We calculated the final features of the test set three times. One time with the full input (light curve and spectra), one time with the light curve replaced with zeros (we call this case spectra-only), and one time with the spectra replaced with zeros (lightcurve-only). We then look at the difference between the full features and spectra-only/lightcurve-only features. The difference is calculated per feature and averaged across all samples. This way, we can trace each individual feature and see for which modality (light curve or spectra only) the difference is smaller. This modality is the more dominant for this specific feature. Figure \ref{fig:ablation_features} in Appendix \ref{appendix:sup_graphs} shows the results of such an experiment. It shows the per-feature difference for both lightcurve-only (gray) and spectra-only (red) features. The blue circles mark the indices where the lightcurve-only features are more dominant. There are $75$ circles which correspond to $\sim 30\%$ of the features, an order of magnitude more than the ratio of lightcurve to spectra samples and number of parameters. This is not negligible and supports the results we find in sections \ref{subsec:binary_res} and \ref{subsec:age prediction}. Interestingly, we see that while there are small areas with clear dominance of one modality (around indices $150$, for example), in most cases both modalities seem important. This is especially true for the last indices that correspond to the added prior
information. We conclude that while there is more spectral information in the final features, the contribution of the light curve encoder is not negligible and especially important in complicated tasks that require combining different information types. We note that while this experiment is convincing, it is not as rigorous as a full ablation study where we completely remove different components and retrain the model. Such an experiment could be explored in future work.

\section{data}\label{sec:data}
We train the full model using low-resolution spectra from LAMOST \citep{Zhao2012, Wang2022} \dataset[10.12149/100632]{https://doi.org/10.12149/100632}, and light curves from Kepler \cite{Mathur_2017} \dataset[10.17909/T9488N]{http://dx.doi.org/10.17909/T9488N}.

\subsection{Spectra preprocessing}\label{subsec:spectra_preprocessing}
\begin{figure*}
    \centering
        \centering
        \includegraphics[width=\textwidth]{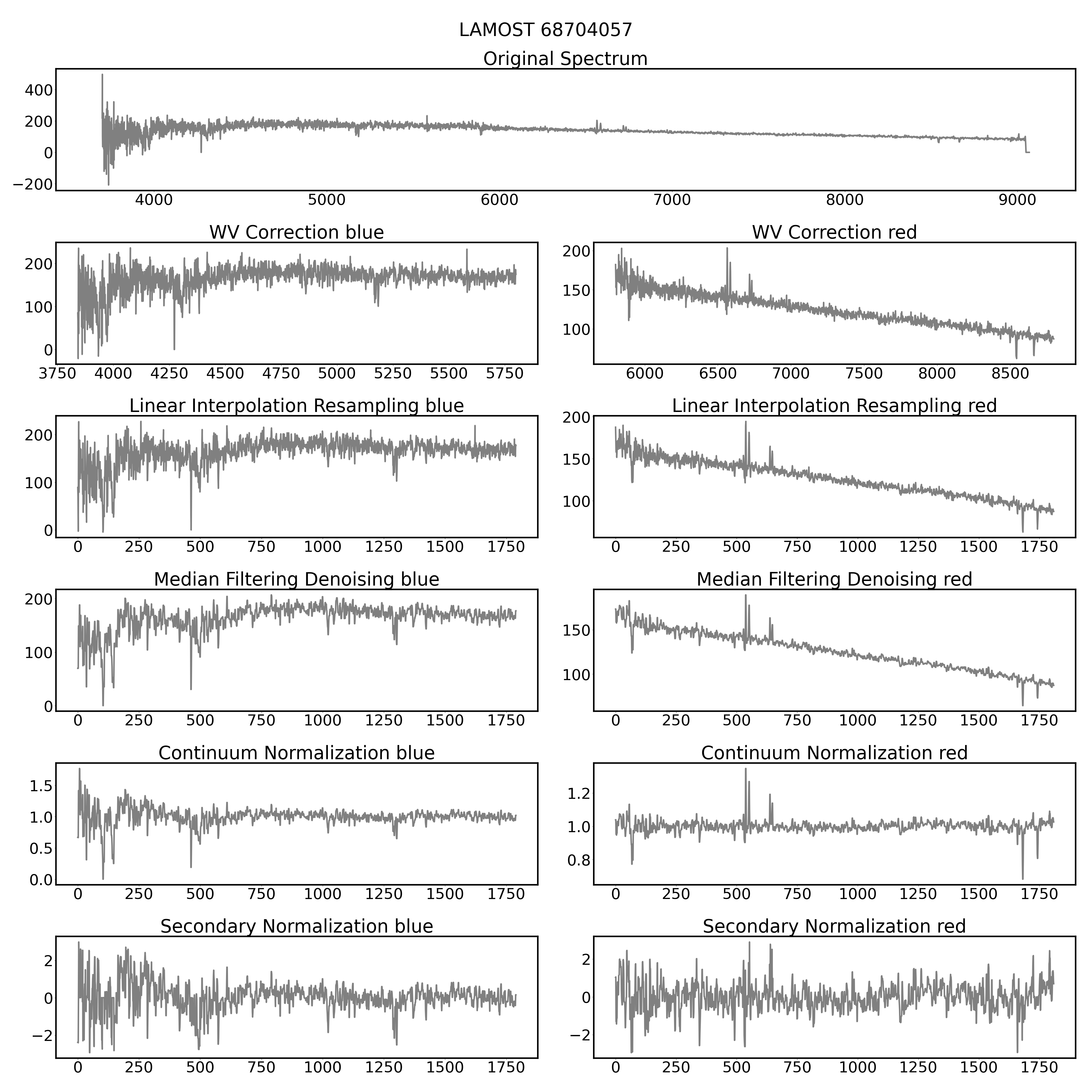}
    \caption{Example of the pre-processing steps for LAMOST spectra. The left column is the blue range, and the right column is the red range. }
    \label{fig:lamost_preprocess}
\end{figure*}

LAMOST is a low-resolution ($\frac{\lambda}{\Delta \lambda} \sim 1800$) spectrometer with two arms: a blue arm that covers a wavelength range of $3700-5900$\AA, and a red arm that covers a wavelength range of $5700-9000$\AA. The LAMOST survey consists of millions of spectra of stars, galaxies, and quasars. We used LAMOST DR8 and downloaded all the spectra of $A,F,G,K,M$ stars with $3000 \leq T_\mathrm{eff} \leq 7500$K. We removed samples without measured $T_\mathrm{eff}, \log g$ or $[\mathrm{Fe} / \mathrm{H}]$ (from the LASP \citep{Wu2014} pipeline), and samples with nonsensical negative SNR. This resulted in about $6.5$M samples. As mentioned in Section \ref{subsec:individual_encoders}, the hybrid approach does not require having supervised labels. Here, we decided to use only samples with labels, given the available large size of the dataset. The preprocessing of spectra is similar to what was suggested in StarGRUNet \citep{Li2023}: we first translate the wavelength to the rest-frame using radial velocity measurements from the LASP pipeline \citep{Wu2014}. Next, we divide the spectrum into blue and red regions and resample each region using linear interpolation. We then apply a median filter to each region, followed by a continuum normalization using a fifth-order polynomial, and a final step where we remove points higher or lower than $\pm 3\sigma$ and normalize the flux to have zero mean and unit standard deviation. For a more detailed explanation of the different pre-processing steps, please refer to \cite{Li2023}. Figure \ref{fig:lamost_preprocess} shows an example of the pre-processing stages of LAMOST spectra. 

\subsection{Light curve preprocessing}\label{subsec:lightcurve_preprocessing}
\begin{figure*}
    \centering
        \centering
        \includegraphics[width=\textwidth]{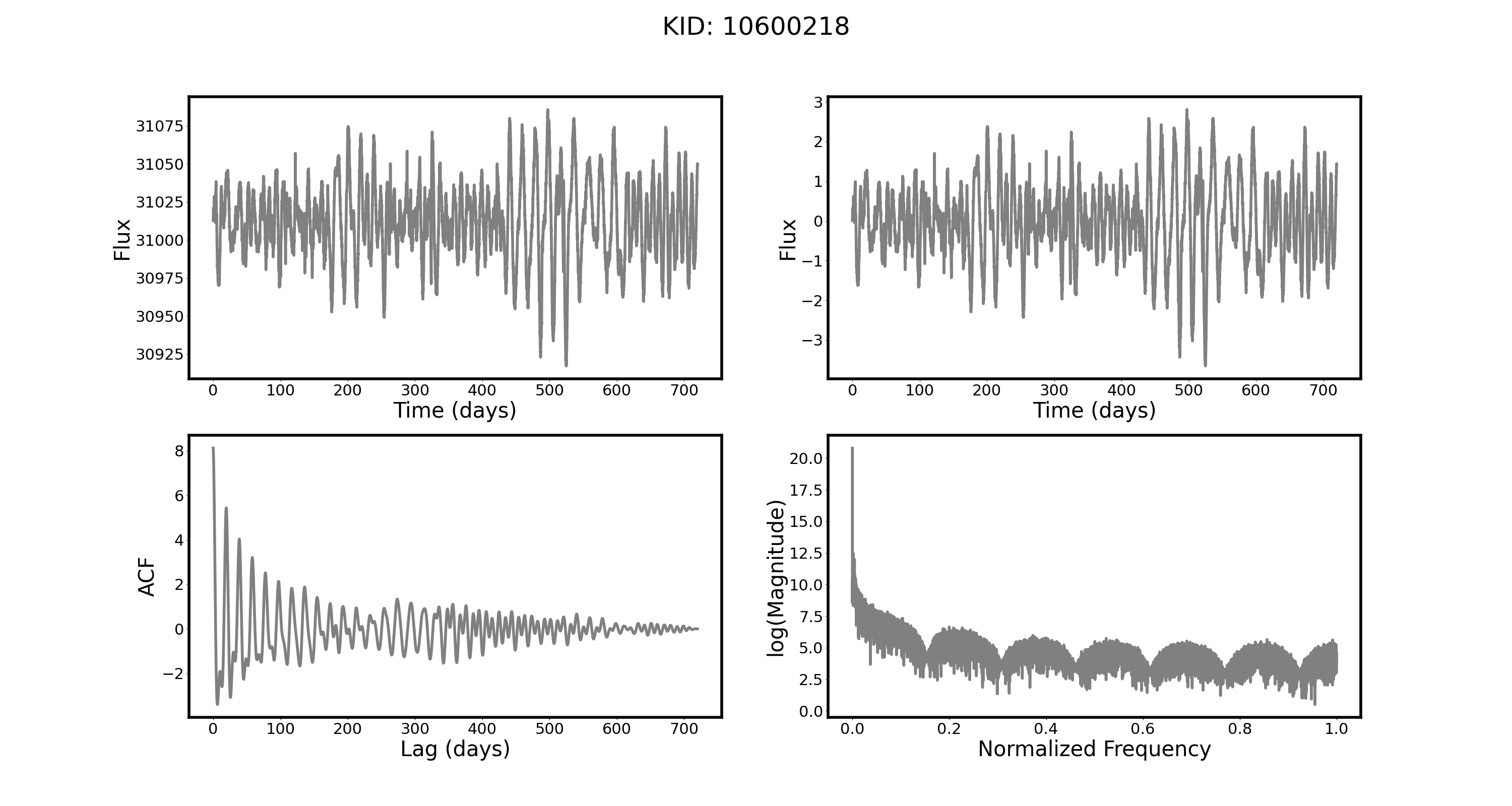}
    \caption{Example of the pre-processing steps for Kepler Light curve. The upper row shows the raw light curve normalized by absolute magnitude (left) and mean and standard deviation (right). The lower row shows the ACF (left) and FFT (right).}
    \label{fig:kepler_preprocess}
\end{figure*}

Kepler was a space mission designed to provide high cadence light curves for stars and for the search of exoplanets through transits. As such, Kepler measured the light curves of around $200000$ main sequence and giant stars for almost $4$ years with a cadence of approx. $30$ minutes. We used Kepler samples with stellar properties from \cite{Berger2020}, resulting in $183,435$ samples. The preprocessing of the light curves follows two different ways: one using the absolute magnitude and the other using the mean and standard deviation. As mentioned in \ref{subsec:lightcurve_encoder}, the reason for the two normalizations is that different properties require different light curve information. The activity and luminosity, for example, are related to the absolute amplitude of the light curve (normalized by the distance). On the other hand, the period is more easily detected using standard normalization, like zero mean and unit standard deviation. The absolute magnitude normalization was done by first dividing the light curve by $2^{-k}$ where $k$ is the absolute magnitude, calculated using the Kepler magnitude, KMAG, and the distance from \cite{Berger2020}. In addition to the raw light curve, as mentioned earlier, we also calculated the FFT and the ACF, which help to determine the period. Figure \ref{fig:kepler_preprocess} shows an example of the pre-processing stages of the Kepler light curve.

\section{Implementation details}\label{sec:implementation}
In big models, like DESA, hyperparameter tuning can be a very challenging task. This becomes even harder when the model consists of two steps - pre-training and alignment. We therefore chose to use a simple heuristic when defining the hyperparameters of our model. As the number of spectra samples is much larger than the number of light curve samples ($6.5$M vs $200$K), we designed the individual encoders such that the spectra encoder has more parameters than the light curve encoder. This is motivated by neural scaling laws, a phenomenological relationship between dataset size, model parameters, and performance that was originally found in the vision and language domains \citep{Kaplan2020, Hoffmann2022}, but recent works showed it also applies to astronomical data \citep{Walmsley2024, Pan2024_scaling}. Therefore, the dimension of the final spectra features was chosen to be $2048$, and that of the light curve was chosen to be $256$. The number of parameters in the spectra and light curve encoders is about $500$M and $11$M, respectively.  During hybrid training, $\lambda$ was chosen arbitrarily to be $0.5$. The embedding dimension in DualFormer was set to $256$, and the number of parameters in this module is also about $11$M. The light curve encoder was trained using a learning rate decay scheduler with the cosine annealing method. The initial learning rate is $2\cdot 10^{-5}$ decreasing to $2\cdot 10^{-6}$. All other modules were trained with a constant learning rate of $2\cdot 10^{-5}$. We trained all modules with AdamW optimizer \citep{Loshchilov2017}. Lastly, we estimated the energy used to train the entire model using CodeCarbon\footnote{\href{https://codecarbon.io/}{https://codecarbon.io/}} package. It is estimated to be $334$ kWh for the entire model, out of which $204$ kWh are for the pretraining stage.  
All the code used for training and experiments is publicly available on \href{https://github.com/IlayMalinyak/DESA} {https://github.com/IlayMalinyak/DESA} and on Zenodo \citep{kamai_zenodo2025}. 

\section{Results}\label{sec:results}
\subsection{Pretraining Results} \label{subsec:pretraining_res}
\begin{figure*}
    \centering
    \setlength{\tabcolsep}{0pt}
    \renewcommand{\arraystretch}{0}
    
    \begin{minipage}[b]{0.32\textwidth}
        \centering
        \includegraphics[width=\textwidth,trim=10 10 10 10,clip]{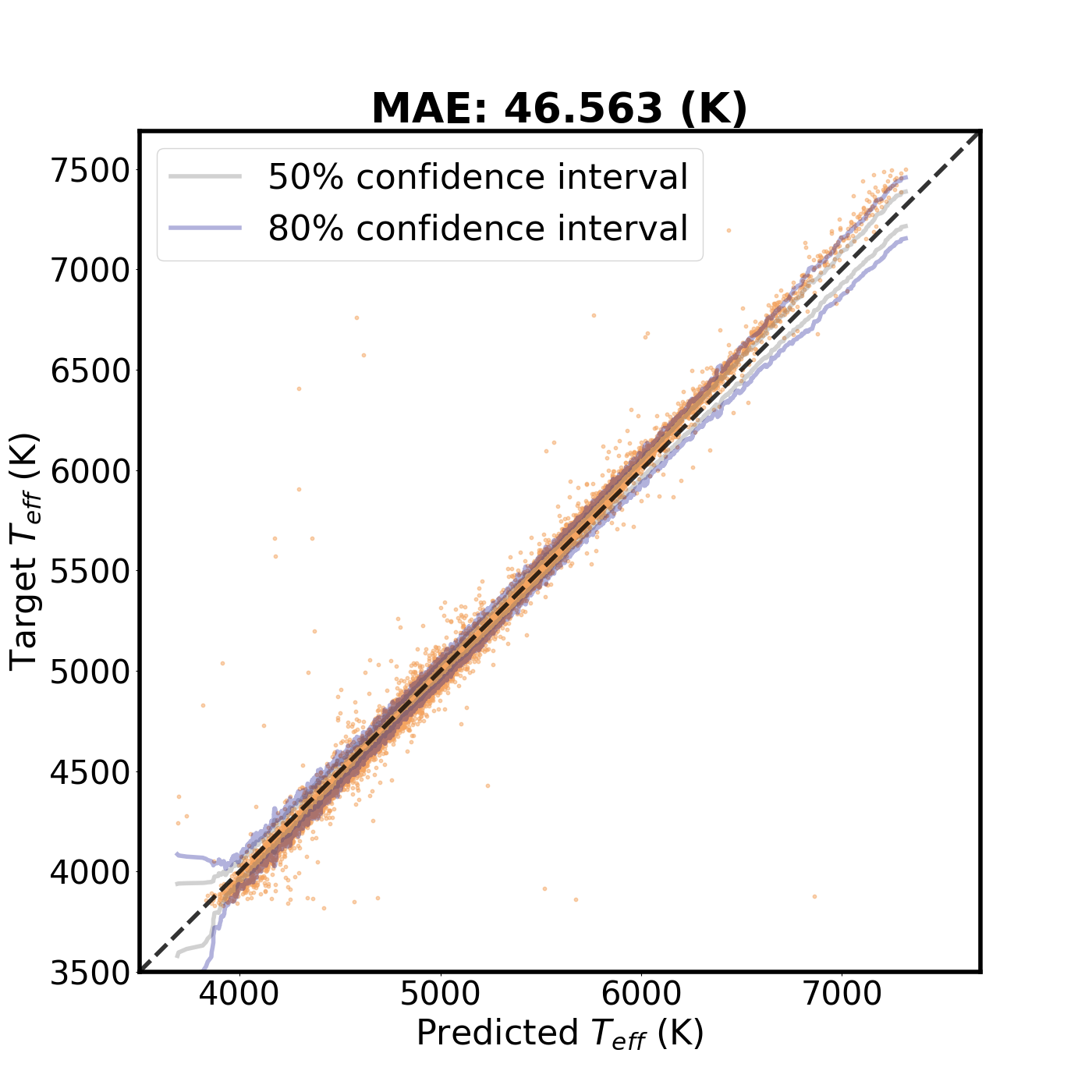}
    \end{minipage}%
    \begin{minipage}[b]{0.32\textwidth}
        \centering
        \includegraphics[width=\textwidth,trim=10 10 10 10,clip]{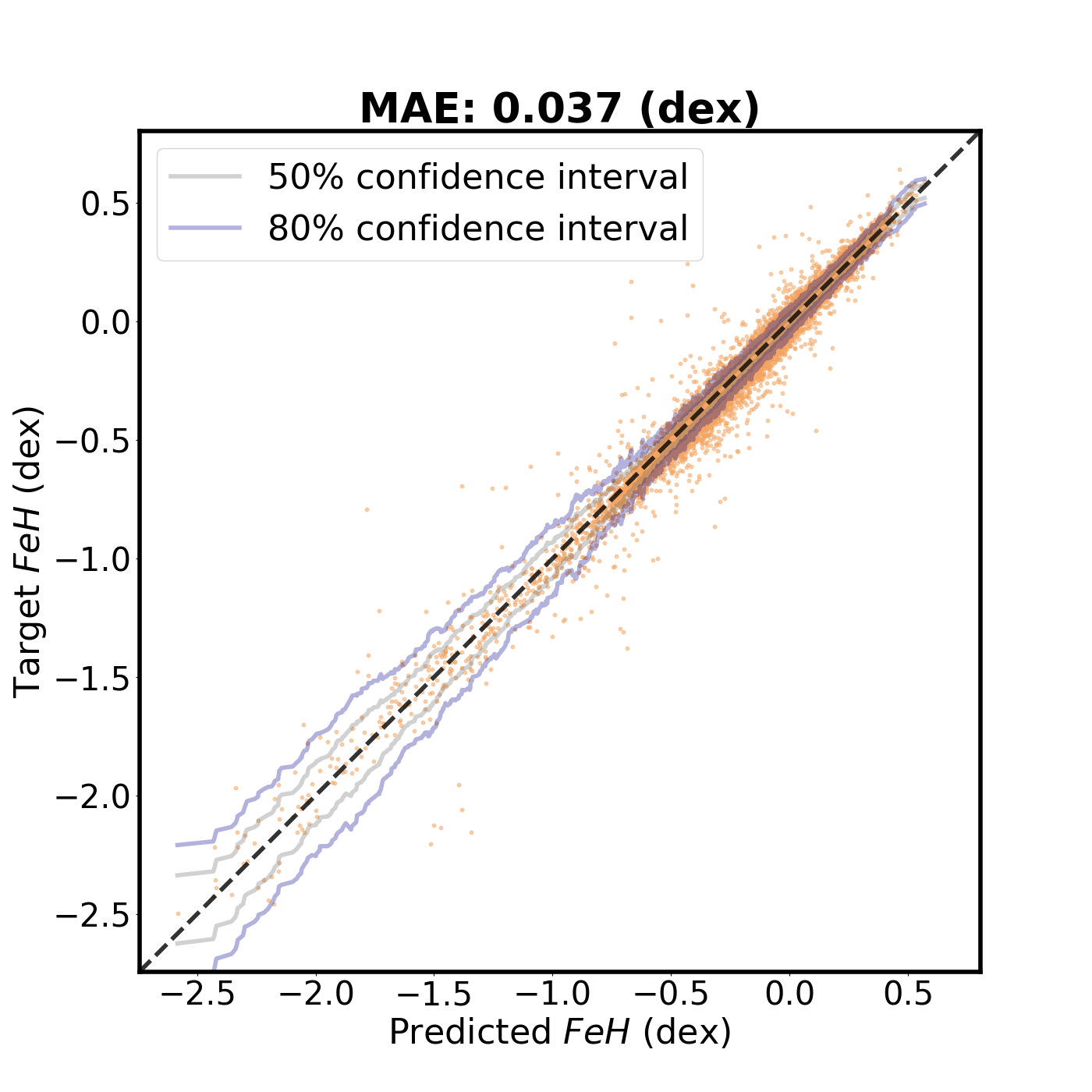}
    \end{minipage}%
    \begin{minipage}[b]{0.32\textwidth}
        \centering
        \includegraphics[width=\textwidth,trim=10 10 10 10,clip]{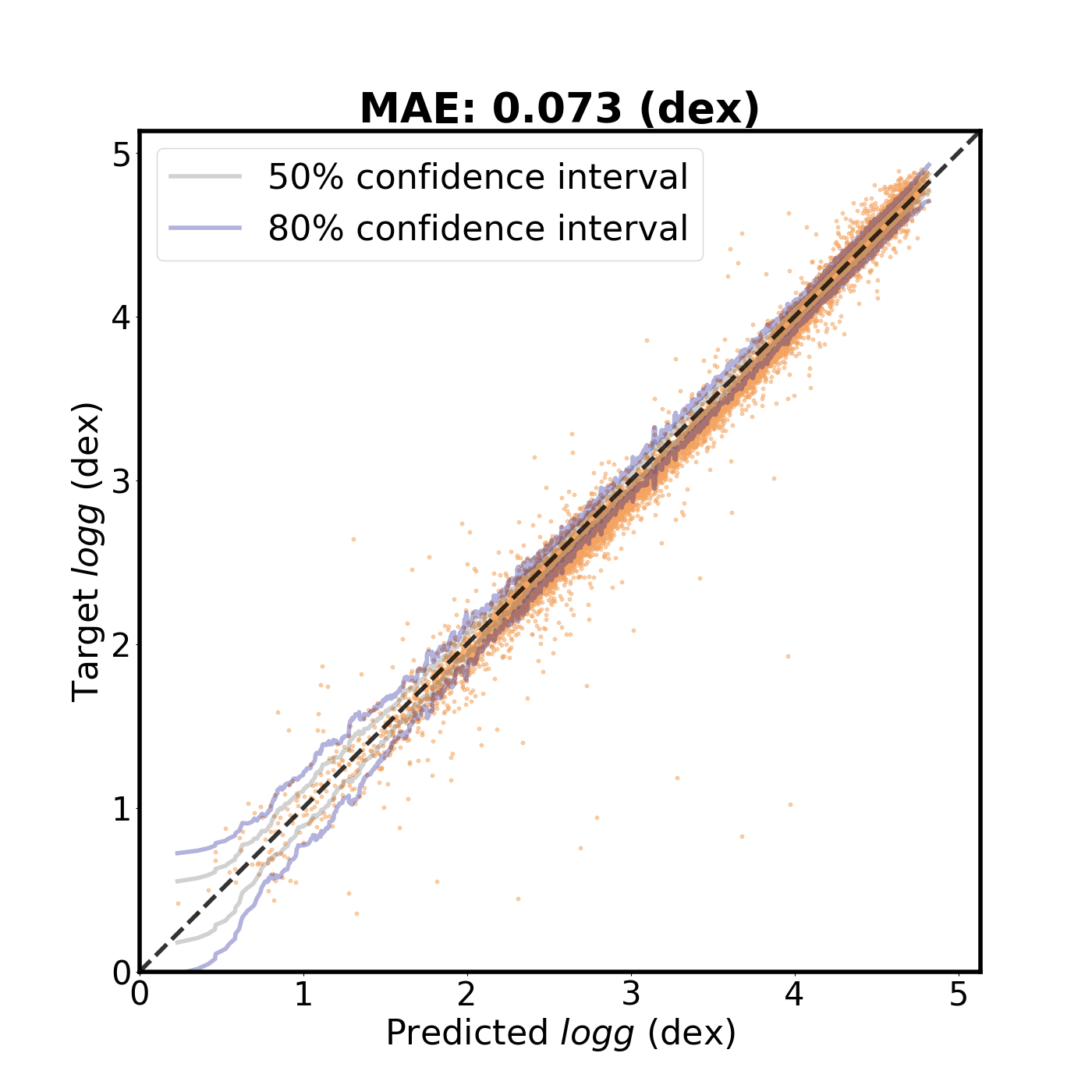}
    \end{minipage} \\[0pt]

    \begin{minipage}[b]{0.32\textwidth}
        \centering
        \includegraphics[width=\textwidth,trim=10 10 10 10,clip]{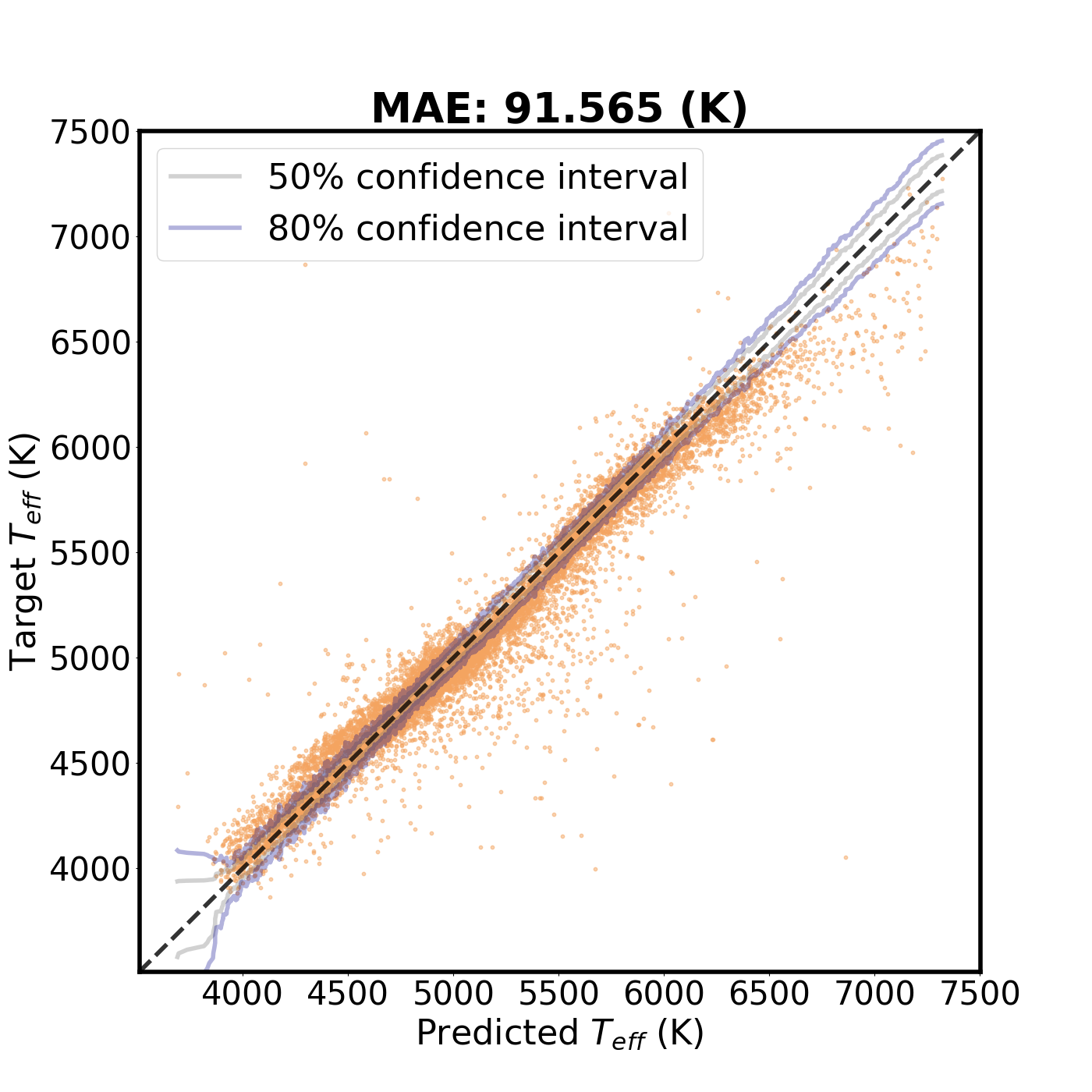}
    \end{minipage}%
    \begin{minipage}[b]{0.32\textwidth}
        \centering
        \includegraphics[width=\textwidth,trim=10 10 10 10,clip]{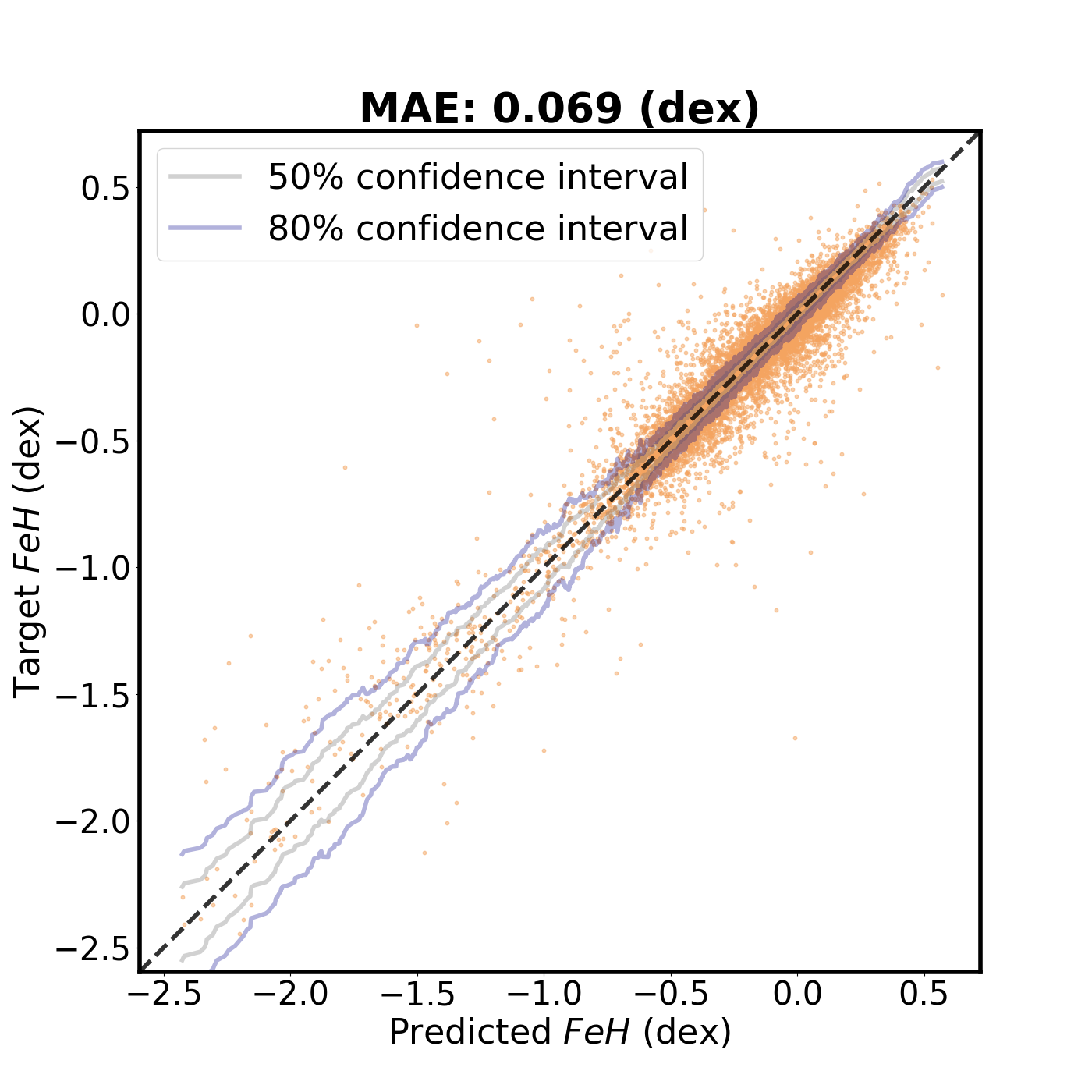}
    \end{minipage}%
    \begin{minipage}[b]{0.32\textwidth}
        \centering
        \includegraphics[width=\textwidth,trim=10 10 10 10,clip]{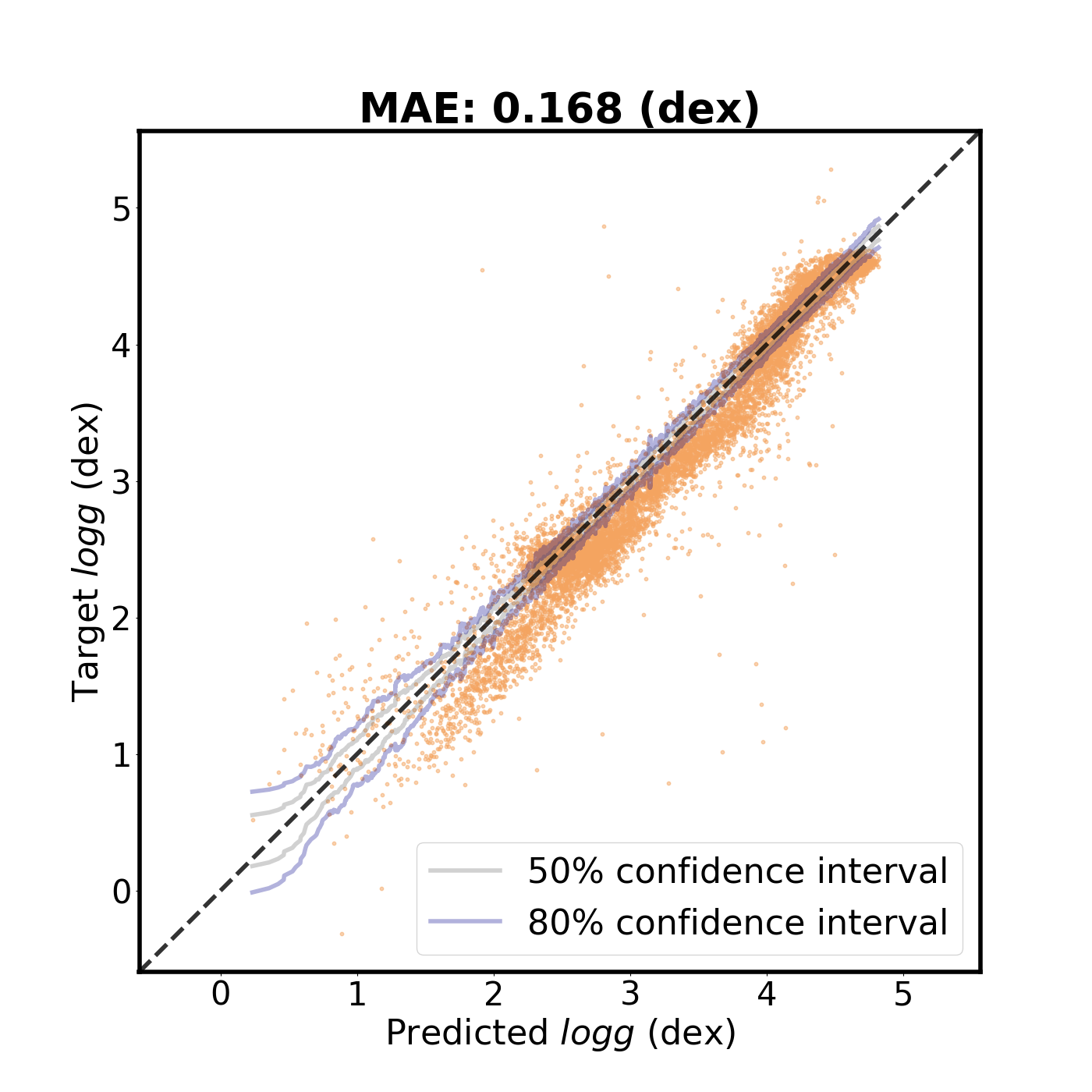}
    \end{minipage}
    
    \caption{Upper panel - results of the spectra encoder on LASP labels. Lower panel - results on the labels from APOGEE. The purple and gray lines represent prediction intervals of $80\%$ and $50\%$.}
    \label{fig:spec_res}
\end{figure*}

First, we present the results of the hybrid pre-training of individual modalities. As mentioned before, the spectra encoder was trained using $T_\mathrm{eff},\log g$ and $[\mathrm{Fe} / \mathrm{H}]$ labels from the LASP pipeline. However, we test the results using labels from the high-resolution APOGEE survey \citep{apogee2022}, at the same temperature range. This is a common practice and was done in previous works such as StarGRUNet \citep{Li2023}. The results on the test set of both LASP labels and APOGEE labels are shown in Figure \ref{fig:spec_res}. For each label, we also plot the $80\%$ and $50\%$ confidence intervals. It can be seen that the mean average error (MAE) on LASP is a factor of $\sim 2$ lower compared to APOGEE for all labels. This is reasonable given the fact that the APOGEE labels come from high-resolution spectra, while the spectra that are given to the model are low resolution. Another interesting difference between the labels is related to the prediction intervals. As mentioned in section \ref{subsec:individual_encoders}, the prediction intervals were calibrated on a held-out validation set, different from the test set. This validation set uses the LASP labels. We see that for the LASP test set, this calibration results in the desired coverage ($80\%$ of the points are inside the $80\%$ interval, for example), while for the APOGEE labels, the coverage is lower. This is another evidence that there is a significant difference between the labels. We also compare our results with StarGRUNet. StarGRUNet \citep{Li2023} used a GRU-based model to predict stellar parameters on a wide range of SNR values (SNR $ > 5$). While there are works with better performance for a specific range of SNR (\cite{Li2022}, for example, who used $10 <$ SNR $<20$), this is the state-of-the-art for a model that uses LAMOST low-resolution spectra and a wide range of SNR values. 

Table \ref{table:spectra_res} shows a comparison between StarGRUNet and DESA. We note that the results are very close. Specifically, our model performs slightly better than StarGRUNet for $T_\mathrm{eff}$ and $[\mathrm{Fe} / \mathrm{H}]$, and slightly worse for $\log g$. However, it is worth mentioning that we used SNR $ > 0$ and StarGRUNet used SNR $ > 5$. This seemed to be a result of the fact that we incorporated the SNR information into the training. In Appendix \ref{appendix:sup_graphs} we show MAE vs different SNRs bins for $T_\mathrm{eff},\log g,$ and $[\mathrm{Fe} / \mathrm{H}]$. We see that the sensitivity of our results to SNR is indeed better than StarGRUNet (see Figure 11 in their paper). However, there is still non-negligible sensitivity for both LASP and APOGEE labels. While it is probably impossible to remove all the dependency on the noise, this might be further investigated and improved in future papers.

\begin{figure}
    \centering
        \centering
        \includegraphics[width=0.8\columnwidth]{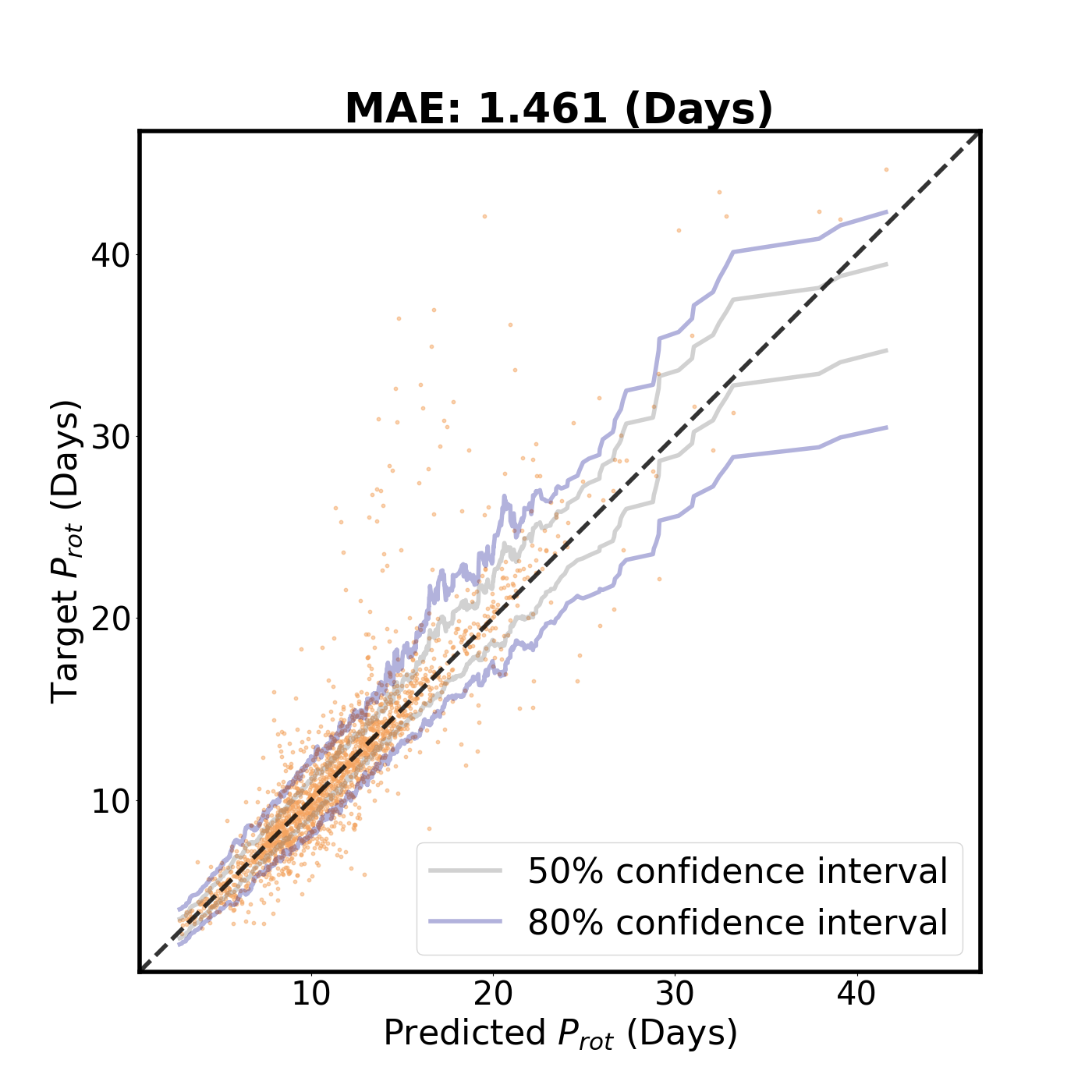}
    \caption{Results of light curve encoder. Gray and purple lines represent $50\%$ and $80\%$ intervals. }
    \label{fig:lightcurve_res}
\end{figure}

The light curve encoder was trained using periods aggregated from all recent catalogs \citep{McQuillan2014, Santos2019, Santos2021, Reinhold2023, Kamai2025}. This results in $104433$ samples with period label. The results are shown in Figure \ref{fig:lightcurve_res}. We can compare the results to the work of \cite{Blancato2020}, which trained a CNN network on data from \cite{McQuillan2014}. In their work, they reported a Root Mean Squared Error (RMSE) of $5.2$ days. Our RMSE is $2.61$ days, a factor of two lower.

\begin{table*}
\centering
\resizebox{0.8\textwidth}{!}{
\begin{tabular}{||c|c|c|c||}
    \hline
    \textit{Model} & \textit{$T_\mathrm{eff}$ MAE (K)} & \textit{$\log g$ MAE (dex)} & \textit{ $[\mathrm{Fe} / \mathrm{H}]$ MAE (dex)} \\ \hline
    \textbf{DESA (ours)} & \textbf{91.56} & 0.168 & \textbf{0.069}  \\ \hline
    StarGRUNet & 93.77 & \textbf{0.162} & 0.070
    \\
    \hline
\end{tabular}
}
\caption{Result of spectra encoder. Ground truth labels are from APOGEE. See text for details.}
\label{table:spectra_res}
\end{table*}

\subsection{Zero- and few-shot results}\label{subsec:zero_few_res}
Next, we test the full DESA model. We train the full model using the pre-trained individual encoders. The conditional labels that were added during training are the labels used for pre training ($T_\mathrm{eff}, \log g, [\mathrm{Fe} / \mathrm{H}], P_{rot}$) with the addition of the radius, $R$, and renormalized unit weighted error (RUWE), which measure the error of the astrometric fit from Gaia. The last two were taken from \cite{Berger2020}. Similar to the hybrid method, we do not require those labels to exist. 

Before discussing the fine-tuning results, we investigate the learned feature space. As mentioned in \ref{subsec:dualformer}, the linear layer $A$ is an effective bottleneck of the network, and we want to utilize the information stored in it in a way that is invariant to the transpose operation. Therefore, our final feature space is the projection of the feature vectors ($z_1,z_2$ in Figure \ref{fig:architecture}) onto the eigenspace of $A$. This can be written as:
\begin{align} \label{eq:final_feautures}
    f = (z_1+z_2)^T V,  
\end{align}
where $z_1, z_2$ are the pre-projection feature vectors and $V$ are the eigenvectors of $A$ after training.

We compare our model with contrastive and regularized self-supervised methods that achieved state-of-the-art results in various tasks: VicReg, SimSiam, and MoCo. Each of the methods represents a different methodology - VicReg is a regularized method, SimSiam is a 'positive-only' contrastive method, and MoCo is a 'positive and negative' contrastive method. The use of positive and negative pairs in our scenario might be challenging because there are many samples with multiple spectra. This means that in a batch of samples, we might have off-diagonal positive pairs, which means that they would count as negative pairs. To overcome this, we created a version of MoCo with a special sampler that ensures the uniqueness of stars in each batch. We call this variant Moco-clean. To make sure that all models get the same information, we added the same conditional labels to all models. 

\begin{figure*}
    \centering
    \begin{minipage}[b]{\textwidth}
        \centering
        \includegraphics[width=\textwidth]{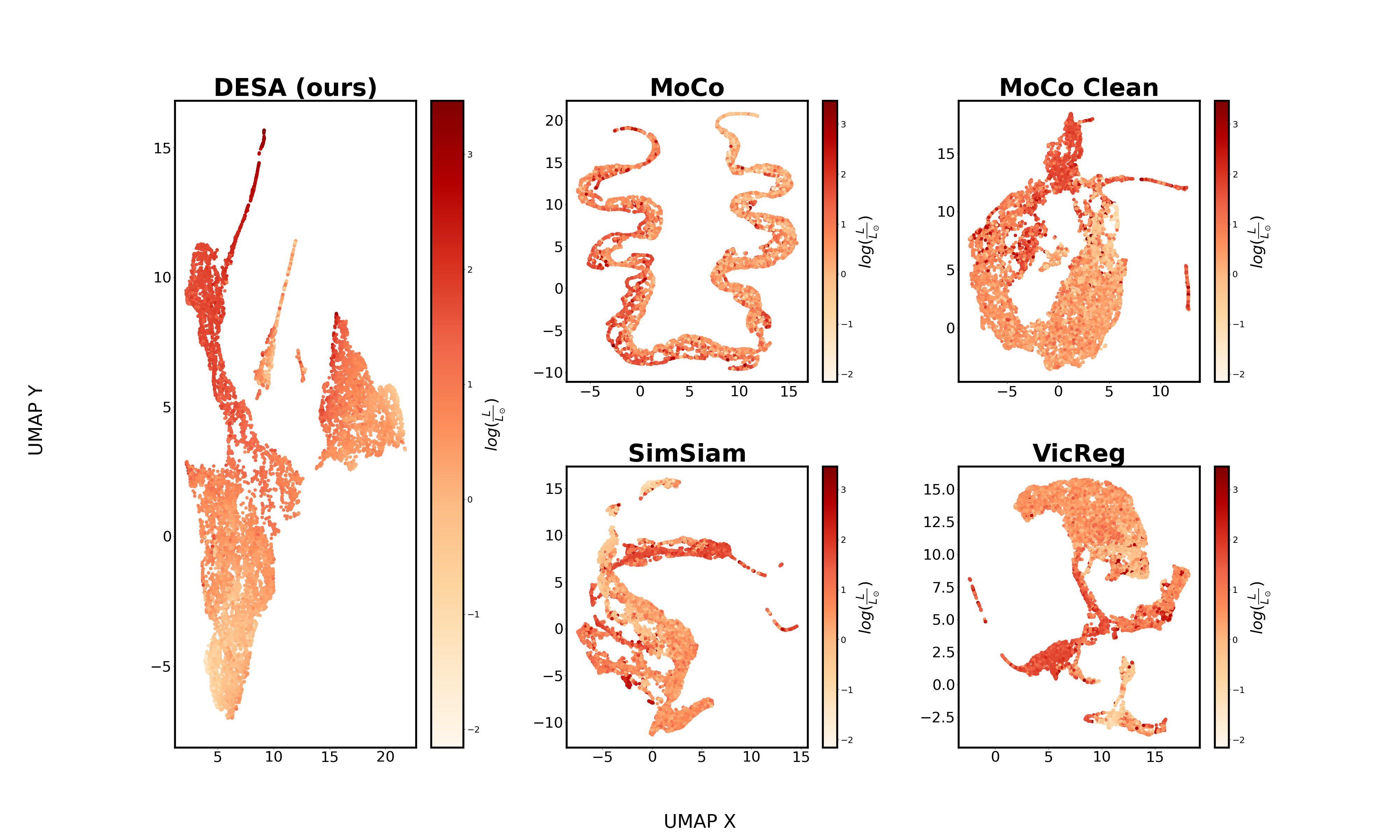}
    \end{minipage}
    \hfill
    \begin{minipage}[b]{\textwidth}
        \centering
        \includegraphics[width=\textwidth]{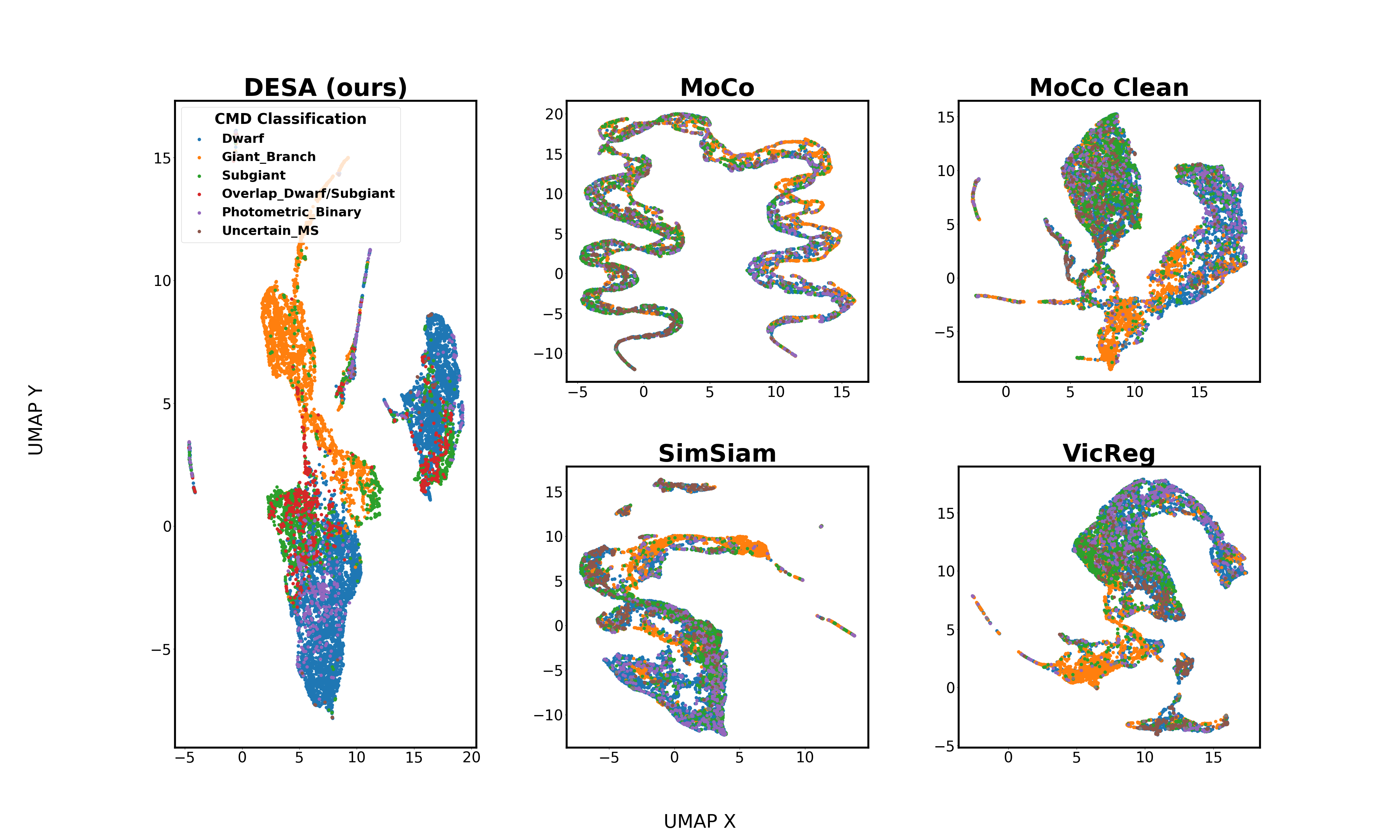}
    \end{minipage}
    \caption{UMAP of final features of different models. In the upper panel, color represents luminosity. In the lower panel, the color represents CMD classes from \cite{Godoy-Rivera2025}.}
    \label{fig:umap}
\end{figure*}

\begin{figure}
    \centering
        \centering
     \includegraphics[width=\columnwidth]{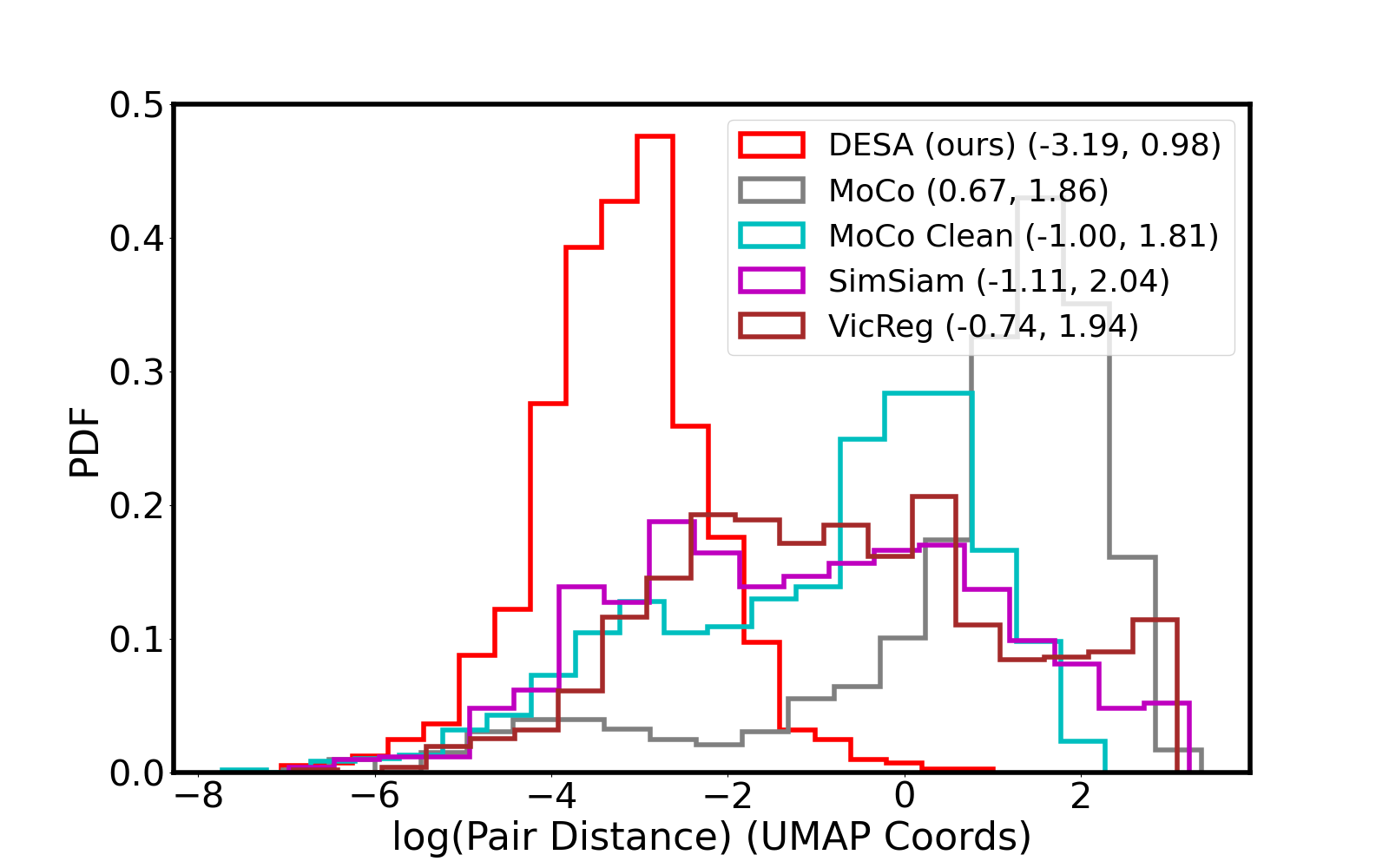}
    \caption{Histograms of UMAP distance of stars with multiple spectra. The legend shows the model name and the mean and standard deviation of the distribution in parentheses.}
    \label{fig:pairs_hist}
\end{figure}

First, we want to compare the final features of the different models. This is done by a UMAP \citep{McInnes2018} dimensionality reduction of the final features. Figure \ref{fig:umap} shows the UMAP of our model and all other alternatives, calculated only on the test set. The UMAP is colored in two ways. The upper panel shows coloring that corresponds to the stellar luminosity from \cite{Berger2020}. It can be seen that our model shows a smooth change in color, which reflects a strong correlation between the latent features and luminosity. Looking at other models, we see that vanilla MoCo shows a thin manifold in UMAP space, which implies a collapse mode. MoCo-clean and SimSiam do not show a strong correlation with luminosity, and VicReg is the only model that shows a smooth correlation. The lower panel shows colors that correspond to stellar classes derived by \cite{Godoy-Rivera2025}, based on a position on a Color-Magnitude Diagram (CMD). The classes are Dwarfs, Giants, Subgiants, overlap Dwarf/Subgiant, photometric binaries, and 'uncertain MS', which correspond to samples that sit below a lower envelope on the CMD. It can be seen that our model creates the best natural clustering with different stellar types lying in different areas of the UMAP space. This is significantly better than all other models, which mix different types, usually Dwarfs and Subgiants. Next, we would like to test the consistency of the final embeddings. This is done using stars with multiple spectra measurements. Those stars have different instances in the test set that reflect the same stellar object. We therefore expect them to have very similar embeddings and to sit close to each other in UMAP space. First, we found all stars with a pair of samples in the test set. Then, for each star, we calculated the UMAP distance between the stars in the pair. Figure \ref{fig:pairs_hist} shows histograms of distances of pairs for all models. It can be seen that the distances of our model are significantly lower compared to all other models. Specifically, the average distance of our model is more than two orders of magnitude lower than all alternatives. This can also be shown visually by looking at the UMAPs and highlighting pairs. Such plots are shown in Appendix \ref{appendix:sup_graphs}, showing $10$ example pairs. We conclude that the latent features of our model are significantly more meaningful and consistent compared to all other models. 

\begin{figure}
    \centering
        \includegraphics[width=\columnwidth]{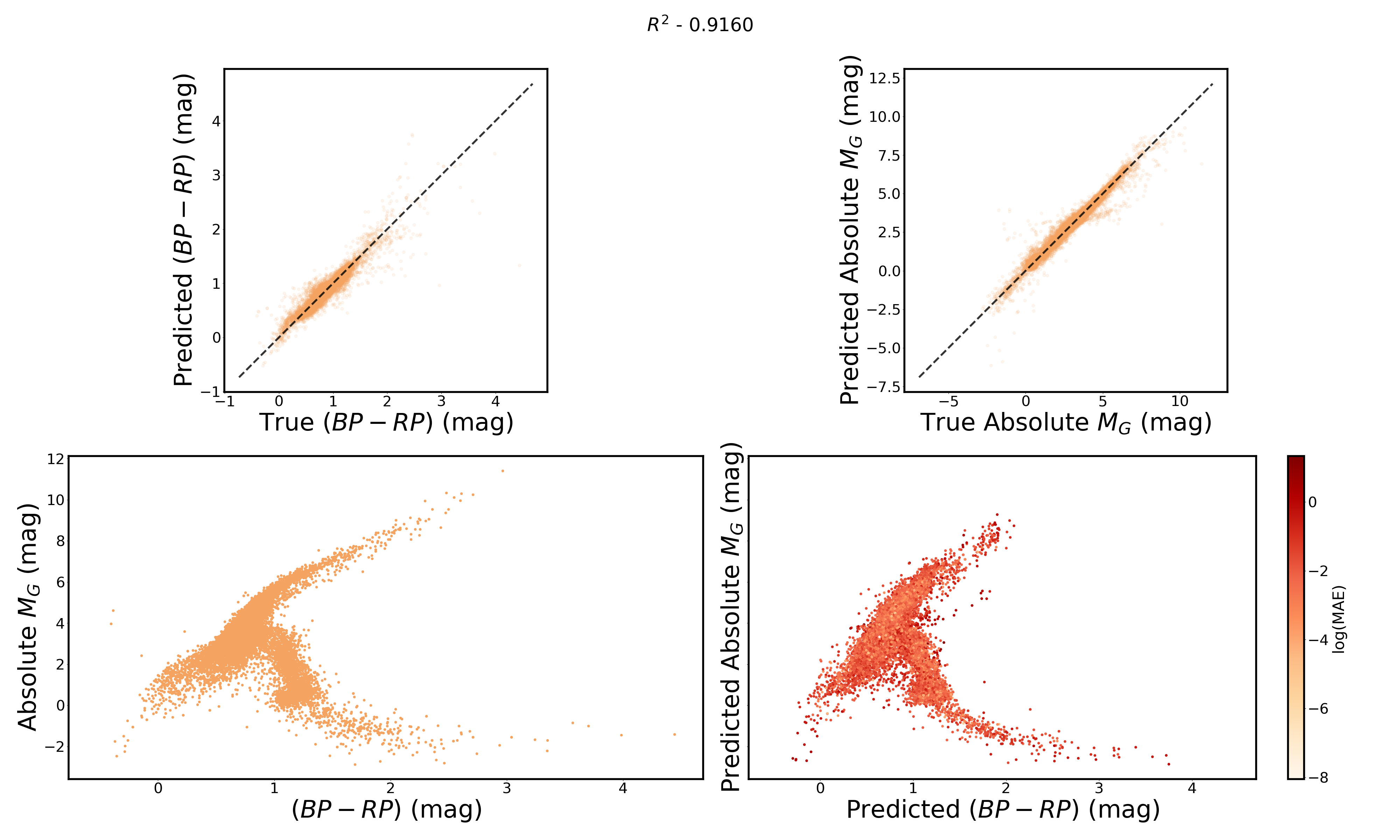}
    \caption{Result of few-shot learning of a linear regression model. The two upper panels show the predicted labels vs. the true labels. The lower left panel shows the true color-magnitude diagram. The lower right panel shows the predicted color-magnitude diagram. Colors on the lower right panel correspond to the mean average error.}
    \label{fig:cmd_diagram}
\end{figure}

\begin{figure}
    \centering
        \includegraphics[width=\columnwidth]{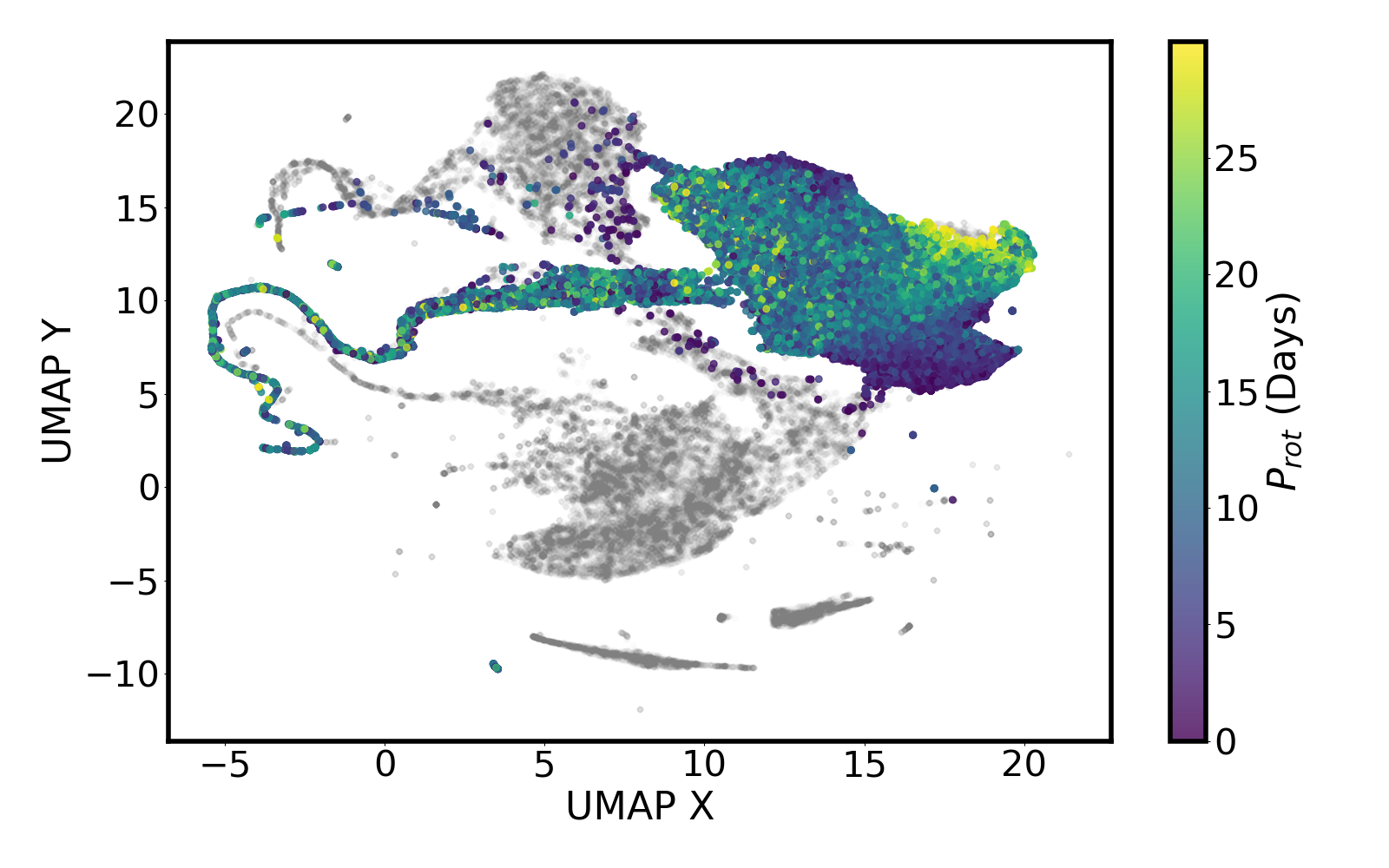}
    \caption{UMAP of final features of the DESA model. All points are in gray. Colored points correspond to samples with $P_{rot} < 30 $ days in one of the latest period catalogs \citep{McQuillan2014, Santos2019, Santos2021, Reinhold2023, Kamai2025}.}
    \label{fig:umap_prot}
\end{figure}

\begin{figure*}
    \centering
    \begin{minipage}[b]{0.45\textwidth}
        \centering
        \includegraphics[width=\textwidth]{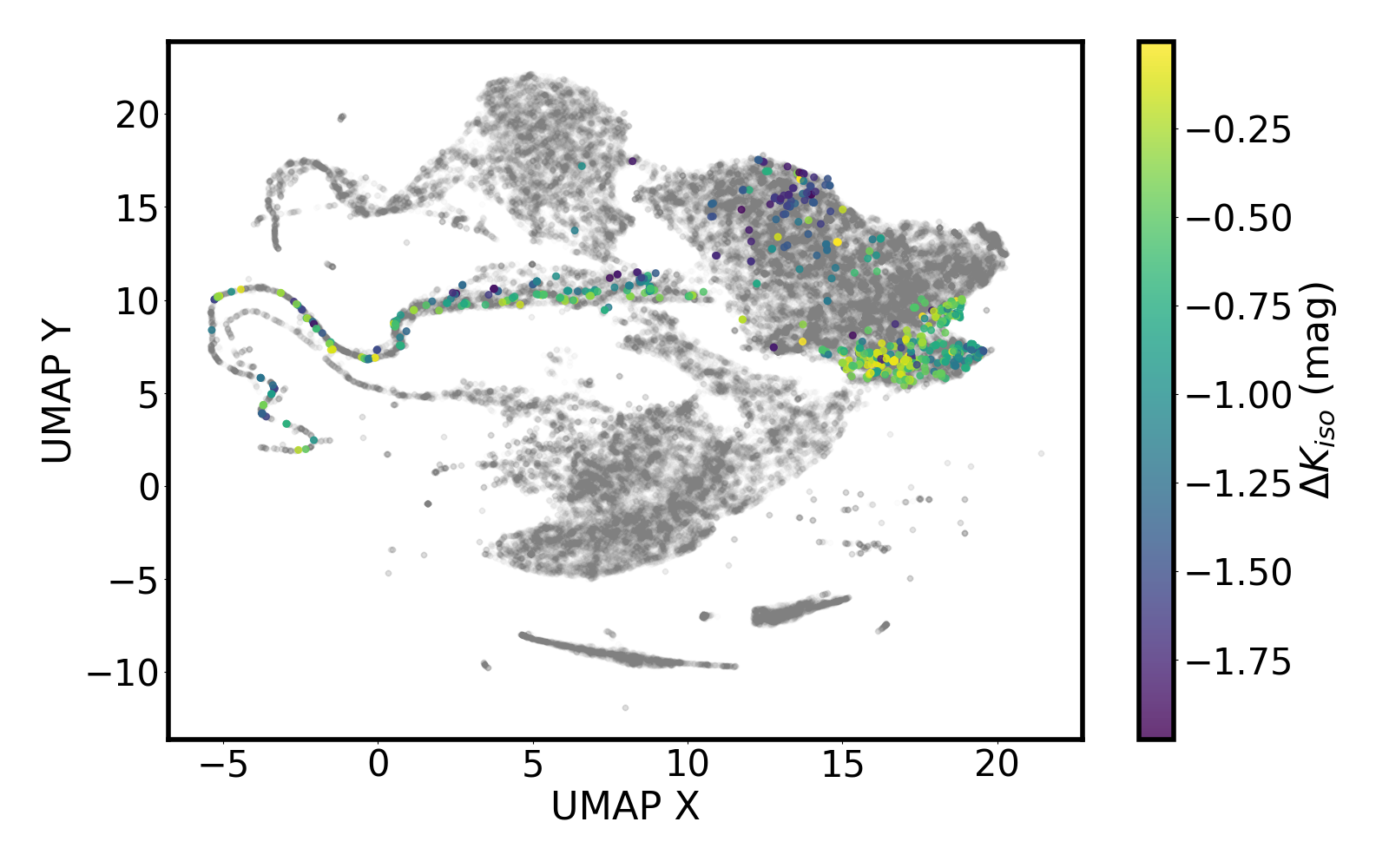}
    \end{minipage}
    \begin{minipage}[b]{0.45\textwidth}
        \centering
        \includegraphics[width=\textwidth]{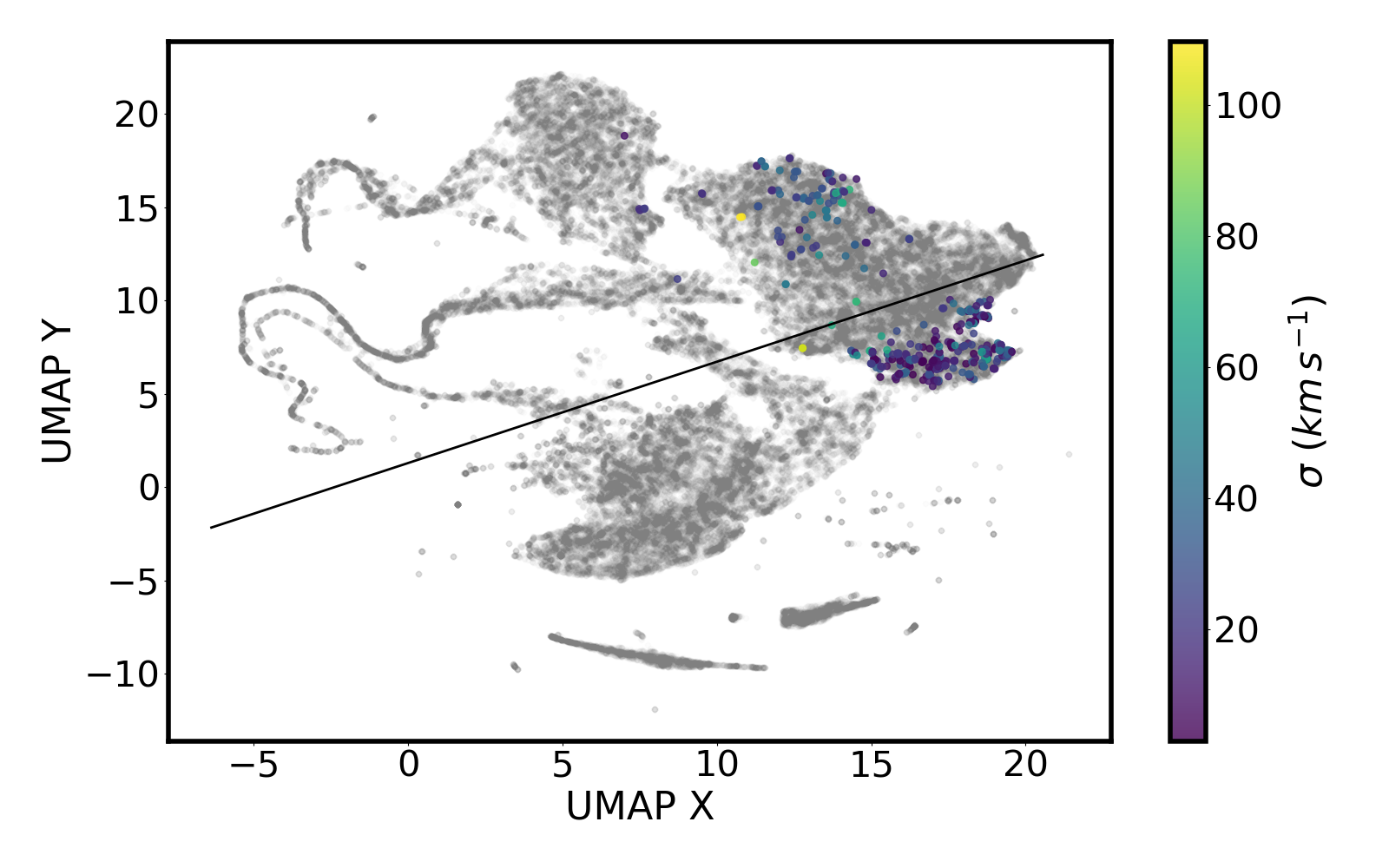}
    \end{minipage}
    \hfill
    \begin{minipage}[b]{0.45\textwidth}
        \centering
        \includegraphics[width=\textwidth]{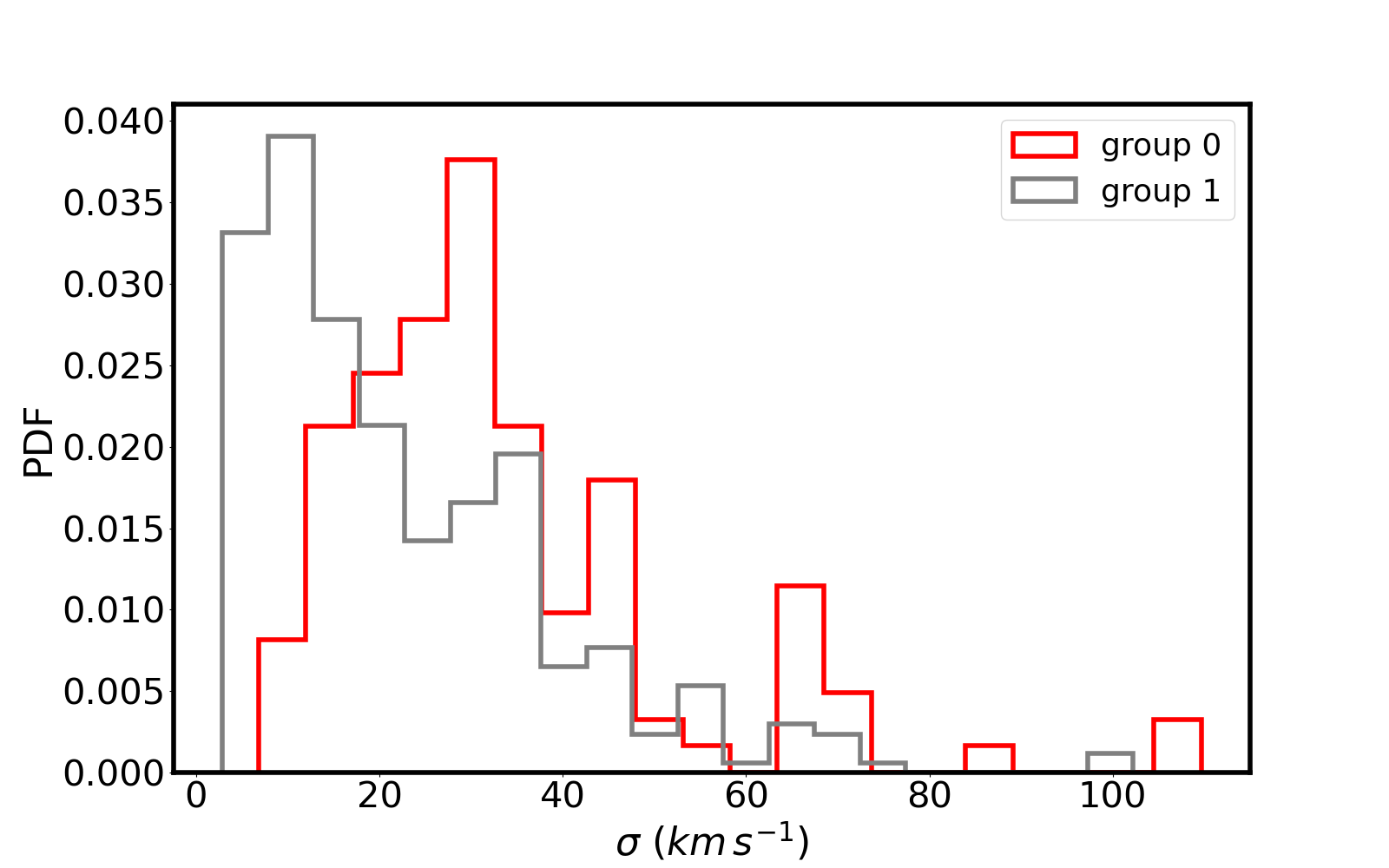}
    \end{minipage}
    \begin{minipage}[b]{0.45\textwidth}
        \centering
        \includegraphics[width=\textwidth]{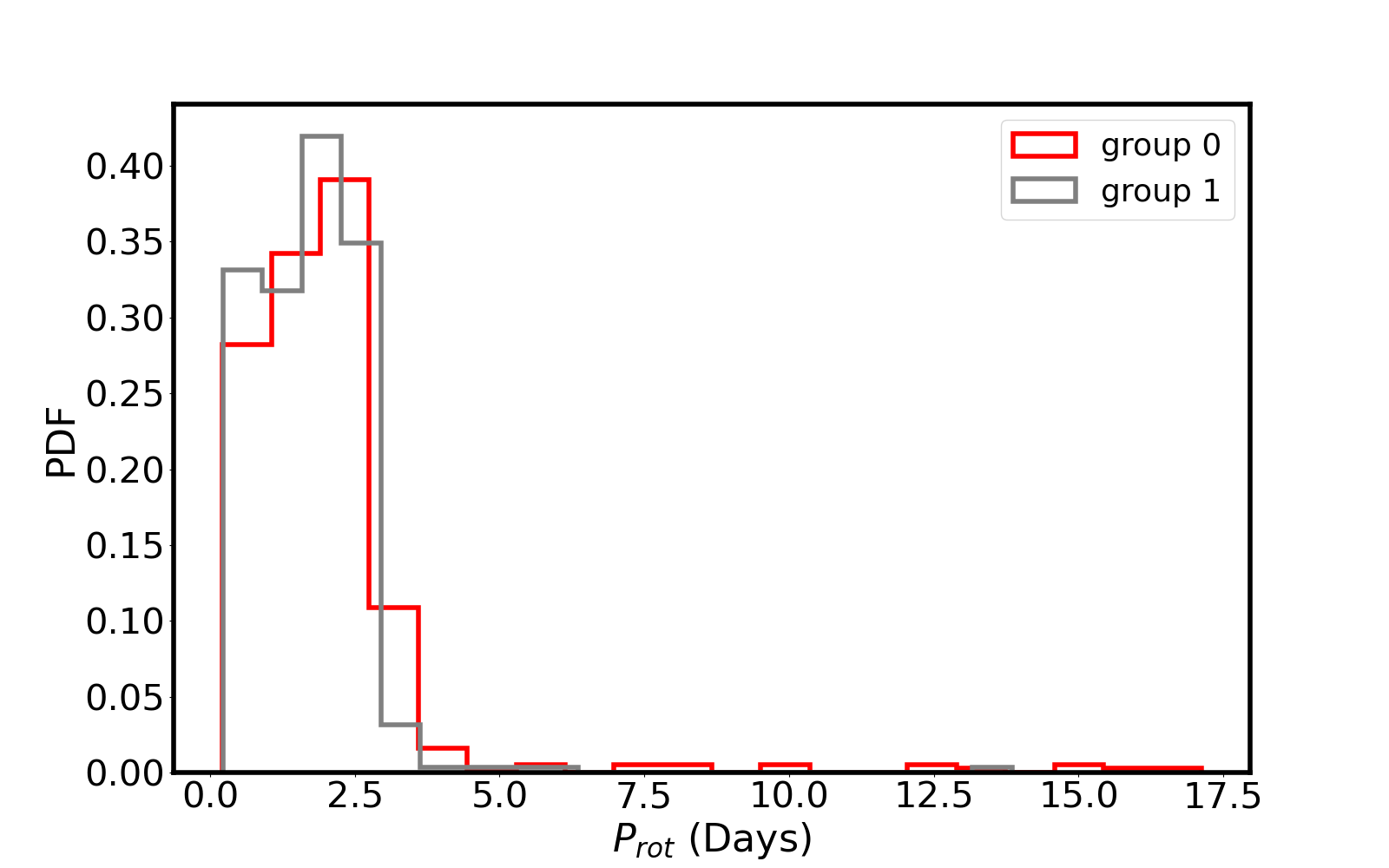}
    \end{minipage}
    \caption{UMAP of fast rotators from \cite{Kamai2025_b}. Upper left - colors correspond to the luminosity excess from a single star model ($\Delta K_\mathrm{iso}$, see \cite{Kamai2025_b} for details.). Upper right - color corresponds to peculiar velocity dispersion ($\sigma$) from \cite{Chen2021}. Note that not all samples have $\sigma$ value. The black line is a separation line found by a linear SVM classifier and a clustering of the points into two classes. Lower left - $\sigma$ histograms of the two clusters from the upper right panel. Lower right - $P_{rot}$ histograms of the same clusters.}
    \label{fig:umap_sync_sigma}
\end{figure*}

We now move to a more quantitative comparison, using zero-shot and few-shot experiments. The zero-shot experiment consists of applying a simple clustering algorithm on the UMAP-reduced features, with classes corresponding to the CMD classes from \cite{Godoy-Rivera2025}. The clustering algorithm is a Gaussian Mixture Model (GMM). The few-shot experiment consists of applying a simple linear regression model on $20\%$ of the test set ($\sim 2500$ samples) to predict color (de-reddened BP-RP) and magnitude (absolute de-reddened G band magnitude), both from \cite{Godoy-Rivera2025}. We measure the accuracy of the zero-shot classifier as well as the $R^2$ and accuracy (defined as the number of points within $10\%$ absolute error) of the few-shot regressor. The results are summarized in Table \ref{table:zero_few_shot}. It can be seen that our model outperforms all other models with a very large margin on all metrics. Specifically, the $R^2$ of all alternative models is around zero. This means that there is no linear relationship between their features and the desired labels. Our model shows a strong relationship of $R^2=0.92$. This point is further demonstrated in Figure \ref{fig:cmd_diagram}, where we plot the prediction results. It can be seen that the model learns not only each parameter alone, but also the correct relationships between them, which create an effective color magnitude diagram. The last point suggests that our final features can be easily fine-tuned, using a few-shot learning, to reproduce known stellar diagrams and can serve as a learned general diagram. We call these diagrams \emph{Neural Diagrams}. Another example of few-shot learning and recovery of the HR diagram can be seen in Appendix \ref{appendix:sup_graphs}. \\ To conclude this subsection, we provide an example of the type of insights that result from explorations of $f$, the final feature space of DESA. For that purpose, we calculate the UMAP of $f$, calculated using Eq. \ref{eq:final_feautures}, on the entire dataset. Figure \ref{fig:umap_prot} shows the UMAP of the entire dataset, with samples that have $P_{rot} < 30$ colored by their period. It is easy to notice a unique structure in the position of samples on the UMAP space, considering their period. There are two main blobs of short periods, located on either side of the main shape. This is interesting given the fact that short-period stars are usually a mixed population of very young stars, which are fast because they are not yet affected by magnetic braking, and synchronized binaries, which are fast because of tidal synchronization. Separating the populations is a hard task and involves different methods. One way is to look at the luminosity excess of stars - we expect binaries to be over-luminous compared to single stars with the same parameters. Another way is by looking at the kinematics of stars - since the peculiar velocities of stars are excited by encounters with spiral arms and molecular clouds, the velocity dispersion of young stars should be lower than that of old stars (i.e. the velocity dispersion provides a "kinematic age"). In our previous paper, \cite{Kamai2025_b}, we analyzed a population of fast-rotating stars and separated them into groups of binaries and young stars using these two approaches. To see if our model creates such separation naturally, we use their data of fast rotators and plot it in UMAP space. Figure \ref{fig:umap_sync_sigma} shows such plots with different coloring. In the upper left panel, we color the fast rotators according to their luminosity excess, compared to a single star with the same parameters. This difference is called $\Delta K_\mathrm{iso}$. The lower the value of $\Delta K_{iso}$, the more over-luminous the star. For more details on how we calculated $\Delta K_\mathrm{iso}$, please refer to \cite{Kamai2025_b}. We can see that the points are separated into three distinct areas - the two blobs on either side of the main structure, which we refer to as the 'mainland', and the long and thin 'tail' that goes out of the main structure. We note that this long tail is an area of stars with large RUWE (figure of UMAP with RUWE colors can be seen in Appendix \ref{appendix:sup_graphs}), and therefore possibly binaries. In the mainland, the two populations are clearly distinguished by their $\Delta K_\mathrm{iso}$. This is further demonstrated in the upper right panel, which shows their kinematics. Namely, their peculiar velocity dispersion, $\sigma$, taken from \cite{Chen2021}. It can be seen that only samples on the mainland have $\sigma$ value, with essentially the same separation as in the upper left panel, suggesting that samples on the upper blob are synchronized binaries and samples on the lower blob are young. To further test this assumption, we separate the two blobs using a GMM clustering and a linear SVM classifier. The separation line can be seen as the black line in the upper right panel. The lower panels show the distributions of $\sigma$ (left) and $P_{rot}$ (right) for the two clusters. We can see that while the two groups have almost identical rotation period distributions, their kinematic distributions are very different, justifying the fact that these are indeed separate populations of synchronized binaries (upper blob with high $\sigma$ and low $\Delta K_\mathrm{iso}$) and young stars (lower panel with low $\sigma$ and high $\Delta K _\mathrm{iso}$). The fact that these populations are naturally separated by the model has far-reaching implications. For example, it means that we do not need $\Delta K_\mathrm{iso}$ and $\sigma$ measurements, which are prone to different errors to separate the groups. All we need is a light curve, spectrum, and a period measurement. Then, we can conclude with high confidence if a star is young, only by looking at its position on the UMAP space. This demonstrates that the embeddings of DESA can be a very powerful and very flexible tool for different stellar population analysis tasks. Next, we test DESA on challenging fine-tuning tasks.

\subsection{Binary Detection}\label{subsec:binary_res}
Here, we test our model in a full fine-tuning scenario. We add a small Transformer prediction head and fine-tune the model on a more challenging task, binary detection. Roughly half of the stars are part of a binary or higher-multipole system \citep{Raghavan2010}. However, detecting them might be very challenging. There are many methods to detect binaries, with different sensitivities. If, for example, the orbital plane of the binary is perpendicular to the line of sight, we would call them 'eclipsing binaries' since we would see the stars eclipse each other. Then it would be easier to detect them using light curve measurement. Another example is the use of spectroscopy to measure the changes in radial velocity that come from the gravitational pull of the companion. This is stronger when the stars are in close orbit. Since every method is sensitive to only a subset of binaries, it is crucial to combine photometric and spectroscopic information to accurately predict a wide range of stars. To create a dataset, we again use \cite{Godoy-Rivera2025}. They reported more than $30000$ binaries in the Kepler field, which results from different detection methods. Specifically, to create a dataset for binary detection, we need a high-confidence sample of binaries and a high-confidence sample of singles. While the first one might be trivial, the second is more challenging, since there are probably many undetected binaries in Kepler. Therefore, to create a sample of single stars, we selected stars satisfying the following four criteria:
\begin{itemize}
    \item Not flagged as binaries by \cite{Godoy-Rivera2025} (Flag Binary Union).
    \item Have RUWE $< 1.2$
    \item Not flagged as potential synchronized binaries by \cite{Kamai2025_b}.
    \item Have $|\Delta K_\mathrm{iso}|<0.3$
\end{itemize}
The last point is motivated by the fact that binaries are expected to be more luminous compared to single stars with the same temperature, and $\Delta K_\mathrm{iso}$ refers to the difference between the expected magnitude of a single star and the measured absolute magnitude as mentioned in \ref{subsec:zero_few_res}. To create a sample of binaries, we took samples that were flagged by \cite{Godoy-Rivera2025} as binaries using the most common methods: RUWE, NSS, RV Variable, and Eclipsing. This results in a dataset of about $14000$ binaries and $4000$ singles.

\begin{figure*}
    \centering
        \centering      \includegraphics[width=0.7\textwidth]{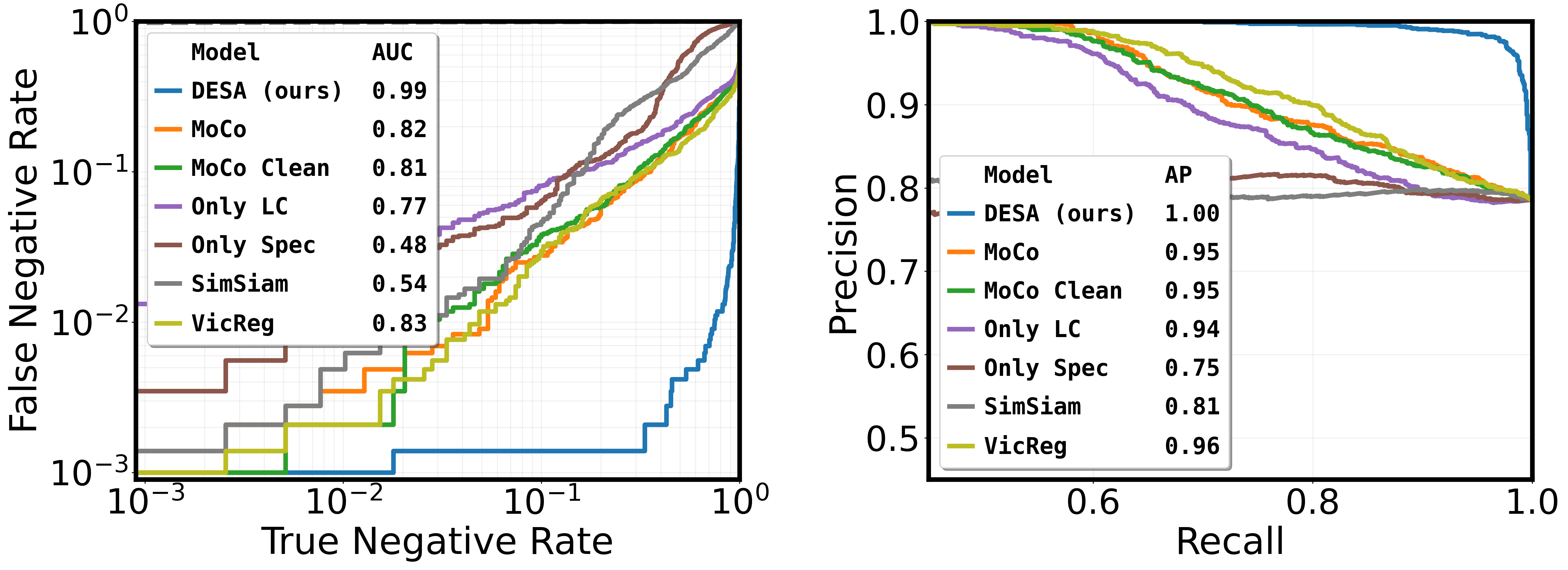}
    \caption{Experimental results of binary detection for different models. The left panel shows True Negative Rate (TNR) vs False Negative Rate (FNR). The right panel shows precision-recall curves.}
    \label{fig:binaries_exp}
\end{figure*}

\begin{figure}
    \centering
        \centering
     \includegraphics[width=\columnwidth]{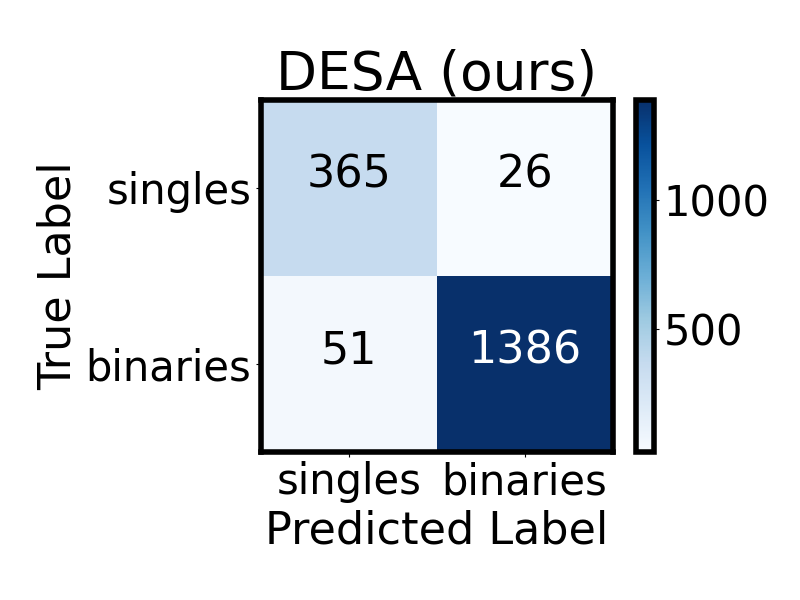}
    \caption{Confusion matrix of our model on the binary detection task.}
    \label{fig:binaries_confusion}
\end{figure}

We fine-tuned our model, with an added classification head, for a binary classification of binarity - all the binaries were assigned as class $1$, regardless of the detection method, and all the singles were assigned as class $0$. As baselines, we trained the same models that were used in \ref{subsec:zero_few_res} as well as unimodal models. The unimodal models are simply the pretrained encoders with an added prediction head. Using the pretrained encoders is important as a sanity check - we want to see if our model is able to extract meaningful information from the combination of modalities and, specifically, better than each modality alone. In all models, we made sure that the number of trainable parameters is not less than that in our DESA model ($11$M). Figure \ref{fig:binaries_exp} compares the results using a precision-recall curve (right panel) and a True Negative Rate (TNR) vs. False Negative Rate (FNR) curve (left panel). Notice that a TNR-FNR curve can be seen as a flipped ROC curve, and therefore, the area under the curve (AUC) metric has the same meaning (with the area calculated in the opposite direction). Note that we opted for the TNR-FNR curve instead of the standard ROC curve for visual clarity. It can be seen that our model performs significantly better compared to all other models, with an AUC of $0.99$ and average precision (AP) of $1.00$. Figure \ref{fig:binaries_confusion} shows the confusion matrix of our model. It is interesting to compare it with the confusion matrices of all other models. Such a comparison revealed that, except for VicReg, all other models collapsed to predict the majority class (binary class). With the exception of the proposed DESA, VicReg was the only model able to predict some single stars correctly, but it did so with much lower precision for that class, in comparison to our model. The precision of VicReg on the single-star class is $48\%$, while our model achieved a precision of $88\%$ on that class. Confusion matrices of all other models are reported in Appendix \ref{fig:appendix_confusion}.

\subsection{Stellar age predicition}\label{subsec:age prediction}
Our last experiment is fine-tuning for stellar age prediction. Age is one of the most challenging stellar properties to detect. Since age is not directly measurable, people use relationships with other properties to infer ages. However, these relationships might not be trivial. One such example is gyrochronology, which started as the simple Skumanich relationship \citep{Skumanich1972}, describing the age as a simple power law of the rotation of the star. Recent works on gyrochronology \citep{barnes_ages_2007, barnes_rotational_2003, mamajek_improved_2008, angus_calibrating_2015, angus_toward_2019, bouma_empirical_2023, lu_age_2024} reveal a much more complicated picture, with dependencies on the temperature and type of stars. This complex relationship, combining photometric and spectroscopic information, is our motivation to use a multi-modal approach for age prediction. In addition, like in binary detection, there are different methods to predict age, and each method is reliable only on a subset of stars. This suggests that combined information might increase the set of predictable stars.

To create an age dataset, we combine two recent age catalogs: \cite{bouma_ages_2024} and \cite{lu_age_2024}. Although both are gyrochronology methods, they are different. The ages in \cite{bouma_ages_2024} were calculated using period and temperature relationship on ages not exceeding $2.7 Gyr$. The ages in \cite{lu_age_2024} also used kinematic information and were calculated on a wider range. We used a combined catalog and cross-matched it with LAMOST spectra. The resulting dataset has around $14000$ samples. We used the same setup as in \ref{subsec:binary_res}, this time for a regression task, and fine-tuned the model to predict the age and the reported error. The loss function was chosen to be CQR. Table \ref{table:age_exp} summarizes the results of age predictions. It shows both the MAE and the RMSE of all models. It can be seen that our model outperforms all other models, being the only model with an RMSE lower than $1$ Gyr, which was reported by \cite{lu_age_2024} as the typical error of their method. Prediction plots can be seen in Appendix \ref{appendix:sup_graphs}.

\begin{table*}
\centering

\begin{tabular}{||c|c|c|c|c||}
    \hline
    \textit{Model} & \textit{zero-shot accuracy} & \textit{Linear Regression $R^2$} & \textit{ $BP-RP$ Accuracy} & \textit{$Gmag$ Accuracy} \\ \hline
    \textbf{DESA (ours)} & \textbf{0.40} & \textbf{0.920} & \textbf{0.677} & \textbf{0.711}  \\ \hline
    MoCo & 0.18 & -0.001 & 0.208 & 0.159  \\ 
    \hline
    MoCo Clean & 0.11 & -0.0004 & 0.202 & 0.160  \\ 
    \hline
    SimSiam & 0.15 & -0.0003 & 0.202 & 0.158
    \\
    \hline
    VICReg & 0.25 & -0.0003 & 0.202 & 0.158
    \\
    \hline
\end{tabular}
\caption{Result of zero-shot CMD clustering (first column), and few-shot regression. See text for details.}
\label{table:zero_few_shot}
\end{table*}

\section{Conclusions} \label{sec:conclusions}
We presented DESA, a new multi-modality model for stellar astrophysics. DESA backbone consists of pre-trained unimodality encoders that were trained in a hybrid approach and shows state-of-the-art performance. The alignment module, DualFormer, is motivated by the observation that astrophysical data is unique and different compared to common multi-modality domains such as vision and NLP. We show that DESA learns a latent space that distills important physical information and can be easily transformed into meaningful diagrams. These diagrams can be used for tasks such as population analysis and outlier detection. The effectiveness of DESA is further demonstrated in various ways, including zero-shot, few-shot, and fine-tuning experiments to predict challenging and important labels such as binarity and age. DESA consistently outperforms all baselines on all experiments with (sometimes very) large margins, proving its superiority in the astronomical domain and marking a new era in data-driven astronomy where multiple measurements are efficiently combined to extract new insights. Each fine-tune task is of great physical importance and deserves an in-depth investigation and separate papers, which we plan for future work. DESA faces two primary limitations: interpretability and single-resolution modeling. Interpretability remains a challenge across machine learning models, lacking robust methods for results interpretation. Regarding resolution, DESA, while multi-modal, was trained on single surveys per modality (Kepler and LAMOST), without explicit consideration for survey-specific resolutions. The last point is related to the applicability of DESA with other surveys. In general, DESA can be applied to different pairs of surveys, such as APOGEE \citep{apogee2022} and TESS \citep{tess2014}, straightforwardly, since nothing in the suggested method itself is specific to LAMOST and Kepler. While the pre-processing is also general, there might be some practical adaptations needed in that part. For example, TESS samples are much shorter compared to Kepler. This means that it might be more challenging to crop each sample into a representative view (in that case, overlapping views can be used). Another example is that the ASPCAP pipeline in APOGEE already moves the wavelength into the rest frame, which makes this step redundant in the spectra preprocessing. However, extending DESA with multiple surveys of the same modality (e.g., APOGEE, LAMOST, TESS, and Kepler together) would require encoding the wavelength and cadence of each survey, which is one of our future directions. Addressing interpretability and multi-resolution capabilities stands as our main goal for future research. Nonetheless, DESA represents a potential groundbreaking advancement in multimodal stellar astronomy. 

Moreover, beyond its architectural innovations, DESA achieves consistently superior performance across diverse tasks, including zero-shot stellar classification, few-shot regression of photometric properties, and fine-tuning for physically complex problems like binary detection and stellar age inference. In particular, DESA distinguishes itself by recovering classical diagrams (e.g., HR and CMD) from raw embeddings and separating physically degenerate populations—such as synchronized binaries and young stars—without requiring external labels. These results demonstrate that DESA is not merely a predictive model, but a foundation model capable of extracting physically meaningful structure from heterogeneous data. We anticipate that DESA will serve as a powerful framework for future data-driven discovery in large stellar surveys, facilitating population studies, anomaly detection, and improved parameter estimation across the HR diagram.

\begin{table*}
\centering
\resizebox{\columnwidth}{!}{
\begin{tabular}{||c|c|c||}
    \hline
    \textit{Model} & \textit{Age MAE (Gyr)} & \textit{Age RMSE (Gyr)} \\ \hline
    \textbf{DESA (ours)} & \textbf{0.61} & \textbf{0.94}\\
    \hline
    MoCo & 0.81 & 1.28   \\ 
    \hline
    MoCo Clean & 0.78 & 1.23  \\ 
    \hline
    SimSiam & 1.24 & 1.81    \\
    \hline
    VICReg & 0.78 & 1.23 \\
    \hline
    Only Spectra & 1.30 & 1.70 \\
    \hline
    Only Light Curve & 0.80 & 1.25 \\
    \hline
\end{tabular}
}
\caption{Result of age prediction finetuning experiment.}
\label{table:age_exp}
\end{table*}


\bibliographystyle{aasjournal}
\bibliography{main}

\begin{thebibliography}{}
\expandafter\ifx\csname natexlab\endcsname\relax\def\natexlab#1{#1}\fi
\providecommand{\url}[1]{\href{#1}{#1}}
\providecommand{\dodoi}[1]{doi:~\href{http://doi.org/#1}{\nolinkurl{#1}}}
\providecommand{\doeprint}[1]{\href{http://ascl.net/#1}{\nolinkurl{http://ascl.net/#1}}}
\providecommand{\doarXiv}[1]{\href{https://arxiv.org/abs/#1}{\nolinkurl{https://arxiv.org/abs/#1}}}

\bibitem[{{Abdurro'uf} {et~al.}(2022){Abdurro'uf}, {Accetta}, {Aerts}, {Silva Aguirre}, {Ahumada}, {Ajgaonkar}, {Filiz Ak}, {Alam}, {Allende Prieto}, {Almeida}, {Anders}, {Anderson}, {Andrews}, {Anguiano}, {Aquino-Ort{\'\i}z}, {Arag{\'o}n-Salamanca}, {Argudo-Fern{\'a}ndez}, {Ata}, {Aubert}, {Avila-Reese}, {Badenes}, {Barb{\'a}}, {Barger}, {Barrera-Ballesteros}, {Beaton}, {Beers}, {Belfiore}, {Bender}, {Bernardi}, {Bershady}, {Beutler}, {Bidin}, {Bird}, {Bizyaev}, {Blanc}, {Blanton}, {Boardman}, {Bolton}, {Boquien}, {Borissova}, {Bovy}, {Brandt}, {Brown}, {Brownstein}, {Brusa}, {Buchner}, {Bundy}, {Burchett}, {Bureau}, {Burgasser}, {Cabang}, {Campbell}, {Cappellari}, {Carlberg}, {Wanderley}, {Carrera}, {Cash}, {Chen}, {Chen}, {Cherinka}, {Chiappini}, {Choi}, {Chojnowski}, {Chung}, {Clerc}, {Cohen}, {Comerford}, {Comparat}, {da Costa}, {Covey}, {Crane}, {Cruz-Gonzalez}, {Culhane}, {Cunha}, {Dai}, {Damke}, {Darling}, {Davidson}, {Davies}, {Dawson}, {De Lee}, {Diamond-Stanic}, {Cano-D{\'\i}az}, {S{\'a}nchez},
  {Donor}, {Duckworth}, {Dwelly}, {Eisenstein}, {Elsworth}, {Emsellem}, {Eracleous}, {Escoffier}, {Fan}, {Farr}, {Feng}, {Fern{\'a}ndez-Trincado}, {Feuillet}, {Filipp}, {Fillingham}, {Frinchaboy}, {Fromenteau}, {Galbany}, {Garc{\'\i}a}, {Garc{\'\i}a-Hern{\'a}ndez}, {Ge}, {Geisler}, {Gelfand}, {G{\'e}ron}, {Gibson}, {Goddy}, {Godoy-Rivera}, {Grabowski}, {Green}, {Greener}, {Grier}, {Griffith}, {Guo}, {Guy}, {Hadjara}, {Harding}, {Hasselquist}, {Hayes}, {Hearty}, {Hern{\'a}ndez}, {Hill}, {Hogg}, {Holtzman}, {Horta}, {Hsieh}, {Hsu}, {Hsu}, {Huber}, {Huertas-Company}, {Hutchinson}, {Hwang}, {Ibarra-Medel}, {Chitham}, {Ilha}, {Imig}, {Jaekle}, {Jayasinghe}, {Ji}, {Johnson}, {Jones}, {J{\"o}nsson}, {Katkov}, {Khalatyan}, {Kinemuchi}, {Kisku}, {Knapen}, {Kneib}, {Kollmeier}, {Kong}, {Kounkel}, {Kreckel}, {Krishnarao}, {Lacerna}, {Lane}, {Langgin}, {Lavender}, {Law}, {Lazarz}, {Leung}, {Leung}, {Lewis}, {Li}, {Li}, {Lian}, {Liang}, {Lin}, {Lin}, {Lin}, {Lintott}, {Long}, {Longa-Pe{\~n}a}, {L{\'o}pez-Cob{\'a}}, {Lu},
  {Lundgren}, {Luo}, {Mackereth}, {de la Macorra}, {Mahadevan}, {Majewski}, {Manchado}, {Mandeville}, {Maraston}, {Margalef-Bentabol}, {Masseron}, {Masters}, {Mathur}, {McDermid}, {Mckay}, {Merloni}, {Merrifield}, {Meszaros}, {Miglio}, {Di Mille}, {Minniti}, {Minsley}, \& {Monachesi}}]{apogee2022}
{Abdurro'uf}, {Accetta}, K., {Aerts}, C., {et~al.} 2022, apjs, 259, 35, \dodoi{10.3847/1538-4365/ac4414}

\bibitem[{Angus {et~al.}(2015)Angus, Aigrain, Foreman-Mackey, \& McQuillan}]{angus_calibrating_2015}
Angus, R., Aigrain, S., Foreman-Mackey, D., \& McQuillan, A. 2015, Monthly Notices of the Royal Astronomical Society, 450, 1787, \dodoi{10.1093/mnras/stv423}

\bibitem[{Angus {et~al.}(2019)Angus, Morton, Foreman-Mackey, van Saders, Curtis, Kane, Bedell, Kiman, Hogg, \& Brewer}]{angus_toward_2019}
Angus, R., Morton, T.~D., Foreman-Mackey, D., {et~al.} 2019, The Astronomical Journal, 158, 173, \dodoi{10.3847/1538-3881/ab3c53}

\bibitem[{Bai {et~al.}(2020)Bai, Liu, Wang, \& Wang}]{Bai_2020}
Bai, Y., Liu, J., Wang, Y., \& Wang, S. 2020, The Astronomical Journal, 159, 84, \dodoi{10.3847/1538-3881/ab63d5}

\bibitem[{{Bailer-Jones}(2000)}]{Bailer-Jones2000}
{Bailer-Jones}, C.~A.~L. 2000, aap, 357, 197, \dodoi{10.48550/arXiv.astro-ph/0003071}

\bibitem[{{Bardes} {et~al.}(2021){Bardes}, {Ponce}, \& {LeCun}}]{Bardes2021a}
{Bardes}, A., {Ponce}, J., \& {LeCun}, Y. 2021, arXiv e-prints, arXiv:2105.04906, \dodoi{10.48550/arXiv.2105.04906}

\bibitem[{Barnes(2003)}]{barnes_rotational_2003}
Barnes, S.~A. 2003, The Astrophysical Journal, 586, 464, \dodoi{10.1086/367639}

\bibitem[{Barnes(2007)}]{barnes_ages_2007}
---. 2007, The Astrophysical Journal, 669, 1167, \dodoi{10.1086/519295}

\bibitem[{{Berger} {et~al.}(2020){Berger}, {Huber}, {van Saders}, {Gaidos}, {Tayar}, \& {Kraus}}]{Berger2020}
{Berger}, T.~A., {Huber}, D., {van Saders}, J.~L., {et~al.} 2020, \aj, 159, 280, \dodoi{10.3847/1538-3881/159/6/280}

\bibitem[{{Blancato} {et~al.}(2020){Blancato}, {Ness}, {Huber}, {Lu}, \& {Angus}}]{Blancato2020}
{Blancato}, K., {Ness}, M., {Huber}, D., {Lu}, Y., \& {Angus}, R. 2020, arXiv e-prints, arXiv:2005.09682, \dodoi{10.48550/arXiv.2005.09682}

\bibitem[{Bouma {et~al.}(2024)Bouma, Hillenbrand, Howard, Isaacson, Masuda, \& Palumbo}]{bouma_ages_2024}
Bouma, L.~G., Hillenbrand, L.~A., Howard, A.~W., {et~al.} 2024, The Astrophysical Journal, 976, 234, \dodoi{10.3847/1538-4357/ad855f}

\bibitem[{Bouma {et~al.}(2023)Bouma, Palumbo, \& Hillenbrand}]{bouma_empirical_2023}
Bouma, L.~G., Palumbo, E.~K., \& Hillenbrand, L.~A. 2023, The Astrophysical Journal Letters, 947, L3, \dodoi{10.3847/2041-8213/acc589}

\bibitem[{{Brown} {et~al.}(2025){Brown}, {Kazmierski}, {Pasquarella}, {Rucklidge}, {Samsikova}, {Zhang}, {Shelhamer}, {Lahera}, {Wiles}, {Ilyushchenko}, {Gorelick}, {Zhang}, {Alj}, {Schechter}, {Askay}, {Guinan}, {Moore}, {Boukouvalas}, \& {Kohli}}]{Brown2025_alpha_earth}
{Brown}, C.~F., {Kazmierski}, M.~R., {Pasquarella}, V.~J., {et~al.} 2025, arXiv e-prints, arXiv:2507.22291, \dodoi{10.48550/arXiv.2507.22291}

\bibitem[{{Chen} {et~al.}(2021){Chen}, {Yang}, {Xie}, {Zhou}, {Dong}, {Zheng}, {Zhang}, {Liu}, {Wang}, {Xiang}, {Zong}, {Huang}, \& {Luo}}]{Chen2021}
{Chen}, D.-C., {Yang}, J.-Y., {Xie}, J.-W., {et~al.} 2021, \aj, 162, 100, \dodoi{10.3847/1538-3881/ac0f08}

\bibitem[{{Chen} {et~al.}(2020){Chen}, {Kornblith}, {Norouzi}, \& {Hinton}}]{Chen_simclr2020}
{Chen}, T., {Kornblith}, S., {Norouzi}, M., \& {Hinton}, G. 2020, arXiv e-prints, arXiv:2002.05709, \dodoi{10.48550/arXiv.2002.05709}

\bibitem[{{Chen} \& {He}(2020)}]{Chen2020}
{Chen}, X., \& {He}, K. 2020, arXiv e-prints, arXiv:2011.10566, \dodoi{10.48550/arXiv.2011.10566}

\bibitem[{{Claytor} \& {Tayar}(2025)}]{Claytor2025}
{Claytor}, Z.~R., \& {Tayar}, J. 2025, arXiv e-prints, arXiv:2506.03248, \dodoi{10.48550/arXiv.2506.03248}

\bibitem[{Claytor {et~al.}(2024)Claytor, van Saders, Cao, Pinsonneault, Teske, \& Beaton}]{Claytor2024}
Claytor, Z.~R., van Saders, J.~L., Cao, L., {et~al.} 2024, The Astrophysical Journal, 962, 47, \dodoi{10.3847/1538-4357/ad159a}

\bibitem[{{Claytor} {et~al.}(2022){Claytor}, {van Saders}, {Llama}, {Sadowski}, {Quach}, \& {Avallone}}]{Claytor2022}
{Claytor}, Z.~R., {van Saders}, J.~L., {Llama}, J., {et~al.} 2022, \apj, 927, 219, \dodoi{10.3847/1538-4357/ac498f}

\bibitem[{Cui {et~al.}(2025)Cui, Tejada-Lapuerta, Brbić, Saez-Rodriguez, Cristea, Goodarzi, Lotfollahi, Theis, \& Wang}]{cui_2025}
Cui, H., Tejada-Lapuerta, A., Brbić, M., {et~al.} 2025, Nature, 640, 623, \dodoi{10.1038/s41586-025-08710-y}

\bibitem[{{Garc{\'\i}a P{\'e}rez} {et~al.}(2016){Garc{\'\i}a P{\'e}rez}, {Allende Prieto}, {Holtzman}, {Shetrone}, {M{\'e}sz{\'a}ros}, {Bizyaev}, {Carrera}, {Cunha}, {Garc{\'\i}a-Hern{\'a}ndez}, {Johnson}, {Majewski}, {Nidever}, {Schiavon}, {Shane}, {Smith}, {Sobeck}, {Troup}, {Zamora}, {Weinberg}, {Bovy}, {Eisenstein}, {Feuillet}, {Frinchaboy}, {Hayden}, {Hearty}, {Nguyen}, {O'Connell}, {Pinsonneault}, {Wilson}, \& {Zasowski}}]{Garcia2016}
{Garc{\'\i}a P{\'e}rez}, A.~E., {Allende Prieto}, C., {Holtzman}, J.~A., {et~al.} 2016, aj, 151, 144, \dodoi{10.3847/0004-6256/151/6/144}

\bibitem[{{Godoy-Rivera} {et~al.}(2025){Godoy-Rivera}, {Mathur}, {Garc{\'\i}a}, {Pinsonneault}, {Santos}, {Beck}, {Grossmann}, {Schimak}, {Bedell}, {Merc}, \& {Escorza}}]{Godoy-Rivera2025}
{Godoy-Rivera}, D., {Mathur}, S., {Garc{\'\i}a}, R.~A., {et~al.} 2025, aap, 696, A243, \dodoi{10.1051/0004-6361/202348735}

\bibitem[{{Grill} {et~al.}(2020){Grill}, {Strub}, {Altch{\'e}}, {Tallec}, {Richemond}, {Buchatskaya}, {Doersch}, {Avila Pires}, {Guo}, {Gheshlaghi Azar}, {Piot}, {Kavukcuoglu}, {Munos}, \& {Valko}}]{Grill2020}
{Grill}, J.-B., {Strub}, F., {Altch{\'e}}, F., {et~al.} 2020, arXiv e-prints, arXiv:2006.07733, \dodoi{10.48550/arXiv.2006.07733}

\bibitem[{{Gulati} {et~al.}(2020){Gulati}, {Qin}, {Chiu}, {Parmar}, {Zhang}, {Yu}, {Han}, {Wang}, {Zhang}, {Wu}, \& {Pang}}]{Gulati2020}
{Gulati}, A., {Qin}, J., {Chiu}, C.-C., {et~al.} 2020, arXiv e-prints, arXiv:2005.08100, \dodoi{10.48550/arXiv.2005.08100}

\bibitem[{{Hattori} {et~al.}(2025){Hattori}, {Angus}, {Foreman-Mackey}, {Yuxi}, {Lu}, \& {Colman}}]{Hattori2025}
{Hattori}, S., {Angus}, R., {Foreman-Mackey}, D., {et~al.} 2025, arXiv e-prints, arXiv:2505.10376, \dodoi{10.48550/arXiv.2505.10376}

\bibitem[{{He} {et~al.}(2019){He}, {Fan}, {Wu}, {Xie}, \& {Girshick}}]{He2019}
{He}, K., {Fan}, H., {Wu}, Y., {Xie}, S., \& {Girshick}, R. 2019, arXiv e-prints, arXiv:1911.05722, \dodoi{10.48550/arXiv.1911.05722}

\bibitem[{{Hoffmann} {et~al.}(2022){Hoffmann}, {Borgeaud}, {Mensch}, {Buchatskaya}, {Cai}, {Rutherford}, {de Las Casas}, {Hendricks}, {Welbl}, {Clark}, {Hennigan}, {Noland}, {Millican}, {van den Driessche}, {Damoc}, {Guy}, {Osindero}, {Simonyan}, {Elsen}, {Rae}, {Vinyals}, \& {Sifre}}]{Hoffmann2022}
{Hoffmann}, J., {Borgeaud}, S., {Mensch}, A., {et~al.} 2022, arXiv e-prints, arXiv:2203.15556, \dodoi{10.48550/arXiv.2203.15556}

\bibitem[{{Kamai} {et~al.}(2025){Kamai}, {Bronstein}, \& {Perets}}]{kamai_zenodo2025}
{Kamai}, I., {Bronstein}, A., \& {Perets}, H. 2025, {Machine-learning inference of stellar properties using integrated photometric and spectroscopic data},  Zenodo, \dodoi{10.5281/zenodo.17088663}

\bibitem[{{Kamai} \& {Perets}(2025{\natexlab{a}})}]{Kamai2025}
{Kamai}, I., \& {Perets}, H.~B. 2025{\natexlab{a}}, aj, 169, 59, \dodoi{10.3847/1538-3881/ad99ab}

\bibitem[{{Kamai} \& {Perets}(2025{\natexlab{b}})}]{Kamai2025_b}
---. 2025{\natexlab{b}}, The Open Journal of Astrophysics, 8, 59, \dodoi{10.33232/001c.138238}

\bibitem[{{Kaplan} {et~al.}(2020){Kaplan}, {McCandlish}, {Henighan}, {Brown}, {Chess}, {Child}, {Gray}, {Radford}, {Wu}, \& {Amodei}}]{Kaplan2020}
{Kaplan}, J., {McCandlish}, S., {Henighan}, T., {et~al.} 2020, arXiv e-prints, arXiv:2001.08361, \dodoi{10.48550/arXiv.2001.08361}

\bibitem[{{Koblischke} \& {Bovy}(2024)}]{Koblischke2024}
{Koblischke}, N., \& {Bovy}, J. 2024, arXiv e-prints, arXiv:2411.04750, \dodoi{10.48550/arXiv.2411.04750}

\bibitem[{{Leung} \& {Bovy}(2019)}]{Leung2019}
{Leung}, H.~W., \& {Bovy}, J. 2019, \mnras, 483, 3255, \dodoi{10.1093/mnras/sty3217}

\bibitem[{Leung \& Bovy(2023)}]{Bovy2023}
Leung, H.~W., \& Bovy, J. 2023, Monthly Notices of the Royal Astronomical Society, 527, 1494, \dodoi{10.1093/mnras/stad3015}

\bibitem[{{Li} {et~al.}(2025){Li}, {Lu}, {Wang}, \& {Wang}}]{Li2025}
{Li}, G., {Lu}, Z., {Wang}, J., \& {Wang}, Z. 2025, arXiv e-prints, arXiv:2502.15300, \dodoi{10.48550/arXiv.2502.15300}

\bibitem[{{Li} \& {Lin}(2023)}]{Li2023}
{Li}, X., \& {Lin}, B. 2023, mnras, 521, 6354, \dodoi{10.1093/mnras/stad831}

\bibitem[{Li {et~al.}(2022)Li, Zeng, Wang, Du, Kong, \& Liao}]{Li2022}
Li, X., Zeng, S., Wang, Z., {et~al.} 2022, Monthly Notices of the Royal Astronomical Society, 514, 4588

\bibitem[{{Loshchilov} \& {Hutter}(2017)}]{Loshchilov2017}
{Loshchilov}, I., \& {Hutter}, F. 2017, arXiv e-prints, arXiv:1711.05101, \dodoi{10.48550/arXiv.1711.05101}

\bibitem[{Lu {et~al.}(2020)Lu, Angus, Agüeros, Blancato, Ness, Rowland, Curtis, \& Grunblatt}]{Lu_2020}
Lu, Y., Angus, R., Agüeros, M.~A., {et~al.} 2020, The Astronomical Journal, 160, 168, \dodoi{10.3847/1538-3881/abada4}

\bibitem[{Lu {et~al.}(2024)Lu, Angus, Foreman-Mackey, \& Hattori}]{lu_age_2024}
Lu, Y., Angus, R., Foreman-Mackey, D., \& Hattori, S. 2024, The Astronomical Journal, 167, 159, \dodoi{10.3847/1538-3881/ad28b9}

\bibitem[{Mamajek \& Hillenbrand(2008)}]{mamajek_improved_2008}
Mamajek, E.~E., \& Hillenbrand, L.~A. 2008, The Astrophysical Journal, 687, 1264, \dodoi{10.1086/591785}

\bibitem[{{Mathur} {et~al.}(2014){Mathur}, {Garc{\'\i}a}, {Ballot}, {Ceillier}, {Salabert}, {Metcalfe}, {R{\'e}gulo}, {Jim{\'e}nez}, \& {Bloemen}}]{Mathur2014}
{Mathur}, S., {Garc{\'\i}a}, R.~A., {Ballot}, J., {et~al.} 2014, aap, 562, A124, \dodoi{10.1051/0004-6361/201322707}

\bibitem[{Mathur {et~al.}(2017)Mathur, Huber, Batalha, Ciardi, Bastien, Bieryla, Buchhave, Cochran, Endl, Esquerdo, Furlan, Howard, Howell, Isaacson, Latham, MacQueen, \& Silva}]{Mathur_2017}
Mathur, S., Huber, D., Batalha, N.~M., {et~al.} 2017, The Astrophysical Journal Supplement Series, 229, 30, \dodoi{10.3847/1538-4365/229/2/30}

\bibitem[{{McInnes} {et~al.}(2018){McInnes}, {Healy}, \& {Melville}}]{McInnes2018}
{McInnes}, L., {Healy}, J., \& {Melville}, J. 2018, arXiv e-prints, arXiv:1802.03426, \dodoi{10.48550/arXiv.1802.03426}

\bibitem[{{McQuillan} {et~al.}(2014){McQuillan}, {Mazeh}, \& {Aigrain}}]{McQuillan2014}
{McQuillan}, A., {Mazeh}, T., \& {Aigrain}, S. 2014, apjs, 211, 24, \dodoi{10.1088/0067-0049/211/2/24}

\bibitem[{{Morvan} {et~al.}(2022){Morvan}, {Nikolaou}, {Yip}, \& {Waldmann}}]{Morvan2022}
{Morvan}, M., {Nikolaou}, N., {Yip}, K., \& {Waldmann}, I. 2022, in Machine Learning for Astrophysics, 11, \dodoi{10.48550/arXiv.2207.02777}

\bibitem[{Olney {et~al.}(2020)Olney, Kounkel, Schillinger, Scoggins, Yin, Howard, Covey, Hutchinson, \& Stassun}]{Olney_2020}
Olney, R., Kounkel, M., Schillinger, C., {et~al.} 2020, The Astronomical Journal, 159, 182, \dodoi{10.3847/1538-3881/ab7a97}

\bibitem[{{Pan} {et~al.}(2024{\natexlab{a}}){Pan}, {Ting}, {Huang}, {Yu}, \& {Liu}}]{Pan2024_scaling}
{Pan}, J.-S., {Ting}, Y.-S., {Huang}, Y., {Yu}, J., \& {Liu}, J.-F. 2024{\natexlab{a}}, arXiv e-prints, arXiv:2405.17156, \dodoi{10.48550/arXiv.2405.17156}

\bibitem[{{Pan} {et~al.}(2024{\natexlab{b}}){Pan}, {Ting}, \& {Yu}}]{Pan2024}
{Pan}, J.-S., {Ting}, Y.-S., \& {Yu}, J. 2024{\natexlab{b}}, mnras, 528, 5890, \dodoi{10.1093/mnras/stae068}

\bibitem[{{Parker} {et~al.}(2024){Parker}, {Lanusse}, {Golkar}, {Sarra}, {Cranmer}, {Bietti}, {Eickenberg}, {Krawezik}, {McCabe}, {Morel}, {Ohana}, {Pettee}, {R{\'e}galdo-Saint Blancard}, {Cho}, {Ho}, \& {Polymathic AI Collaboration}}]{Parker2024}
{Parker}, L., {Lanusse}, F., {Golkar}, S., {et~al.} 2024, mnras, 531, 4990, \dodoi{10.1093/mnras/stae1450}

\bibitem[{{Radford} {et~al.}(2021){Radford}, {Kim}, {Hallacy}, {Ramesh}, {Goh}, {Agarwal}, {Sastry}, {Askell}, {Mishkin}, {Clark}, {Krueger}, \& {Sutskever}}]{Radford2021}
{Radford}, A., {Kim}, J.~W., {Hallacy}, C., {et~al.} 2021, arXiv e-prints, arXiv:2103.00020, \dodoi{10.48550/arXiv.2103.00020}

\bibitem[{{Raghavan} {et~al.}(2010){Raghavan}, {McAlister}, {Henry}, {Latham}, {Marcy}, {Mason}, {Gies}, {White}, \& {ten Brummelaar}}]{Raghavan2010}
{Raghavan}, D., {McAlister}, H.~A., {Henry}, T.~J., {et~al.} 2010, \apjs, 190, 1, \dodoi{10.1088/0067-0049/190/1/1}

\bibitem[{{Reinhold} {et~al.}(2013){Reinhold}, {Reiners}, \& {Basri}}]{Reinhold2013}
{Reinhold}, T., {Reiners}, A., \& {Basri}, G. 2013, aap, 560, A4, \dodoi{10.1051/0004-6361/201321970}

\bibitem[{{Reinhold} {et~al.}(2023){Reinhold}, {Shapiro}, {Solanki}, \& {Basri}}]{Reinhold2023}
{Reinhold}, T., {Shapiro}, A.~I., {Solanki}, S.~K., \& {Basri}, G. 2023, aap, 678, A24, \dodoi{10.1051/0004-6361/202346789}

\bibitem[{{Ricker} {et~al.}(2014){Ricker}, {Winn}, {Vanderspek}, {Latham}, {Bakos}, {Bean}, {Berta-Thompson}, {Brown}, {Buchhave}, {Butler}, {Butler}, {Chaplin}, {Charbonneau}, {Christensen-Dalsgaard}, {Clampin}, {Deming}, {Doty}, {De Lee}, {Dressing}, {Dunham}, {Endl}, {Fressin}, {Ge}, {Henning}, {Holman}, {Howard}, {Ida}, {Jenkins}, {Jernigan}, {Johnson}, {Kaltenegger}, {Kawai}, {Kjeldsen}, {Laughlin}, {Levine}, {Lin}, {Lissauer}, {MacQueen}, {Marcy}, {McCullough}, {Morton}, {Narita}, {Paegert}, {Palle}, {Pepe}, {Pepper}, {Quirrenbach}, {Rinehart}, {Sasselov}, {Sato}, {Seager}, {Sozzetti}, {Stassun}, {Sullivan}, {Szentgyorgyi}, {Torres}, {Udry}, \& {Villasenor}}]{tess2014}
{Ricker}, G.~R., {Winn}, J.~N., {Vanderspek}, R., {et~al.} 2014, in Society of Photo-Optical Instrumentation Engineers (SPIE) Conference Series, Vol. 9143, Space Telescopes and Instrumentation 2014: Optical, Infrared, and Millimeter Wave, ed. J.~M. {Oschmann}, Jr., M.~{Clampin}, G.~G. {Fazio}, \& H.~A. {MacEwen}, 914320, \dodoi{10.1117/12.2063489}

\bibitem[{{Rizhko} \& {Bloom}(2025)}]{Rizhko2025}
{Rizhko}, M., \& {Bloom}, J.~S. 2025, \aj, 170, 28, \dodoi{10.3847/1538-3881/adcbad}

\bibitem[{{Romano} {et~al.}(2019){Romano}, {Patterson}, \& {Cand{\`e}s}}]{Romano2019}
{Romano}, Y., {Patterson}, E., \& {Cand{\`e}s}, E.~J. 2019, arXiv e-prints, arXiv:1905.03222, \dodoi{10.48550/arXiv.1905.03222}

\bibitem[{{Santos} {et~al.}(2021){Santos}, {Breton}, {Mathur}, \& {Garc{\'\i}a}}]{Santos2021}
{Santos}, A.~R.~G., {Breton}, S.~N., {Mathur}, S., \& {Garc{\'\i}a}, R.~A. 2021, apjs, 255, 17, \dodoi{10.3847/1538-4365/ac033f}

\bibitem[{{Santos} {et~al.}(2019){Santos}, {Garc{\'\i}a}, {Mathur}, {Bugnet}, {van Saders}, {Metcalfe}, {Simonian}, \& {Pinsonneault}}]{Santos2019}
{Santos}, A.~R.~G., {Garc{\'\i}a}, R.~A., {Mathur}, S., {et~al.} 2019, apjs, 244, 21, \dodoi{10.3847/1538-4365/ab3b56}

\bibitem[{{Santos} {et~al.}(2024){Santos}, {Godoy-Rivera}, {Finley}, {Mathur}, {Garc{\'\i}a}, {Breton}, \& {Broomhall}}]{Santos2024}
{Santos}, {\^A}. R.~G., {Godoy-Rivera}, D., {Finley}, A.~J., {et~al.} 2024, Frontiers in Astronomy and Space Sciences, 11, 1356379, \dodoi{10.3389/fspas.2024.1356379}

\bibitem[{{Skumanich}(1972)}]{Skumanich1972}
{Skumanich}, A. 1972, \apj, 171, 565, \dodoi{10.1086/151310}

\bibitem[{{Su} {et~al.}(2021){Su}, {Lu}, {Pan}, {Murtadha}, {Wen}, \& {Liu}}]{Su2021}
{Su}, J., {Lu}, Y., {Pan}, S., {et~al.} 2021, arXiv e-prints, arXiv:2104.09864, \dodoi{10.48550/arXiv.2104.09864}

\bibitem[{{Walmsley} {et~al.}(2022){Walmsley}, {Slijepcevic}, {Bowles}, \& {Scaife}}]{Walmsley2022}
{Walmsley}, M., {Slijepcevic}, I., {Bowles}, M.~R., \& {Scaife}, A. 2022, in Machine Learning for Astrophysics, 29, \dodoi{10.48550/arXiv.2206.11927}

\bibitem[{{Walmsley} {et~al.}(2024){Walmsley}, {Bowles}, {Scaife}, {Shingirai Makechemu}, {Gordon}, {Ferguson}, {Mann}, {Pearson}, {Popp}, {Bovy}, {Speagle}, {Dickinson}, {Fortson}, {G{\'e}ron}, {Kruk}, {Lintott}, {Mantha}, {Mohan}, {O'Ryan}, \& {Slijepevic}}]{Walmsley2024}
{Walmsley}, M., {Bowles}, M., {Scaife}, A. M.~M., {et~al.} 2024, arXiv e-prints, arXiv:2404.02973, \dodoi{10.48550/arXiv.2404.02973}

\bibitem[{{Wang} {et~al.}(2022){Wang}, {Huang}, {Yuan}, {Zhang}, {Xiang}, \& {Liu}}]{Wang2022}
{Wang}, C., {Huang}, Y., {Yuan}, H., {et~al.} 2022, apjs, 259, 51, \dodoi{10.3847/1538-4365/ac4df7}

\bibitem[{{Wu} {et~al.}(2014){Wu}, {Du}, {Luo}, {Zhao}, \& {Yuan}}]{Wu2014}
{Wu}, Y., {Du}, B., {Luo}, A., {Zhao}, Y., \& {Yuan}, H. 2014, in IAU Symposium, Vol. 306, Statistical Challenges in 21st Century Cosmology, ed. A.~{Heavens}, J.-L. {Starck}, \& A.~{Krone-Martins}, 340--342, \dodoi{10.1017/S1743921314010825}

\bibitem[{{Zbontar} {et~al.}(2021){Zbontar}, {Jing}, {Misra}, {LeCun}, \& {Deny}}]{Zbontar2021}
{Zbontar}, J., {Jing}, L., {Misra}, I., {LeCun}, Y., \& {Deny}, S. 2021, arXiv e-prints, arXiv:2103.03230, \dodoi{10.48550/arXiv.2103.03230}

\bibitem[{Zhang {et~al.}(2024)Zhang, Helfer, Gagliano, Mishra-Sharma, \& Ashley~Villar}]{Zhang_2024}
Zhang, G., Helfer, T., Gagliano, A.~T., Mishra-Sharma, S., \& Ashley~Villar, V. 2024, Machine Learning: Science and Technology, 5, 045069, \dodoi{10.1088/2632-2153/ad990d}

\bibitem[{{Zhang} {et~al.}(2021){Zhang}, {Wu}, {Yan}, {Wipf}, \& {Yu}}]{zhang2021}
{Zhang}, H., {Wu}, Q., {Yan}, J., {Wipf}, D., \& {Yu}, P.~S. 2021, arXiv e-prints, arXiv:2106.12484, \dodoi{10.48550/arXiv.2106.12484}

\bibitem[{{Zhao} {et~al.}(2012){Zhao}, {Zhao}, {Chu}, {Jing}, \& {Deng}}]{Zhao2012}
{Zhao}, G., {Zhao}, Y.-H., {Chu}, Y.-Q., {Jing}, Y.-P., \& {Deng}, L.-C. 2012, Research in Astronomy and Astrophysics, 12, 723, \dodoi{10.1088/1674-4527/12/7/002}

\bibitem[{{Zuo} {et~al.}(2025){Zuo}, {Tao}, {Huang}, {Kang}, {Chen}, {Cui}, {Pan}, {Kong}, {Tang}, {Han}, {Mu}, {Xu}, {Fan}, {Xue}, {Luo}, \& {Liu}}]{Zuo2025}
{Zuo}, X., {Tao}, Y., {Huang}, Y., {et~al.} 2025, arXiv e-prints, arXiv:2504.20290, \dodoi{10.48550/arXiv.2504.20290}

\end{thebibliography}

\clearpage

\appendix

\section{Supplementary graphs}\label{appendix:sup_graphs}

\begin{figure}[H]
    \centering
    \begin{minipage}[b]{0.33\textwidth}    
        \centering
        \includegraphics[width=\textwidth]{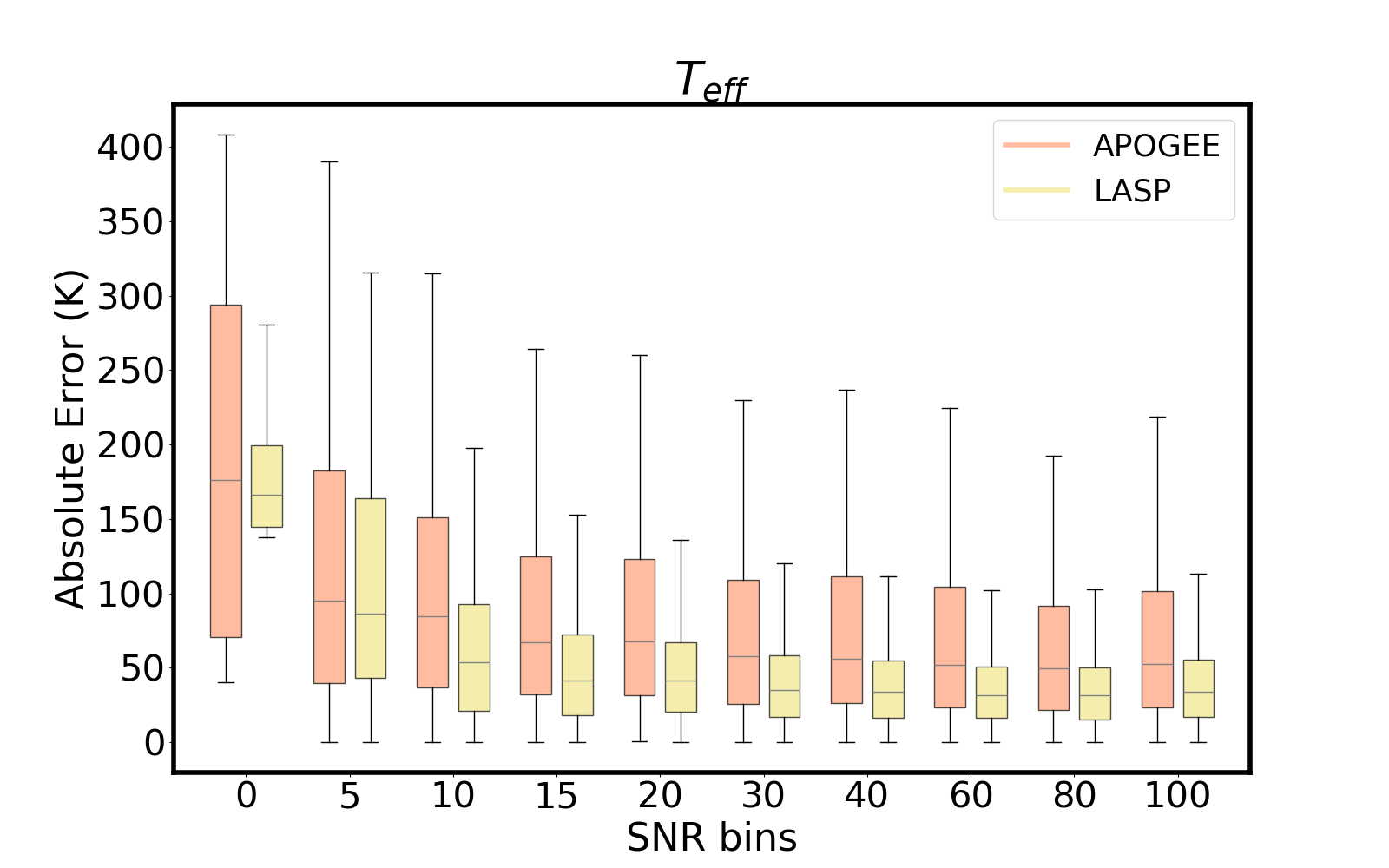}
    \end{minipage}
    \begin{minipage}[b]{0.33\textwidth}    
        \centering
        \includegraphics[width=\textwidth]{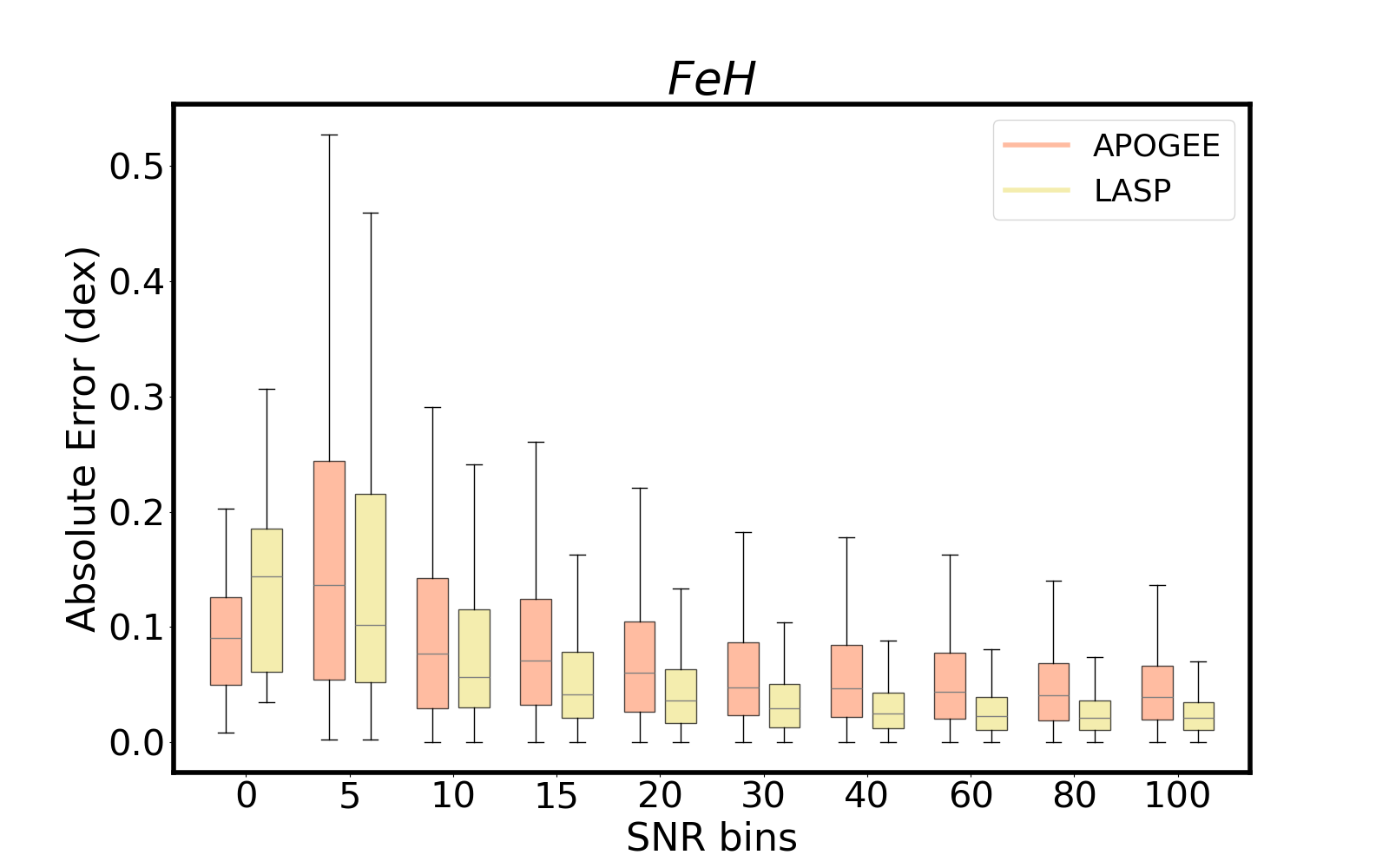}
    \end{minipage}
    \begin{minipage}[b]{0.33\textwidth}    
        \centering
        \includegraphics[width=\textwidth]{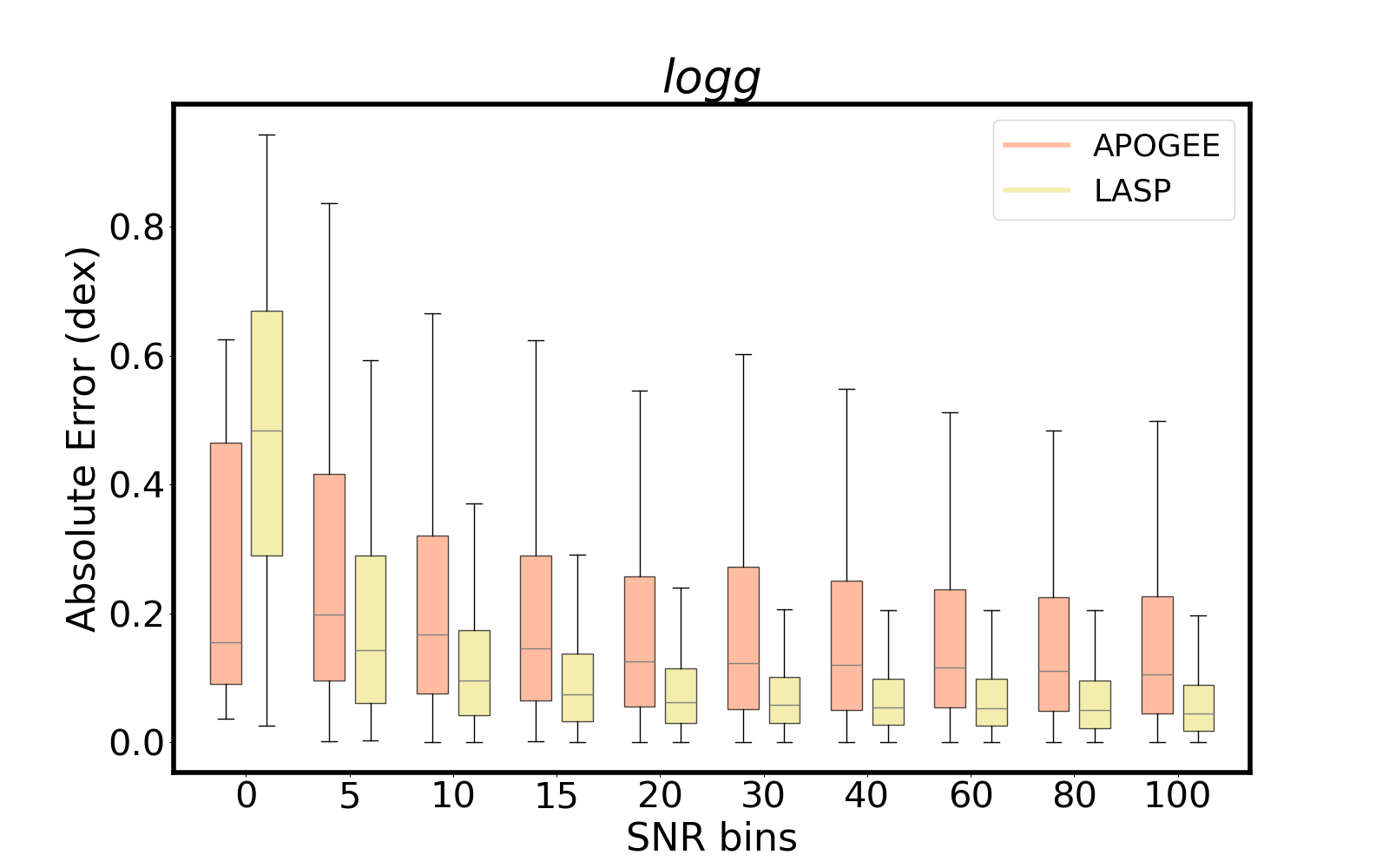}
    \end{minipage}
    \caption{Box plots of MAE vs SNR for APOGEE test set and LAMOST test set.}
    \label{fig:snr_vs_error}
\end{figure}

\begin{figure}[H]
    \centering
    
    \includegraphics[width=0.6\textwidth,trim=10 10 10 10,clip]{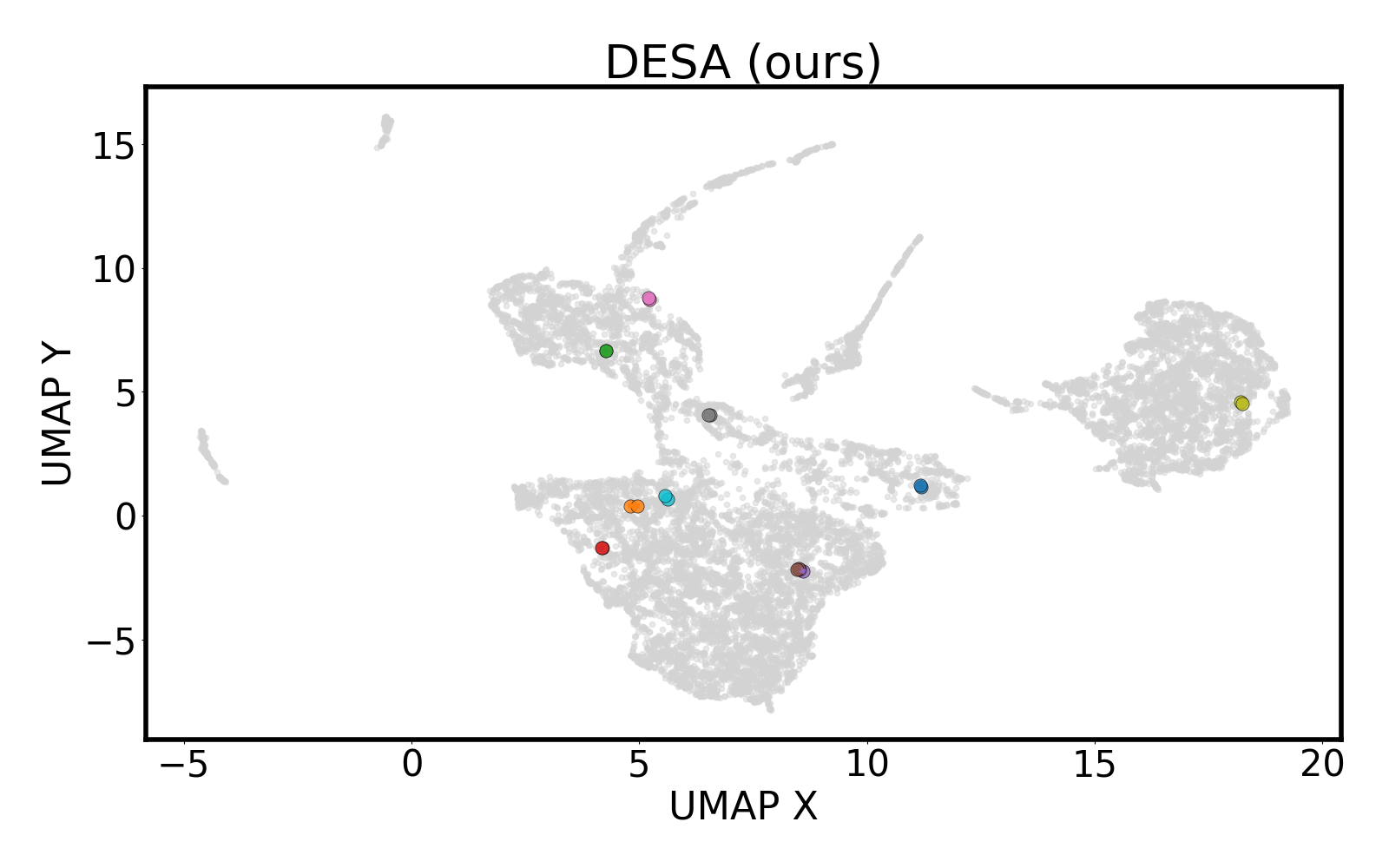}
    
    \vspace{10pt} 
    
    \setlength{\tabcolsep}{0pt}
    \renewcommand{\arraystretch}{0}
    
    \begin{tabular}{@{}cc@{}}
        \includegraphics[width=0.45\textwidth,trim=10 10 10 10,clip]{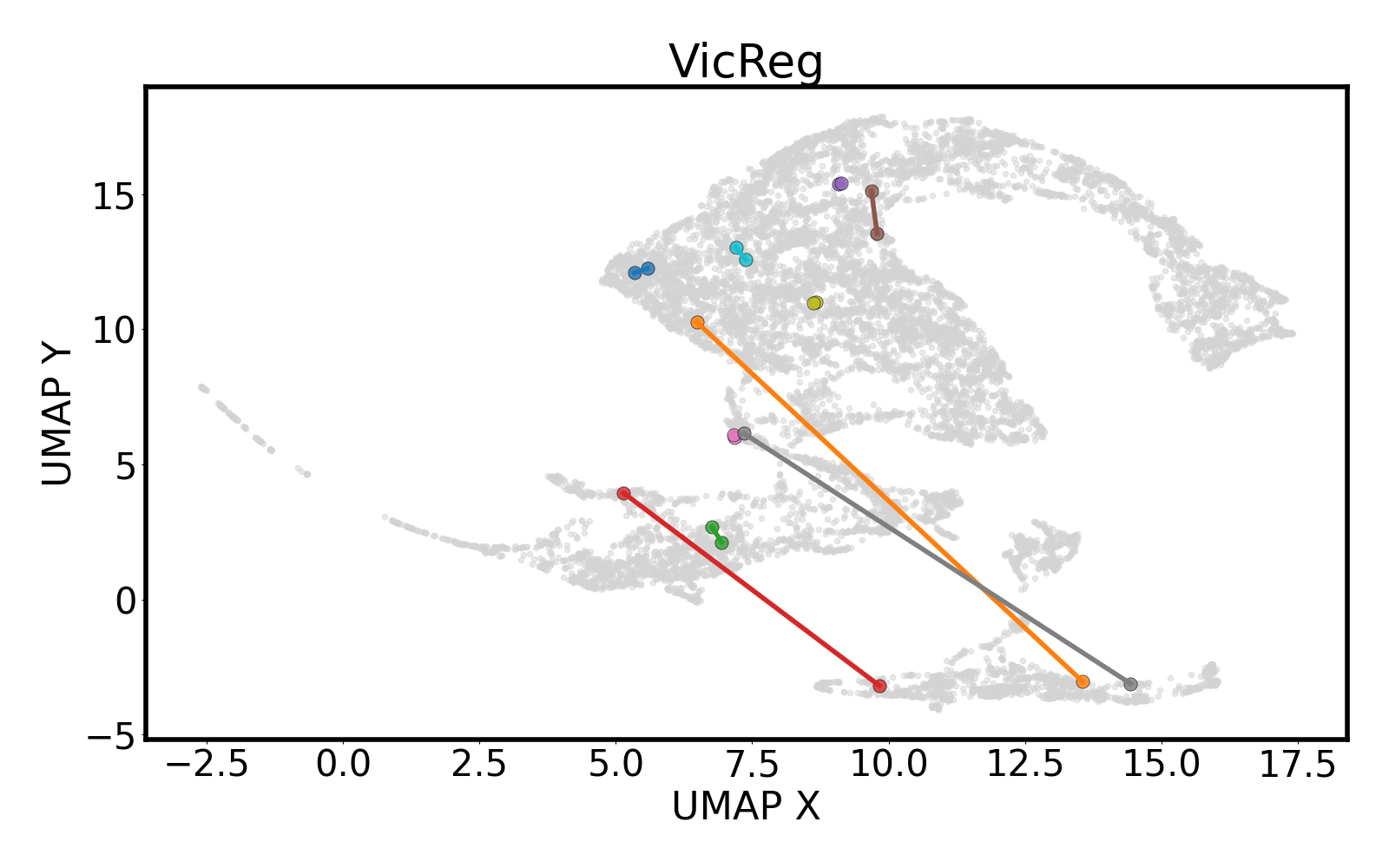} &
        \includegraphics[width=0.45\textwidth,trim=10 10 10 10,clip]{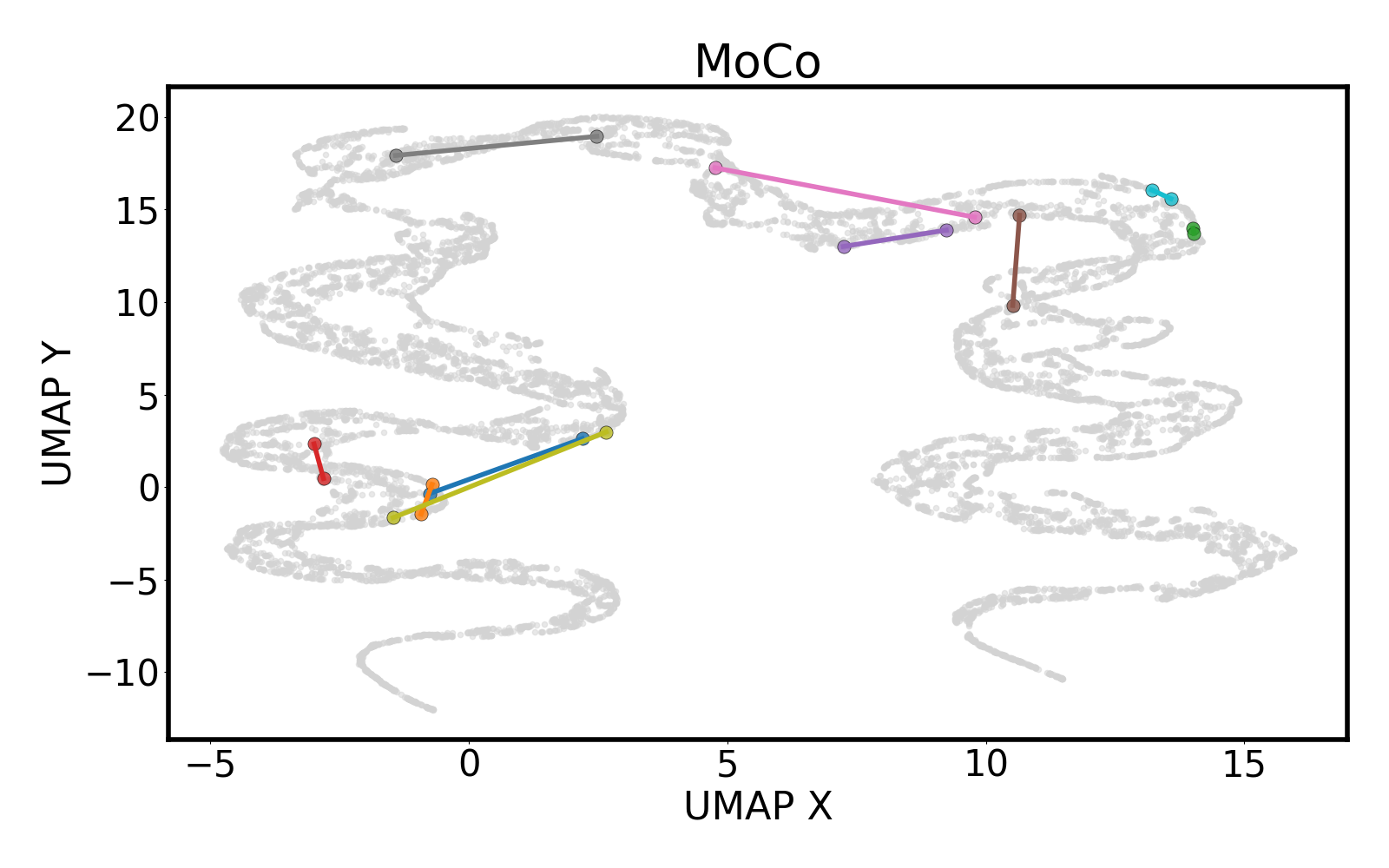} \\[5pt]
        \includegraphics[width=0.45\textwidth,trim=10 10 10 10,clip]{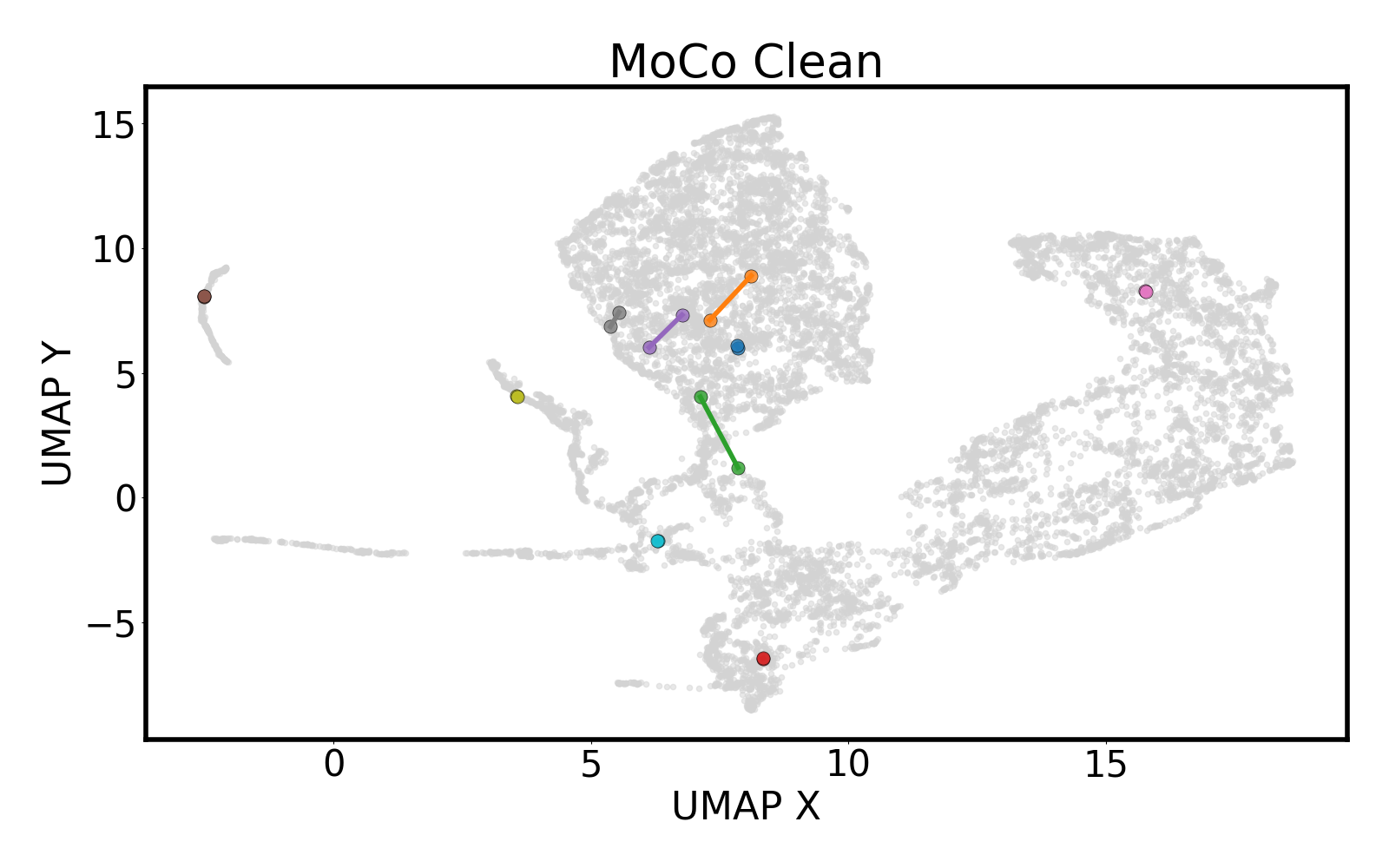} &
        \includegraphics[width=0.45\textwidth,trim=10 10 10 10,clip]{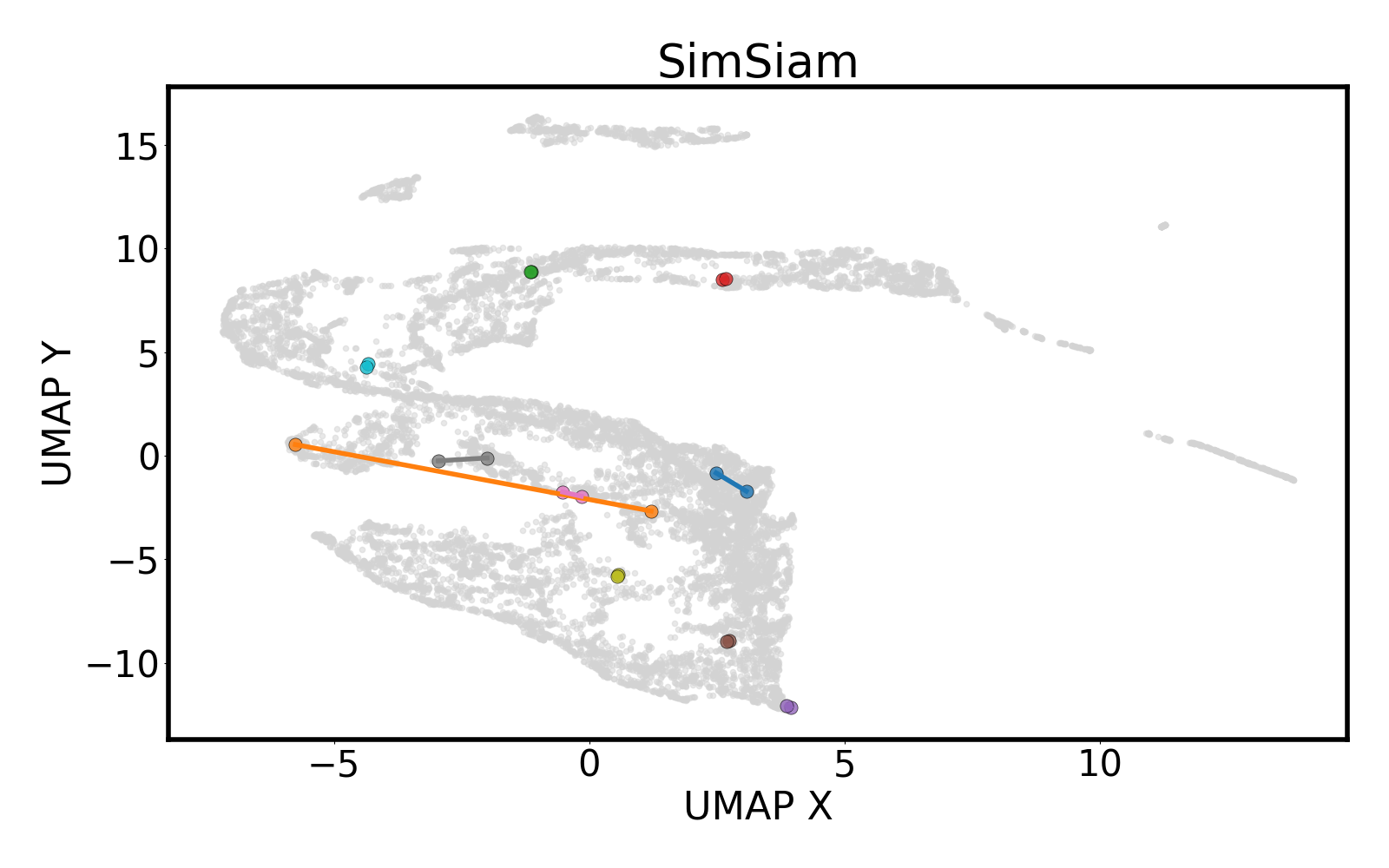}
    \end{tabular}
    
    \caption{UMAP of all models. All points are in shaded gray, and 10 stars with pairs of spectra are colored on top. For each pair, all samples are colored with the same color and connected by a line. We can see the visual difference between DESA (upper figure) and all other models -- in DESA, pairs are arranged very close to each other, implying that they have very similar embeddings. In other models, there are samples of the same star that are very distant, which implies very different embeddings.}
\end{figure}

\begin{figure}[H]
    \centering
        \includegraphics[width=0.6\textwidth]{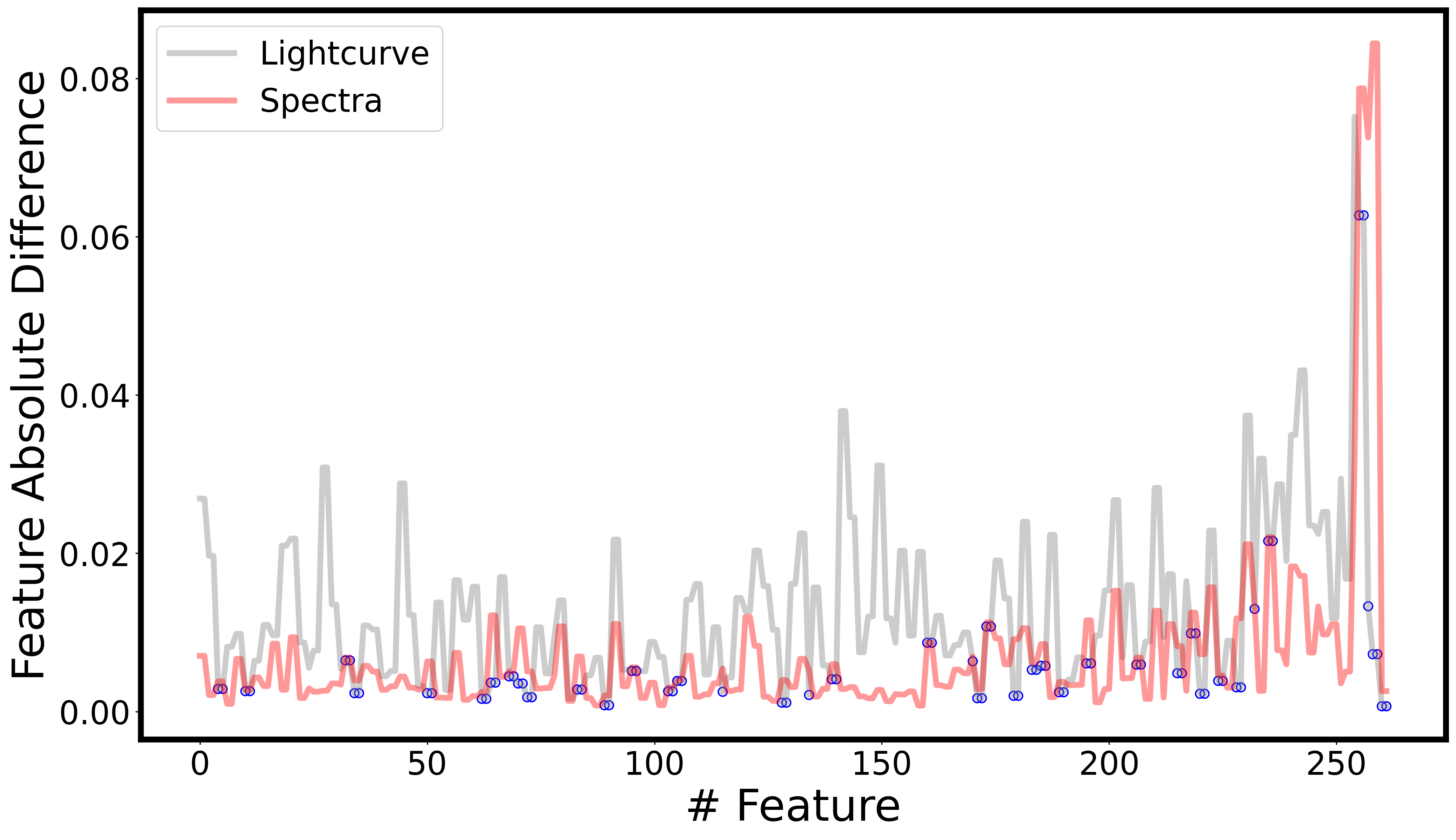}
    \caption{Feature sensitivity study. The x-axis shows the feature index in the final features of DESA on the test set. The y-axis shows the difference between the original features and features that were generated with only spectral input (red) or only light curve input (gray). The blue circles mark indices where the lightcurve-only features are closer to the original features compared to spectra-only features.}
    \label{fig:ablation_features}
\end{figure}

\begin{figure}[H]
    \centering
        \includegraphics[width=0.6\textwidth]{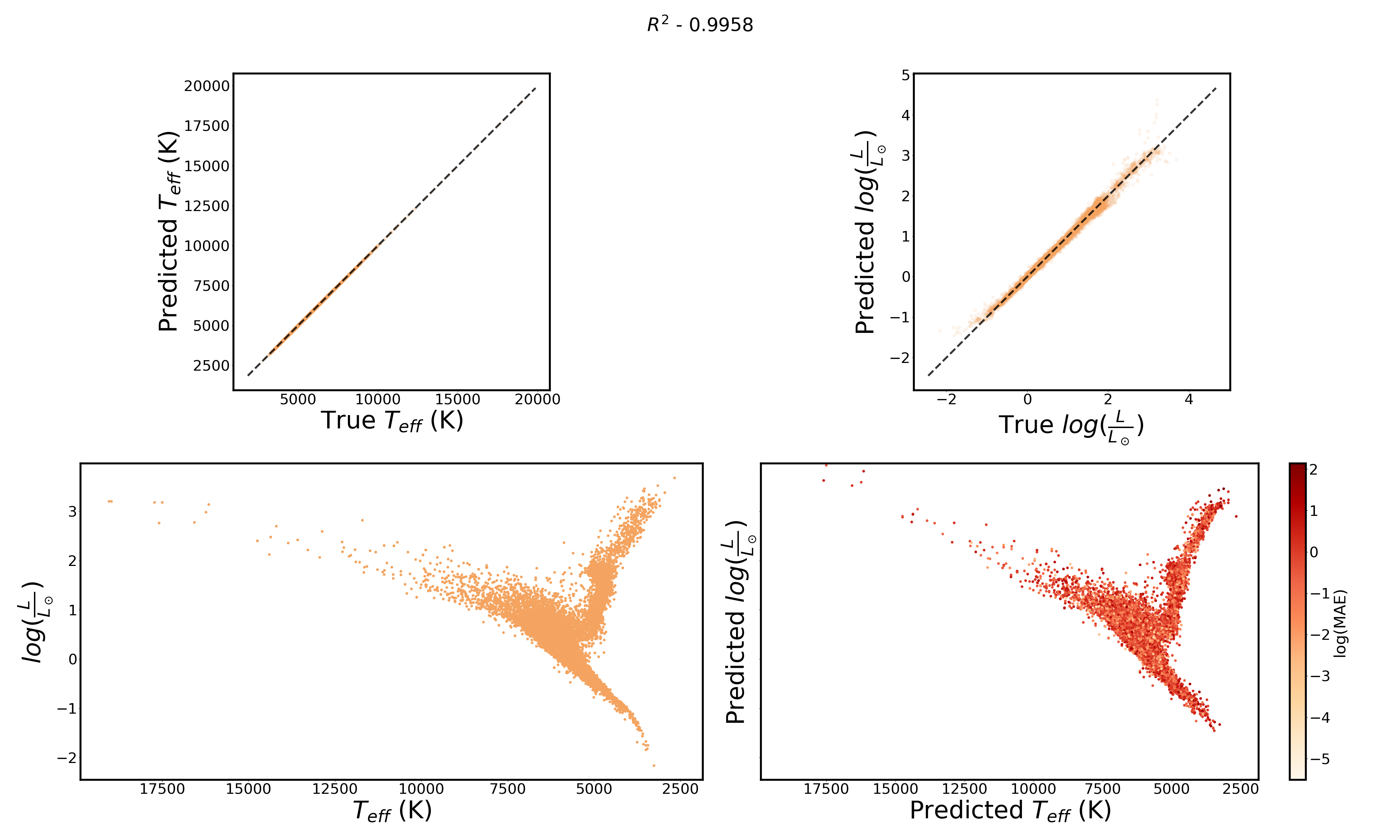}
    \caption{Upper panel - few-shot learning of $T_\mathrm{eff}$ and $\log(L / L_{\odot})$ using linear regression on $20\%$ of the test set. Lower panel -- the resulting HR diagram (right) and true HR diagram (left). Colors proportional to the mean average error (MAE).}
    \label{fig:diagram_hr}
\end{figure}

\begin{figure}[H]
    \centering
        \includegraphics[width=0.6\textwidth]{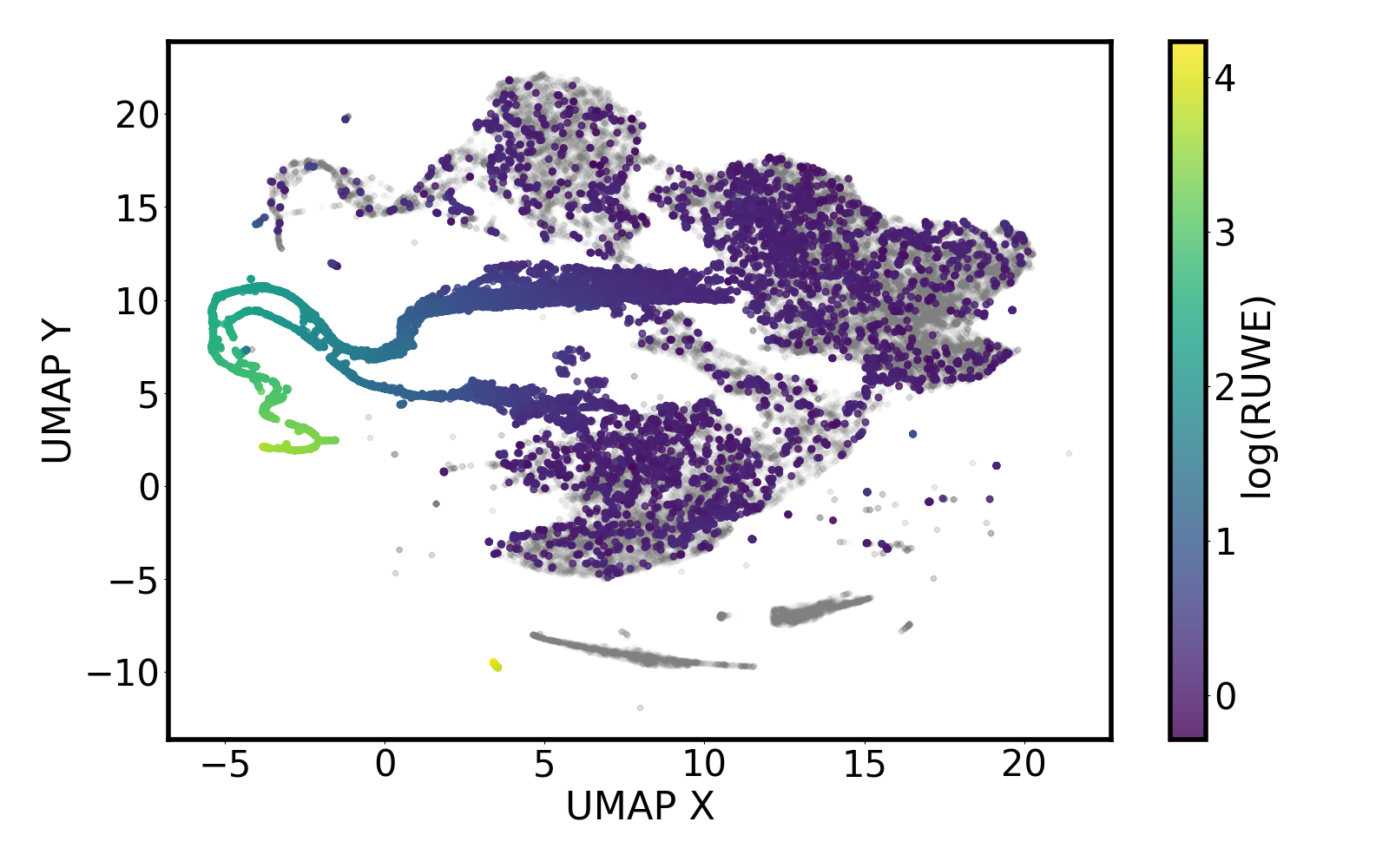}
    \caption{UMAP of DESA final embeddings. All points are in gray, and points with RUWE $> 1.4$ are colored according to their $RUWE$ value.}
    \label{fig:umap_ruwe}
\end{figure}

\begin{figure}[H]
    \centering
    \setlength{\tabcolsep}{0pt}
    \renewcommand{\arraystretch}{0}
    
    \begin{tabular}{@{}ccc@{}}
        \includegraphics[width=0.333\textwidth,trim=10 10 10 10,clip]{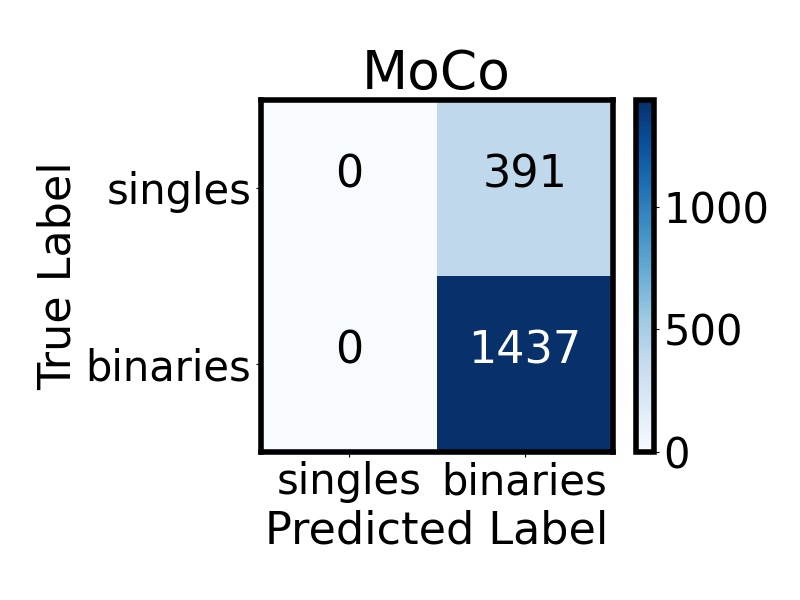} &
        \includegraphics[width=0.333\textwidth,trim=10 10 10 10,clip]{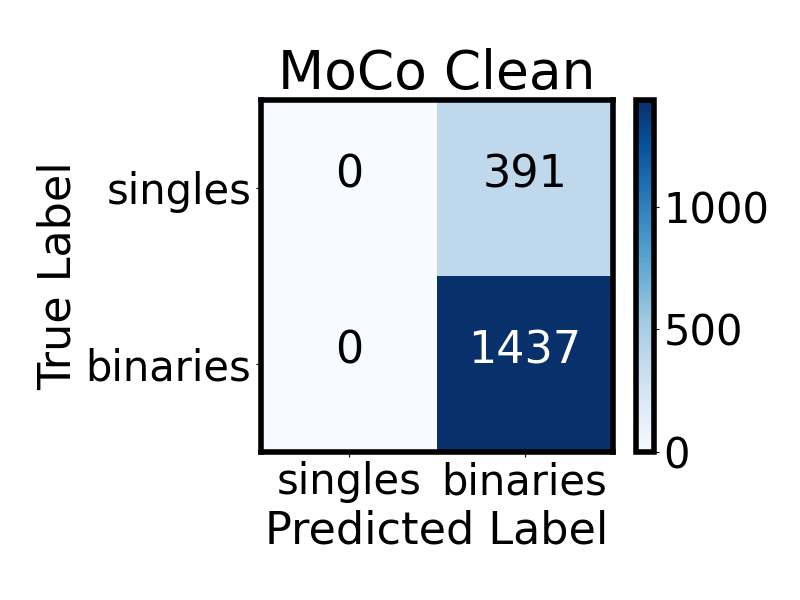} &
        \includegraphics[width=0.333\textwidth,trim=10 10 10 10,clip]{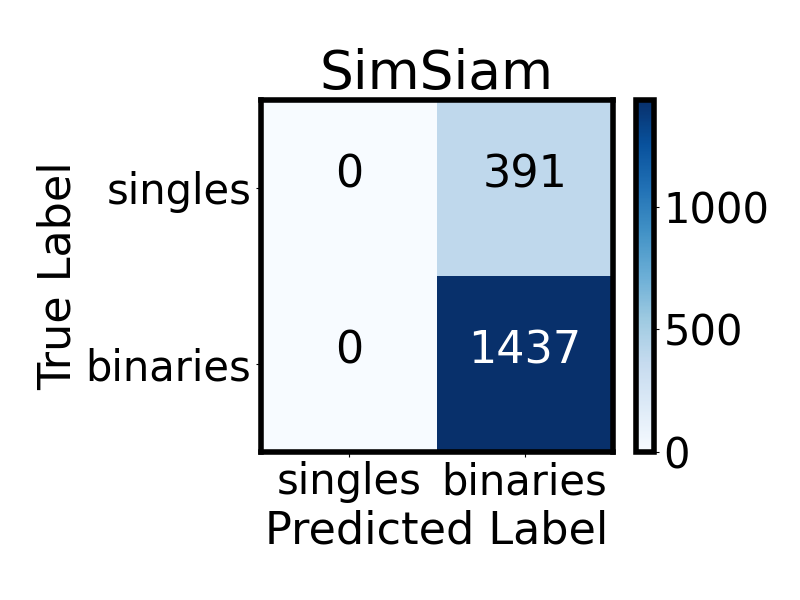} \\[0pt]
        \includegraphics[width=0.333\textwidth,trim=10 10 10 10,clip]{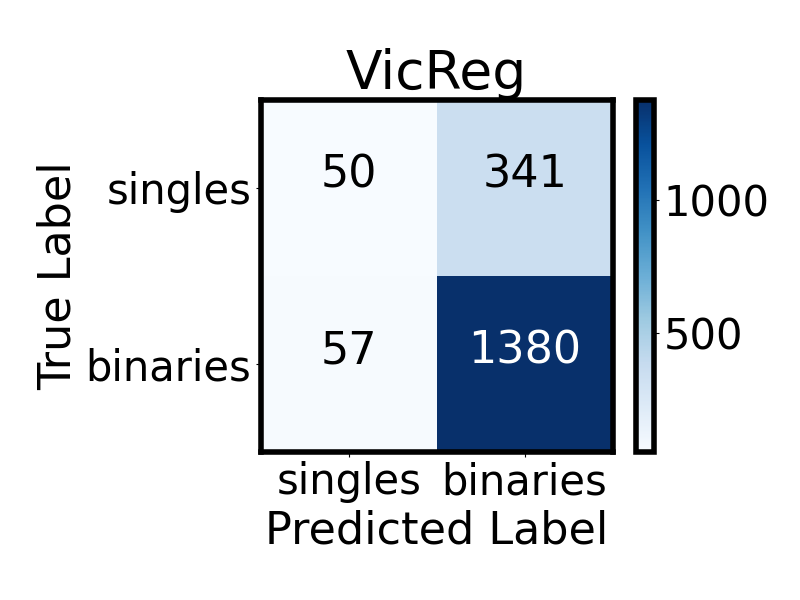} &
        \includegraphics[width=0.333\textwidth,trim=10 10 10 10,clip]{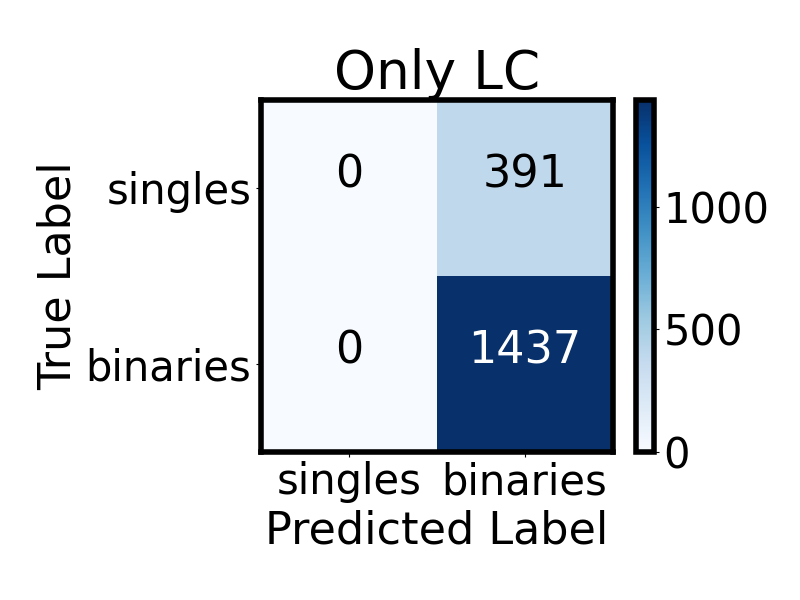} &
        \includegraphics[width=0.333\textwidth,trim=10 10 10 10,clip]{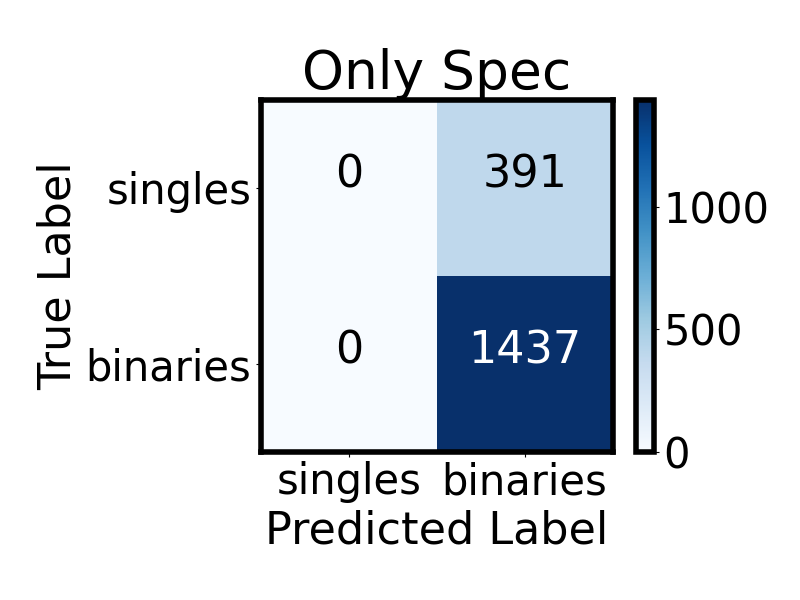}
    \end{tabular}
    
    \caption{Confusion matrix of alternative models on binary prediction task.}
\end{figure} \label{fig:appendix_confusion}

\begin{figure}[H]
    \centering
    \begin{minipage}[b]{0.5\textwidth}
        \centering
        \includegraphics[width=\textwidth]{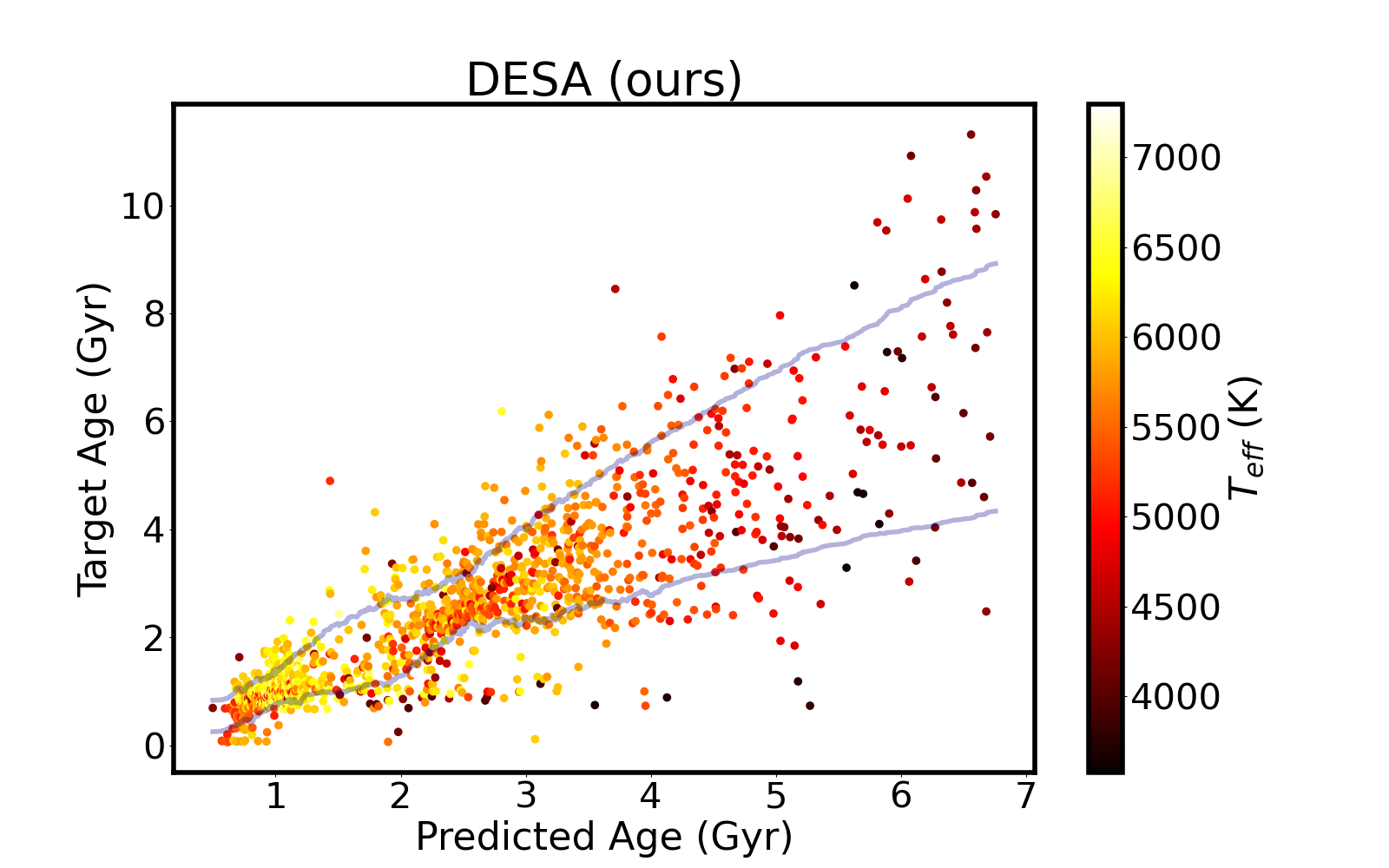}
    \end{minipage}
    \hfill
    \begin{minipage}[b]{0.7\textwidth}
        \centering
        \includegraphics[width=\textwidth]{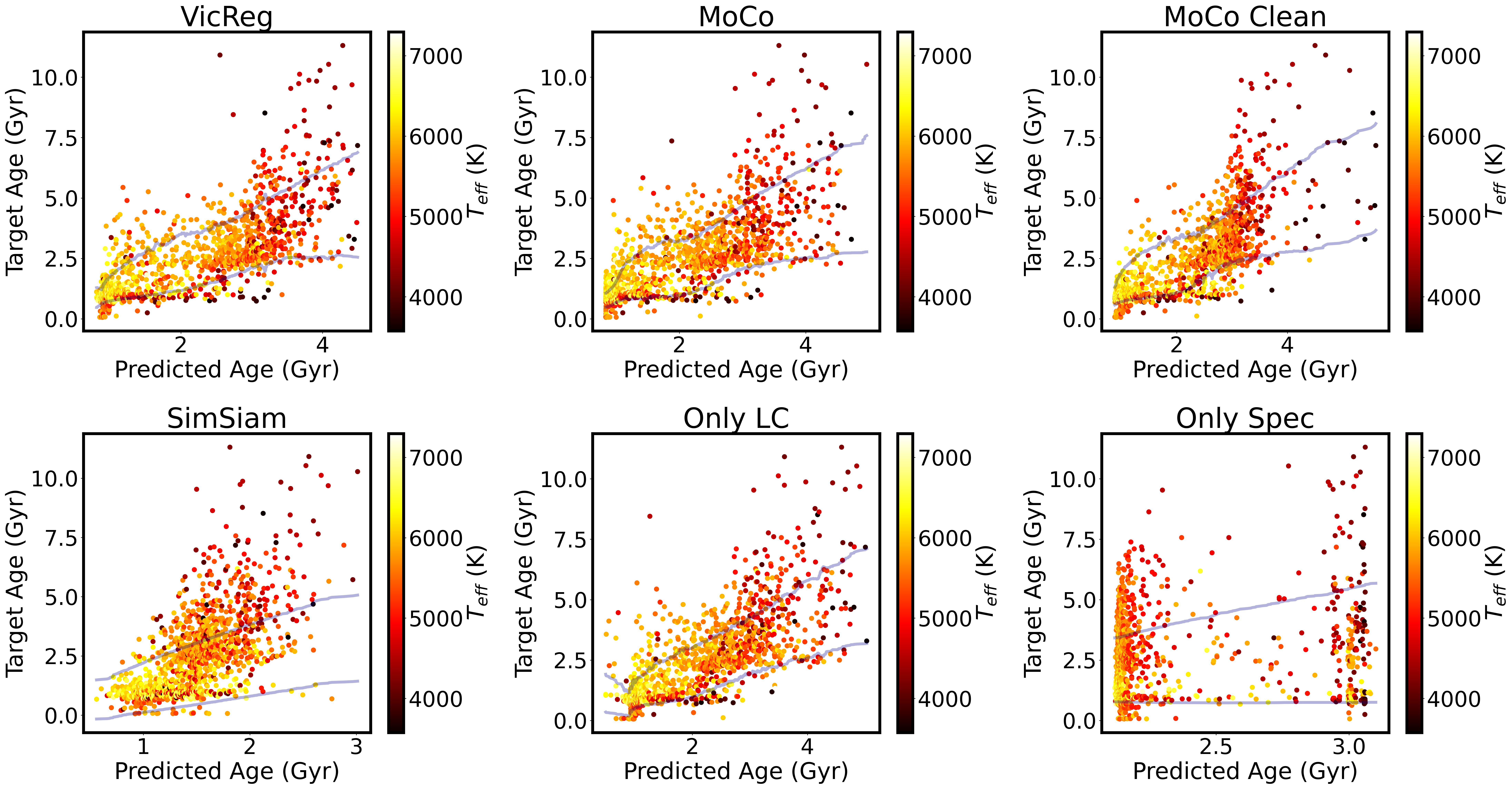}
    \end{minipage}
    \caption{Results of age prediction. The dark blue lines represent a $72\%$ confidence interval. Colors represent $T_\mathrm{eff}$.}
    \label{fig:age_res}
\end{figure}

\end{document}